\documentclass[11pt]{article}
\pdfoutput=1
\usepackage{jheppub}
\usepackage[usenames,dvipsnames,table]{xcolor}
\usepackage{graphicx,amsmath,amssymb,amsthm,multirow,array}
\usepackage{subfigure}

\usepackage[numbers,sort&compress]{natbib}
\usepackage{ccaption}
  \captionnamefont{\bfseries}
  \captiontitlefont{\small\sffamily}
  \captiondelim{: }
  \hangcaption
  \renewcommand{\figurename}{Fig.}

\newcount\xrefpos \xrefpos=0
\newcount\yrefpos \yrefpos=0
\newcount\xput \xput=0
\newcount\yput \yput=0

\def\refpos#1 #2 #3{\global\xrefpos=#1 \global\yrefpos=#2
                         \rlap{$\smash{#3}$}}
\def\put #1 #2 #3{\xput=#1 \yput=#2
                  \advance\xput by -\xrefpos
                  \advance\yput by -\yrefpos
                  \rlap{\kern\the\xput truebp
                        \vbox to 0pt{\vss\hbox{$\displaystyle #3$}
                        \kern\the\yput truebp}}}
\def\beginlabels\refpos#1\endlabels{\hbox{$\refpos#1$}}

\definecolor{rust}{rgb}{0.8,0.2,0.2} 
\definecolor{purple}{rgb}{0.8,0.1,0.9}

\def\AdS#1{AdS$_{#1}$}
\title{Spatial Modulation and Conductivities in Effective Holographic Theories}

\author[a]{Mukund Rangamani}
\author[b]{\!\!, Moshe Rozali}
\author[b]{\!\!, Darren Smyth}

\affiliation[\,a]{Centre for Particle Theory \& Department of Mathematical Sciences,\\
Durham University, South Road, Durham DH1 3LE, UK.}
\affiliation[\,b]{Department of Physics and Astronomy, \& University of British Columbia,\\
Vancouver, BC V6T 1Z1, Canada.}
\emailAdd{mukund.rangamani@durham.ac.uk}
\emailAdd{rozali@phas.ubc.ca}
\emailAdd{dsmyth@phas.ubc.ca}

\abstract
{We analyze a class of bottom-up holographic models for low energy thermo-electric transport. The models we focus on belong to a family of Einstein-Maxwell-dilaton theories parameterized by two scalar functions, characterizing the dilaton self-interaction and the gauge coupling function. We impose spatially inhomogeneous lattice boundary conditions for the dilaton on the AdS boundary and study the resulting phase structure attained at low energies. We find that as we dial the scalar functions at our disposal (changing thus the theory under consideration), we obtain either (i) coherent metallic, or (ii) insulating, or (iii) incoherent metallic phases. We chart out the domain where the incoherent metals appear in a restricted parameter space of theories. We also analyze the optical conductivity, noting that non-trivial scaling behaviour at intermediate frequencies appears to only be possible for very narrow regions of parameter space. }

\preprint{DCPT-15/29}

\begin{document}

\maketitle

\section{Introduction and Outline}

The holographic AdS/CFT correspondence provides a valuable theoretical framework for exploring the dynamics of strongly coupled field theories. In recent years this approach has been found to be quite useful in building qualitative models for a variety of phenomena encountered in many-body physics, ranging from superconductors and superfluids, to non-Fermi liquids \cite{Hartnoll:2009sz, Herzog:2009xv, Iqbal:2011ae}.  The general thrust of the initial discussion has been to consider relativistic quantum field theories which are deformed by relevant operators to access non-trivial phases in the infra-red (IR). In interest of computational simplicity, much of the discussion was focused on situations where spatial homogeneity is retained. 

Given the progress in these homogeneous models it is natural to consider less symmetric systems as a means to accessing more realistic scenarios.  For instance, consider the linear response functions which may be efficiently computed in the holographic context. In the absence of dissipation the low frequency behaviour of correlation functions often display IR divergences.  While some of these can be cured by examining the theory at finite temperature (effectively using a thermal IR regulator), there are some issues that are incurable by this simple approach. 

A simple case in point is the physics of electrical conductivity. Naively, we might try to compute this quantity in a translationally invariant system by monitoring the response to turning on an external, time dependent electric field. Using the Kronig-Kramers relations one would then find that momentum conservation results in a divergent zero frequency value of the imaginary part of the conductivity. As a consequence recent efforts have been focused on constructing  models with more realistic conductivity behaviour. This in turn is motivated by the need to better understand the experimental observation of metal/incoherent-metal/insulator transitions in the phase diagram of high temperature superconductor materials \cite{2011RvMP...83..471B}.\footnote{ A nice summary of the issues involved in characterizing such behaviour in many-body systems can be found in \cite{Hartnoll:2015aa}.}  

The first models for realistic behaviour of (thermo)-electrical conductivities in holography was accomplished in 
\cite{Horowitz:2012ky,Horowitz:2012gs}. These authors considered situations where the spatial homogeneity is explicitly broken by a background lattice of sources,  and studied thermo-electric transport in the resulting holographic dual background. By use of the lattice, they were able to demonstrate that the low frequency conductivity was finite, and well fitted in their set-up by the Drude form. More curiously, their analysis revealed a mid-range scaling of the AC conductivity with frequency, which furthermore agreed with experimental results in cuprate systems \cite{van2006scaling}. 

Since this seminal work, various groups have attempted to understand the thermo-electric properties of holographic systems. One of the key aspects to understanding this phenomenon is to ascertain the potential low energy behaviour of strongly coupled systems subjected to inhomogeneous sources for relevant operators. A classification of emergent IR phases crucially provides one with a picture of what to expect for the transport properties of the system at low energies.  Extensive work has already been carried out in characterizing the possible IR geometries which may emerge when translational invariance is relaxed; see for example \cite{Donos:2012js, Davison:2014lua} where the IR flow of perturbative inhomogeneous modes was analyzed. We now have a reasonable understanding for the spectrum of possibilities that can occur in holographic systems, with low energy behaviour ranging from metallic (with various dressing) to insulating. Part of the motivation for the present work was to get a better picture of the landscape of possibilities within the framework of bottom-up holography.

We should also note other innovative approaches towards modelling aspects of inhomogeneous systems. These include the techniques of massive gravity \cite{Vegh:2013sk,Blake:2013bqa, Davison:2013jba, Blake:2013owa,Amoretti:2014zha} and Q-lattices \cite{Andrade:2013gsa, Donos:2013eha,Gouteraux:2014hca,Donos:2014uba, Donos:2014cya,Donos:2014oha} (cf., also the memory function approach \cite{Lucas:2015pxa,Lucas:2015vna}).  The Q-lattices are particularly appealing as, though translational invariance is broken, the metric remains homogeneous and the equations of motion remain relatively tractable ODEs. Unfortunately the fine tuned nature of this symmetry raises the possibility that some of the behaviour observed in the resulting solutions may be equally non-generic. Likewise while much progress has been made using the massive gravity approach, questions remain about the interpretation of such models in the dual field theory.  

 With these issues in mind let us turn back to  explicit breaking of translational invariance in one of the spatial directions of the dual field theory, as described in \cite{Horowitz:2012ky,Horowitz:2012gs} (see also \cite{Ling:2013nxa}). As mentioned above, the introduction of the holographic lattice and solutions of the resulting PDEs mitigates the zero frequency delta function in the conductivity, leading to a conventional Drude form for low frequencies.   An extremely thorough examination of the subject in \cite{Donos:2014yya} combines technology gained from Q-lattice calculations with the numerical approach of \cite{Horowitz:2012ky,Horowitz:2012gs,Ling:2013nxa} to examine the electric, thermoelectric and thermal conductivity. Among the results of this analysis were the confirmation of the Drude regime for the AC conductivity and the existence intermediate resonance peaks. It was also noted that in the model under consideration no evidence of a mid-IR scaling regime was discovered.

In this note we aim to build on the work of \cite{Horowitz:2012ky,Ling:2013nxa,Donos:2014yya} by exploring a particular class of Einstein-Maxwell-Dilaton models. These models are characterized by two functions of the scalar dilaton field $\Phi$ -- these parameterize the scalar self-interaction through a scalar potential and a gauge coupling function, $V(\Phi)$ and $Z(\Phi)$ respectively. This parameterization allows us to access a range of effective field theories which are of interest in various condensed matter applications.  For example, examination of the homogeneous solutions has shown that these models are good candidates for metal/insulator transitions upon breaking of translational invariance \cite{Charmousis:2010zz}. The range of allowed IR phases can be understood in terms of an effective near horizon potential, obtained by approximating $V(\Phi)$. One finds that this potential  modulates the spectrum of ingoing excitations near the horizon from continuous (metallic) to discrete (insulating) as a function of parameters. Qualitatively, the physical features of these models can be understood by noting that scalar potential interpolates between zero at the conformal boundary and a runaway behaviour in the near-horizon region. Similarly the gauge coupling starts at some finite value near the conformal boundary and undergoes exponential growth or decay towards the horizon. 

We will examine the electric conductivity of these models as a function of parameters. We employ two lines of attack, both of which involve first numerically finding the bulk solution to sourced inhomogeneities in the UV. This holographic lattice solution can then be explored for transport. The simplest analysis is then to extract the DC conductivities via a membrane paradigme-esque formula using the techniques of  \cite{Donos:2014yya} (cf., \S\ref{subsec:anal_cond}). This approach has the advantage that we only need knowledge of the bulk solution and therefore gives us a simple way to probe the IR phase by analyzing the DC conductivity \cite{Horowitz:2012ky}. Computing the AC conductivity however requires that we also have the solution of the linearized perturbation equations around the numerically constructed background \cite{Horowitz:2012ky}. We carry out this exercise to obtain the full frequency dependent transport, in the process using the aforementioned DC conductivity as a non-trivial check on the results we obtain.
 We discover the existence of metal-insulator transitions as a function of parameters and also, in finely tuned cases, the potential presence of a mid-IR scaling regime. We then examine the persistence of these behaviours as a function of parameters. We find that that the phase changes we discover via monitoring the behaviour of the response functions correspond to inhomogeneity mediated changes in the form of the scalar potential in the IR. 

The outline of the paper is as follows. In \S\ref{sec:section1} we present a basic review of the holographic set-up, pausing to note the ingredients we pick in our model and their potential effects on the low energy dynamics of the system. We also give a short synopsis of the numerical scheme we employ to study the physical transport. In \S\ref{subsec:anal_cond} we revisit the arguments of \cite{Donos:2014yya} to directly extract the zero frequency conductivities in terms of horizon data. In course of this analysis we take the opportunity to explain the relation between the various conductivities in the hydrodynamic limit (which corresponds to $\omega \ll T$ and a suitably dilute lattice). We illustrate that in this regime there is a single transport coefficient which can be taken to be the electrical conductivity. In \S\ref{subsec:num_results} we present our results for the various transport coefficients. We also comment on the various phases and scaling regimes which we observe as we explore a representative subspace of configurations for the scalar potential and Maxwell coupling. We conclude with some discussion in \S\ref{sec:discuss}. Some technical points about our numerics are collated in Appendix \ref{sec:Appendix1}.

\section{Setup} 
\label{sec:section1}
 
 We start by outlining the holographic  set-up we use to explore the  physics of conductivity in 2+1 dimensional quantum critical systems. We explain the basic ingredients we employ in our phenomenological modelling as well as the salient features of the control functions we introduce in parameterizing the holographic system. Following this discussion we go on to describe the numerical scheme used to construct the gravitational solutions of interest and the computation of the physical conductivities therein.

\subsection{Background}
\label{sec:bg}

We take a bottom up, phenomenological approach to holography, and following \cite{Horowitz:2012ky,
Ling:2013nxa,Donos:2014yya} we choose a model with the minimal ingredients necessary to calculate conductivities in a 2+1 dimensional quantum critical theory. We assume that the field theory is holographically dual to gravitational dynamics in an asymptotically \AdS{4} spacetime. The model for the holographic dynamics is simply Einstein-Maxwell theory, which we couple to an additional neutral scalar field, with potential term, $V(\Phi)$, and a gauge coupling function, $Z(\Phi)$. The scalar field allows us to discuss models with more general behaviour in the IR than the local criticality characterized by the Reissner-Nordstr\"om-\AdS{4} (RN-\AdS{4}) black hole. This results in an action of the form (setting $\ell_\text{AdS} =1$): 
\begin{align}\label{eq:action}
S=\int \mathrm{d}^4 x \, \sqrt{-g} \, 
\left(R+6-\frac{1}{4} Z(\Phi) \,  F_{ab}\, F^{ab} - \nabla_{a} \Phi   \nabla^{a} \Phi - \frac{1}{2} \,V(\Phi) \right).
 \end{align}
The dynamical equations of motion which we will solve are then simply
\begin{align}\label{eq:background_eqns}
& \nabla_{a} \left(Z(\Phi)F^{a}_{b}\right)=0 \,,  \\ \nonumber
&  \nabla_{a}  \nabla^{a} \Phi -V'(\Phi)=0 \,, \\ \nonumber
& R_{ab} + 3\, g_{ab}-\bigg( \nabla_{a} \Phi  \nabla_{b} \Phi 
+\frac{1}{2} \left[\frac{1}{2}\, g_{ab}\, V(\Phi) +Z(\Phi) \left(F_{ac}F_{b}^{\;c}-\frac{1}{4}\,g_{ab} F_{cd} F^{cd} \right)
\right] \bigg)  =0 \,.
 \end{align}
 The field theory will be taken to live in Minkowski spacetime and we shall thus work with the conventional parameterization of AdS geometry in Poincar\'e-like coordinates. The radial coordinate is taken to be  $z \in {\mathbb R}_+$ and the conformal boundary located at $z=0$.

 The potential and the gauge coupling functions are our control functions which allow us to modulate the IR dynamics. In \cite{Charmousis:2010zz} these were parameterized to be of the form $V(\Phi)=e^{-\delta  \Phi }$, and $Z(\Phi)=e^{\gamma  \Phi }$ respectively. It was argued that these choices allow for a range of locally critical IR behaviours as one scans over the parameters $\{\gamma,\delta\}$. As we are interested not only in the IR behaviour, but also in translating the local critical dynamics therein onto the AdS boundary, we need to ensure that any such IR geometry can be patched to the asymptotically \AdS{4} region. We must therefore generalize the form of these functions. In doing so we choose to fix the potential to have a Taylor expansion around the origin of field space of the form $V(\Phi) = \Phi^2 + \cdots$. This choice corresponds to an effective  conformal mass term $m^2 = -2$ for the scalar ensuring that we have simple fall-offs (with non-normalizable and normalizable being $z$ and $z^2$ respectively asymptotically). A convenient choice for the functions which respects these constraints turns out to be
\begin{align}\label{eq:potential_form}
V(\Phi) &= \frac{4}{\upsilon^2} (1-\cosh (\upsilon\  \Phi ))\\ \nonumber
Z(\Phi) &= e^{\upsilon\, \Phi }
 \end{align}
 where we have chosen to focus on a one-parameter family of theories, parametrized by $\upsilon$. This corresponds to the case $\gamma+\delta =0$ in the notation of  \cite{Charmousis:2010zz}. We leave exploration of the full parameter space for future study, and focus on this subset henceforth.

For computational simplicity we introduce inhomogeneity in our model by choosing to break translational invariance in one-direction of ${\mathbb R}^{2,1}$ as in \cite{Horowitz:2012ky,Ling:2013nxa,Donos:2014yya}. We construct a lattice in the $x$-direction while maintaining translational invariance in the $y$-direction.\footnote{ We use the word ``lattice" loosely for we choose not to impose any commensurability conditions between the charge density and the unit cell.} The $x$-translation breaking boundary conditions are imposed  by choosing an inhomogeneous normalizable mode for the scalar field of the form 
\begin{equation*}
\Phi_{1}(x)=C \, \cos(k x)\,,
\end{equation*} 
 where $C$ is the amplitude of the inhomogeneity and the $k$ the wavenumber of the lattice. 

 The manner in which this sourced inhomogeneity deforms the near horizon geometry from its homogeneous behaviour, and the effect that this has on the conductivity is the principal focus of this work. From the dual field theory point of view this corresponds to determining how the UV parameters of the theory change the trajectory of the renormalization group flow to create different phases of matter in the IR.  As we explain below the presence of radially conserved quantities in the theory mean that a great deal about the linear response functions, and therefore the phase of the dual field theory, may be extracted simply from knowledge of the near horizon geometry. 

 Given these boundary conditions, a suitable metric ansatz for the investigation of these inhomogeneous phases is  \cite{Horowitz:2012ky}:
\begin{align}\label{eq:background_metric}
 \mathrm{d}s^2=\frac{1}{z^2}\left(-f(z) \,Q_{tt}\, dt^2+ \frac{Q_{zz}}{f(z)} \, dz^2
 + Q_{xx}\, \left( dx+z^2 \,Q_{xz}\,dz\right)^2 + Q_{yy}\,dy^2 \right)
 \end{align}
with the functions $Q_{ab}(z,x)$  depending both on $z$ and $x$, thanks to the inhomogeneity. In addition we introduce a redshift factor $f(z)$ which, upon exploiting the scaling symmetries in the problem to fix the horizon to be at $z=1$, can be chosen to be: 
\begin{equation}
f(z)=(1-z)P(z)=(1-z)(1+z+z^2 -\mu_{1} \frac{z^3}{2})
\end{equation}  
The factor $\mu_{1}$ can be thought of as a convenient parameterization of the temperature, $T$; they are related via
\begin{equation}
T=\frac{P(1)}{4 \pi}= \frac{6-\mu_{1}^2}{8 \pi} \,.
\label{eq:Tdef}
\end{equation}
This choice is useful as in the homogeneous limit the standard RN-\AdS{4} black hole may be recovered by setting $Q_{tt}=Q_{zz}=Q_{xx}=Q_{yy}=1$, $Q_{xz}=\Phi=0$, $A=(1-z)\,\mu$, and $\mu=\mu_{1}$.  We choose to measure all physical quantities relative to a fixed relation between $T$ and $\mu$ which we fix by taking $\mu=\mu_{1}$ throughout.  Also, for convenience, we choose $A= (1-z)A_0(z,x)\, dt$  ensuring thereby that the timelike component of the gauge field vanishes at the horizon. 

In order to ensure that our problem is well posed we must supply both consistent boundary conditions and an appropriate gauge condition to remove the gauge redundancy of the Einstein equations. We choose to work in DeTurck (or harmonic) gauge  --  this is achieved by modifying the Einstein tensor via the addition of a new term involving the so-called DeTurck vector field $\xi^a$:
\begin{align}\label{eq:DeTurck_form}
G^{H}_{ab} &= G_{ab} - \nabla_{(a} \xi_{b)} \\ \nonumber
\xi^{a} & = g^{cd}\left(\Gamma^{a}_{cd}(g)-\Gamma^{a}_{cd}(\bar{g})\right)
 \end{align}
Here $G_{ab}$ is our original Einstein tensor, $\xi^{a}$ is the DeTurck vector, and $G^{H}_{ab}$ is the modified tensor appearing in the DeTurck equations. The DeTurck vector is defined using the difference in the Christoffel symbols, $\Gamma$, associated with our metric of interest, $g$, and a suitably chosen reference metric, $\bar{g}$. The reference metric should have the same asymptotic and conformal structure as the metric we are attempting to solve for. In our case we have found it convenient to use the RN-\AdS{4} metric as the reference metric. It can be shown that for the metric ansatz \eqref{eq:background_metric},  the DeTurck equations are elliptic \cite{Headrick:2009pv} and therefore can be solved as a boundary value problem. 

As we are  interested in solving the original Einstein equations, we must ensure that the DeTurck vector vanishes on-shell. Thus we choose boundary conditions such that the the DeTurck vector vanishes on the boundary. Provided the solutions to our  problem are unique and smoothly dependent on the choice of boundary conditions, this should ensure the vanishing of the DeTurck vector. In practice, we check that the DeTurck vector is zero to a high numerical precision, so we may be confident in the veracity of our solutions. 

After the gauge fixing procedure has been completed we obtain seven independent equations in the seven unknowns, $(Q_{tt}, Q_{zz}, Q_{xx}, Q_{yy}, Q_{xz}, \Phi, A_0)$. It can be checked that taking appropriate linear combinations of the equations decouples their principal parts, and each has an elliptic form.
 
We next turn to the discussion of the boundary conditions. At each boundary we require one boundary condition for each of the seven dynamic fields. We choose the following conditions at the conformal boundary:
 \begin{align}
Q_{tt}(0,x)& =Q_{zz}(0,x)=Q_{xx}(0,x)=Q_{yy}(0,x)=1,  \quad Q_{xz}(0,x)=0 \\ \nonumber
\Phi'(0,x) &=\Phi_{1}(x),  \quad A_0(0,x)= \mu
 \end{align}
 The motivation for the metric boundary conditions is simply that one obtain \AdS{4} spacetime at $z=0$.  Labeling the Dirichlet boundary condition on $A_0(0,x)$ as $ \mu$ is consistent with its interpretation as the chemical potential in the dual field theory. The condition on the scalar field fixes the non-normalizable mode to be inhomogeneous and thus sources inhomogeneity in the system. 

In the IR we must first impose the regularity of the black hole horizon. This is done by requiring $Q_{tt}(1,x)=Q_{zz}(1,x)$ which ensures the surface gravity is constant along the horizon. The remaining boundary conditions may then be found by expanding the equations of motion to leading order in $(1-z)$. As two of the equation expansions are degenerate at this order we obtain the correct number of horizon boundary conditions.  An intuitive understanding of these conditions can be found by solving these leading order expansions for the radial derivatives of the fields at the horizon. We then see that these equations fix the radial derivatives of the fields in terms of their horizon values, and first and second spatial derivatives of the same fields. Therefore our horizon conditions consist of one Dirichlet condition, necessary to define a regular horizon, and six (mixed) Robin-type conditions.  

At the spatial directions we impose that all fields are periodic, with period $L=\frac{2 \pi}{k}$. This allows for the sourced inhomogeneity to be a linear combination of the basic harmonic $\cos(k x)$, and any of the higher harmonics $\cos( n k x)$ for any integer $n$. For example, \cite{Donos:2014yya} and, in a different context, \cite{Arean:2013mta, Hartnoll:2014cua} have constructed solutions with sourced inhomogeneity consisting of multiple modes with random relative phases in order to represent a `dirty' lattice. While such investigations are interesting we postpone the construction of such solutions and the investigation of their thermodynamic and transport properties to future work, and focus below on constructing solutions sourced by the  single harmonic $\cos(k x)$.
        
\subsection{Perturbations and linear response}
\label{subsec:perturbations}

We now turn to analysis of linearized perturbations, needed to extract information about the conductivity of the QFT in the  linear response regime.  We  will primarily be interested in computing the AC (optical) electrical conductivity as a function of frequency and temperature.  The results have an intrinsic interest, and can also provide a point of comparison to the DC conductivity formula which we will discuss in \S\ref{subsec:anal_cond}. They thus  provide a non-trivial check on our results. Combined with the DC conductivities, this provides a comprehensive picture of the behaviour of the theory at the temperature regimes we probe.

The calculation of the conductivity in a holographic theory entails solving the linearized perturbation equations derived from Eqs.~\eqref{eq:background_eqns}.  To derive these we perform the following expansion of the fields:
\begin{align}\label{eq:pert_basic}
g_{ab}=\hat{g}_{ab}+ \epsilon\,  h_{ab}, \quad A_{a}=\hat{A}_a+\epsilon \, b_{a}, \quad  
\Phi=\hat{\Phi}+\epsilon\, \eta
\end{align}
with $h_{ab}$, $b_{a}$ and $\eta$ being the metric, gauge and scalar perturbations, respectively, and the hats indicating background fields.  We work in leading order in $\epsilon$ and use as the background the solutions to Eqs.~\eqref{eq:background_eqns}, subject to the boundary conditions described in \S\ref{sec:bg}. The symmetries of our background allow us to set $b_{y}, h_{ty}, h_{zy}, h_{xy}$ to zero. Therefore the non-trivial components  our metric and gauge perturbations are restricted to $(h_{tt},h_{zz},h_{xx},h_{yy},h_{tx},h_{tz},h_{zx},b_{x},b_{y},b_{z})$. In addition since our background is static we may Fourier decompose the time dependence of the perturbations:
\begin{align}\label{eq:form_pert}
h_{ab}= \tilde{h}_{ab}(z,x)\, e^{-i\, \omega \,t} , \quad b_{a}= \tilde{b}_{a}(z,x)\, e^{-i\, \omega \,t}  , \quad \eta= \tilde{\eta}(z,x)\, e^{-i\, \omega \,t} 
\end{align}
As our equations are linear the time dependence is encapsulated in the factors of the frequency $\omega$ present in the equations. 

In analogy to the background equations we must impose appropriate gauge constraints on our problem in order to remove spurious degrees of freedom and obtain a match between the number of fields and independent equations of motion. In the case of the perturbation equations the gauge redundancy is more complex as, in addition to the diffeomorphism invariance of the gravity sector, the $U(1)$ invariance of the gauge sector must also be taken into account. In order to remove these gauge redundancies we choose to impose the deDonder and Lorentz gauge conditions: 
\begin{align}\label{eq:deDonder_Lorentz}
 \tau_{b}=\nabla^{a} \left(h_{ab}- h \frac{\hat{g}_{ab}}{2}\right)=0, \quad \chi=\nabla^{a} b_{a}=0 
\end{align}
The procedure we use to impose  the gauge conditions is very similar to that used to form the DeTurck equations in equation \eqref{eq:DeTurck_form}.  For example, in order to impose the deDonder gauge we use a gauge fixing term constructed from a gauge transformation, $\tau_b$.  Using the expression for the variation of the Ricci tensor and subtracting the gauge fixing term $\nabla_{(a} \tau_{b)}=\nabla_{(a} \nabla_{c} h^{c}_{b)}-\frac{1}{2} \nabla_{a} \nabla_{b} h$  gives us:
\begin{align}
\delta R_{a b} -\nabla_{(a} \tau_{b)} &=  -\frac{1}{2} \nabla^{c}\nabla_{c} h_{a b} + \frac{1}{2} \nabla_{c}\nabla_{a} h^{c}_{b} + \frac{1}{2}  \nabla_{c}\nabla_{b} h^{c}_{a} - \frac{1}{2} \nabla_{b}  \nabla_{c} h^{c}_{a} - \frac{1}{2}\nabla_{a}  \nabla_{c} h^{c}_{b} 
\\ \nonumber
& = -\frac{1}{2}  \nabla^{c}\nabla_{c} h_{a b} + \frac{1}{2} [\nabla_{c},\nabla_{a}] h^{c}_{b} + \frac{1}{2} [\nabla_{c},\nabla_{b}] h^{c}_{a} 
\end{align}
Thus we see that the principal part of the equation is hyperbolic and the goal of the gauge fixing procedure is achieved. Similarly, in the case of the gauge field equations we add a gauge fixing term of the form $\alpha(z,x) \nabla_{b} \chi$. Here $\alpha(z,x)$ is some combination of the background fields which is determined by requiring that the principal part of each of the gauge equations takes the appropriate hyperbolic form.

The end result of the gauge fixing procedure is $11$ independent equations in $11$ dynamic fields whose principal parts are decoupled and of a hyperbolic form. We can now consider the boundary conditions.  At  the conformal boundary, these are simple -- we require $b_{x}$ to source the external external electric field and all other perturbations to vanish suitably quickly such that asymptotically \AdS{4} is maintained. Therefore we choose the following:
 \begin{align}\label{eq:form_pert}
 \tilde{h}_{tt}(0,x)=0, \quad  \tilde{h}_{zz}(0,x)=0, \quad  \tilde{h}_{xx}(0,x)=0, \quad  \tilde{h}_{yy}(0,x)=0, \quad  \tilde{h}_{tx}(0,x)=0 \\ \nonumber
 \tilde{h}_{tz}(0,x)=0, \quad  \tilde{h}_{zx}(0,x)=0, \quad \tilde{b}_{t}(0,x)=0, \quad \tilde{b}_{z}(0,x)=0, \quad \tilde{b}_{x}(0,x)=1
\end{align}
Here we have used the linearity of the equations to choose a convenient scale for the magnitude of the external electric field. 

The situation at the horizon is more complicated. Physically our boundary conditions must reflect the fact that all excitations are in-falling at the horizon. This is most readily observed by passing over to (the regular) ingoing coordinates Eddington-Finkelstein coordinates and ensuring that both the stress-energy and Einstein tensors are regular at the horizon. This will determine the leading scalings of the fields at the horizon. Our results, which of course agree with those of \cite{Horowitz:2012ky}, are:
 \begin{align}
 \tilde{h}_{tt}(z,x) & ={\cal P}(z) \, \tilde{h}^{reg}_{tt}(z,x), 
 \qquad\quad \tilde{h}_{yy}(z,x) ={\cal P}(z) \, \tilde{h}^{reg}_{yy}(z,x) \\ \nonumber
\tilde{h}_{xx}(z,x) & ={\cal P}(z) \,  \tilde{h}^{reg}_{xx}(z,x), 
\qquad \quad \tilde{h}_{tx}(z,x) ={\cal P}(z) \, \tilde{h}^{reg}_{tx}(z,x) \\ \nonumber
\tilde{h}_{tz}(z,x) & =\frac{{\cal P}(z)}{1-z} \,\tilde{h}^{reg}_{tz}(z,x), 
\qquad \quad \tilde{h}_{xz}(z,x) =\frac{{\cal P}(z)}{1-z} \, 
\tilde{h}^{reg}_{xz}(z,x) \\ \nonumber
\tilde{h}_{zz}(z,x) & =\frac{{\cal P}(z)}{(1-z)^2} \,  \tilde{h}^{reg}_{zz}(z,x), 
\qquad \tilde{b}_{z}(z,x) =\frac{{\cal P}(z)}{1-z} \, 
\tilde{b}^{reg}_{z}(z,x) \\ \nonumber
\tilde{b}_{t}(z,x) & ={\cal P}(z) \,  \tilde{b}^{reg}_{t}(z,x), 
\qquad \qquad
\tilde{b}_{x}(z,x) ={\cal P}(z) \,  \tilde{b}^{reg}_{x}(z,x) \\ \nonumber
\tilde{\eta}(z,x) & ={\cal P}(z) \,  \tilde{\eta}^{reg}(z,x)
\end{align}
where 
\begin{align}
{\cal P}(z) = (1-z)^{-\frac{i\,{\mathfrak w}}{4\pi} }\,, \qquad {\mathfrak w} = \frac{\omega}{T}
\end{align}
captures the leading non-analytic behaviour of the fields near horizon and the remaining analytic part is indicated by the superscript ``reg".  Those regular fields may be expanded in a power series in $(1-z)$ near the horizon. Including the leading and subleading orders in this expansion,  one identifies $15$ possible non-degenerate boundary conditions -- $4$ from the leading order and $11$ from the subleading order. Only $11$ of these, chosen for numerical convenience, are imposed. Consistency of the equations of motion demands that the remaining $4$  conditions vanish on-shell. The vanishing of these constraints provides a non-trivial check on the accuracy of our numerical solutions. Finally, we will also demand that the solutions obey appropriate periodicity conditions in the $x$-direction.

Once the solutions of the perturbation equations are available the AC conductivity can be calculated as \cite{Donos:2014yya,Horowitz:2012ky} as:
 \begin{align}
 \sigma(\omega,x)= \frac{j_{x}(x)}{i \,\omega}, \qquad \text{with} 
 \quad \tilde{b}_{x}(z,x)=1+j_{x}(x)z +\mathcal{O}(z^2)
 \end{align}
From this the DC conductivity may be extracted by taking the $\omega \rightarrow 0$ limit of the $\mathrm{Re} (\sigma)$. We now turn to describing an alternative way of extracting the DC conductivities, involving knowledge of the background solution alone. 

\section{Analytic Expressions for the DC Conductivities}
\label{subsec:anal_cond}

We now describe how to extract information regarding the electric, thermoelectric, and thermal conductivities from knowledge of the background fields alone. This discussion closely follows that of \cite{Donos:2014yya} 
(adapted to our coordinates). 

We begin by briefly reminding the reader of the form of the linear response transport equations in a $2+1$ dimensional theory. At finite chemical potential, and therefore finite density, the heat and electric currents may mix. Therefore Ohm's Law takes the more general form of:
 \begin{align}\label{eq:ohms_law}
  \begin{pmatrix}
  J \\
  Q
 \end{pmatrix} 
 =  \begin{pmatrix}
  \sigma & \alpha T \\
   \bar{\alpha}\, T  & \bar{\kappa}
 \end{pmatrix}   \begin{pmatrix}
  E\\
{\sf T} 
 \end{pmatrix}  
 \end{align}
In our case $J=J^x$ is the electric current and $Q=T^{tx}-\mu J^x$ is the heat current and ${\sf T} = -\nabla_x T$ is the thermal gradient. Our approach will be to perturb our holographic system such that $E$ and ${\sf T}$ are introduced, consecutively, in the dual field theory. We will then find that certain radially conserved currents will allow us to read off the conductivity pairs of $(\bar{\alpha},\sigma)$ and  $(\bar{\kappa},\alpha)$ from the near horizon geometry of the background. 

\subsection{Response from hydrodynamic perspective}
\label{sec:hydro}

At the outset however, we should remark the following. The DC response of the system comprises of two components: a response due to impurity scattering which leads to momentum relaxation and Drude peak behaviour and a more primitive hydrodynamic response present even in translationally invariant systems. Specifically, we may write as in \cite{Hartnoll:2007ih} the low frequency conductivity in the form
\begin{align}
\sigma(\omega) = \frac{K_\sigma \,\tau }{1-i\,\omega\,\tau} +\sigma_h(T,\mu) \,,
\label{eq:omegah}
\end{align}
to emphasize that there is a calculable contribution to transport even when momentum dissipation is swtiched off. Of course, this contribution has to be extracted after subtracting off the divergent DC conductivity arising from the delta function contribution at $\omega =0$ in the $\tau \to \infty$ limit. 

In the hydrodynamic limit\footnote{ We use the phrase `hydrodynamic limit' to refer to the low energy description of translationally invariant systems which is traditionally well described by relativistic hydrodynamics.  Our aim here is to illustrate the fact that underneath the Drude peak, there is a single frequency independent `hydrodynamic conductivity'  captured by $\sigma_h$.} there is only a single response coefficient given by  $\sigma_h$. To see this, note that the limit involves frequencies and momenta which are much smaller than the characteristic thermal scale. The lattice if present is treated  as a spatial long-wavelength perturbation about a homogeneous background. The hydrodynamic energy-momentum and charge currents to first order in spatio-temporal gradients take the form
\begin{align}
T^{\mu\nu} &= \varepsilon\, u^\mu\, u^\nu + p\, P^{\mu\nu} - 2\, \eta(T,\mu) \, \sigma^{\mu\nu} - \zeta(T,\mu) \, \Theta\, P^{\mu\nu}
\nonumber \\
J^\mu &= \rho\, u^\mu + \sigma_h(T,\mu)\, \left(E^\mu - T \, P^{\mu\alpha} \nabla_\alpha\left(\frac{\mu}{T} \right)\right) .
\label{eq:hydroTJ}
\end{align}
where $P^{ab} = g^{\mu\nu} + u^\mu\, u^\nu$, is the spatial projector, $\sigma^{\mu\nu}$ is the shear tensor and 
$\Theta$ is the fluid expansion. The thermodynamic data is encoded in the energy density $\varepsilon$, pressure $p$ and charge density $\rho$ and we have assumed that the underlying system is relativistic with Lorentz invariance being only broken by the choice of inertial frame picked by the fluid (through $u^\mu$).\footnote{ In writing this expression we have chosen to fix some field redefinition ambiguity inherent in hydrodynamics by demanding that $u^\mu$ be a timelike eigenvector of the stress tensor with eigenvalue being the energy density (Landau frame).} 

The shear and expansion contributions are tensor and scalar modes in the hydrodynamic expansion, leaving the contribution coming from the electric field and the gradient of the chemical potential and temperature to be the only vectorial part of transport. The existence of a single vectorial transport encoded in $\sigma_h$ is related to the fact that fluids are required to satisfy the second law of thermodynamics. Allowing for three independent vector transport coefficients is inconsistent with the existence of an entropy current with non-negative definite divergence, cf., \cite{Hartnoll:2007ih}. Translating this observation we require that in the hydrodynamic limit:\footnote{  A version of this relation was derived without invoking the second law but using holography in \cite{Hartnoll:2007ip}. They however derive a relation that mixes zeroth and first order in gradients, which is at odds with interpreting hydrodynamics as a low energy effective theory. Disentangling this leads to the relation we quote in \eqref{eq:hydroaks}. }
\begin{align}
\sigma = \sigma_h \,, \qquad \alpha =\bar{\alpha}= - \frac{\mu}{T}\, \sigma_h \,, \qquad \bar{\kappa} = - \alpha \, \mu =  \frac{\mu^2}{T}\, \sigma_h
\label{eq:hydroaks}
\end{align}
This can be used to check some aspects of numerics in the high temperature regime (where all our models exhibit metallic behaviour).
\subsection{Membrane paradigm for response}
\label{sec:}

Having understood what the relations we expect are, we can now turn to asking what the holographic modeling has to say about the transport coefficients of interest. 

We start by considering the electric and thermoelectric conductivities, $\sigma$, and $\bar{\alpha}$. To extract the DC conductivities we modify the perturbation ansatz:
 $(h_{tt},h_{zz},h_{xx},h_{yy},h_{tx},h_{tz},h_{zx},b_{y},b_{z}, \eta)$ are time-independent and 
 ${b}_{x}=b_{x}(z,x)- \epsilon \,E \,t\ $. We may then proceed as described in  \S\ref{subsec:perturbations} and derive the corresponding equations of motion.\footnote{ For the purpose of the argument that follows the details of  gauge-fixing of the equations is inessential.} It may then be easily checked that the linearized perturbation equations for the gauge field imply $\partial_{x} (\sqrt{-g} \,Z \,F^{xr})=0$ and $\partial_{r} (\sqrt{-g} \,Z \,F^{rx})=0$, and therefore\footnote{ We will use ${\cal J}$ and ${\cal Q}$ to denote the bulk conserved quantities, which will of course agree with the boundary charge and thermal currents.} 
\begin{equation}
 {\cal J}=\sqrt{-g} \,Z \,F^{xr} = \text{constant}. 
\label{eq:J1}
\end{equation}  

As this quantity is a constant it can be evaluated anywhere including at the horizon. In order to derive an explicit form for such an expression we wish to extract the leading scaling of the fields near the horizon. This may be done, as above, by transforming into ingoing coordinates and demanding the regularity of the stress-energy and Einstein tensor. The results are as follows:
\begin{equation}
\label{eq:elec_pert_exp}
\begin{split}
h_{tt}(z,x) & \simeq 4 \pi h_{tt}^{0}(x) (1-z)+\mathcal{O}(1-z)^2, \quad h_{tz}(z,x)  \simeq h_{tz}^{0}(x)+\mathcal{O}(z-1)  \\ 
h_{tx}(z,x) & \simeq \sqrt{\frac{Q_{xx}(1,x)}{Q_{yy}(1,x)}} h_{tx}^{0}(x)+\mathcal{O}(1-z), \quad h_{zx}(z,x) \simeq \frac{\sqrt{\frac{Q_{xx}(1,x)}{Q_{yy}(1,x)}}}{4 \pi T(1-z)}  h_{rx}^{0}(x) +\mathcal{O}(1-z) \\  
h_{xx}(z,x) & \simeq h_{xx}^{0}(x) +\mathcal{O}(1-z), \quad h_{yy}(z,x)  \simeq h_{yy}^{0}(x) +\mathcal{O}(1-z) \\ 
\eta(z,x) & \simeq \eta^{0}(x) +\mathcal{O}(1-z), \quad b_{t}(z,x)  \simeq b^{0}_{t}(x) +\mathcal{O}(1-z) \\ 
b_{z}(z,x) & \simeq \frac{1}{4 \pi T (1-z)}b^0_{z}(x) +\mathcal{O}(1-z), \quad b_{x}(z,x)  \simeq \log{(4 \pi T (1-z))} b^0_{x}(x) +\mathcal{O}(1-z)
\end{split}
\end{equation}
and with the restrictions that:
 \begin{align}\label{eq:elec_ingoing_res}
h_{zx}^{0}(x)= h_{tx}^{0}(x), \quad h_{zz}^{0}(x)= 2 h_{tz}^{0}(x)- h_{tz}^{0}(x), \quad b_{x}^{0}(x)=\frac{E}{4 \pi T}, \quad b_{z}^{0}(x)= b_{t}^{0}(x)
\end{align}
in order that Einstein and stress energy tensor be regular in this coordinate system. Note that for convenience we have chosen to include factors of the background metric fields evaluated at the horizon in our definition of 
$h^{0}_{tx},h^{0}_{zx}$.

Expanding our expression for ${\cal J}$ and utilizing these near horizon expansions we obtain the following equation:
\begin{align}\label{eq:Jhor_elec}
\sqrt{\frac{Q_{xx}}{Q_{yy}}} \left(\frac{{\cal J}}{Z(\phi )}-A_0 \frac{h^0_{tx}(x)}{Q_{zz}}\right)+b_t^{0\prime}(x)-E=0
\end{align}
where all background fields have been evaluated at the horizon. 

The next step is derive a similar expression for another conserved quantity which, as explained in \cite{Donos:2014yya}, corresponds to the thermoelectric conductivity. To do so we note that if $\xi$ is a Killing vector satisfying $\mathcal{L}_{\xi} F=0$ then we may define:
 \begin{align}
 {\cal G}^{\alpha \beta}= \nabla^{\alpha} \xi^{\beta} + \frac{1}{2} \, \xi^{[\alpha}F^{\beta] \sigma} \,A_{\sigma} 
 + \frac{1}{4} \, (\psi-2 \theta) \, F^{\alpha \beta}
\end{align}
with $\psi$ and $\theta$ defined as $\mathcal{L}_{\xi} A=  d \psi$ and $i_{\xi} F=  d\theta$.\footnote{ Here ${\mathcal L}$ denotes the Lie derivative along the indicated field and $i_\xi$ indicated contraction with the vector field.} This has the important property that 
\begin{align}
\nabla_{\alpha} {\cal G}^{\alpha \beta}= 3 \,\xi^{\beta}
\end{align}
Choosing the Killing vector, $\xi= \partial_t$ and utilizing the above identity and the equations of motion it can be shown that $\nabla_{x} {\cal G}^{xr}=\partial_{x}(\sqrt{-g} \, {\cal G}^{xr})=0$ and 
$\nabla_{r} {\cal G}^{rx}=\partial_{r}(\sqrt{-g} \, {\cal G}^{rx})=0$ and that therefore that\footnote{ The derivation of the conserved charges, ${\cal J}$, and, ${\cal Q}$, is presented in full generality in the appendix of \cite{Donos:2014cya}. We have also explicitly checked that the derivation of the conserved currents and charges holds for the model we are considering.} 
\begin{equation}
{\cal Q}=\sqrt{-g} \, {\cal G}^{rx} = \text{constant}.
\label{eq:Q1}
\end{equation}  
 If we now choose that
$\theta= -\left(-\epsilon \, E \, x +A_0(z,x)+ \epsilon \,b_{t}(z,x)\right)$ and $\psi= \epsilon\, E\, x$, and expand 
${\cal Q}$ around the horizon as we did  ${\cal J}$ , we obtain at leading order:
\begin{align}\label{eq:Q_form_elec}
{\cal Q}=-4 \pi \,T \,h^{0}_{tx}(x)\,.
\end{align}
This indicates that $h^{0}_{tx}(x)=h^{0}_{tx}$ is a constant. Working at subleading order we obtain a constraint expression which, with a little rearranging, can be written as:
\begin{align}\label{eq:Q_constraint_elec}
\sqrt{\frac{Q_{xx}}{Q_{yy}}} \left(\frac{h_{tx}^0}{2 \pi  T Q_{zz}} \left(\pi  \,T \,\partial_z 
\log \left(\frac{Q_{xx} \,Q_{zz}}{Q_{tt}^3\, Q_{yy}}\right)-8 \pi  \,T+6\right)-\frac{ {\cal J}\,A_0}{4  \pi \, T \,Q_{zz}}\right)+\frac{\partial }{\partial x}\frac{h_{tz}^0(x)}{Q_{zz}}=0
\end{align}
where again all background fields are being evaluated at the horizon. Integrating this expression over one period of the background, the last term vanishes as a total derivative whose boundary contributions cancel as a result of periodicity. We may then solve for the only remaining perturbative field, $h^{0}_{tx}$, in terms of ${\cal J}$. Performing the same trick with equation 
\eqref{eq:Jhor_elec} we see that the $\partial_{x} b^{0}_{t}(x)$ disappears under integration, and the result gives us 
${\cal J}={\cal J}(h^{0}_{tx},E)$. Combining these two results we obtain an expression for the conductivity. To write the expressions compactly it is useful to introduce some notation; let
\begin{equation}
\begin{split}
I_1  &=  \int  dx\,  \left(6-8 \pi \, T+\pi \, T \,\partial_z
        \log \left(\frac{Q_{xx} \,Q_{zz}}{Q_{tt}^3 \,Q_{yy}}\right)\right)
        \, \sqrt{\frac{Q_{xx}}{Q_{yy}\, Q_{zz}^2}}\,,   \\
I_2 &= \int dx\,  \frac{1}{Z(\phi)} \, \sqrt{\frac{Q_{xx}}{Q_{yy}}} \,, \\
I_3 &=  \int dx\,  \frac{A_0}{Q_{zz}} \; \sqrt{\frac{Q_{xx}}{Q_{yy}}} \,.
\end{split}
\label{}
\end{equation}
In terms of these integrals we find that the DC electric conductivity and the thermoelectric conductivity are given as
\begin{equation}\label{eq:elec_anal_cond}
\begin{split}
\sigma &= \frac{2\, I_1  }{ 2\, I_1\, I_2 - I_3^2}  \,,\\
\bar{\alpha}&=\frac{Q}{T E}= - \frac{4\pi\, I_3}{2\,I_1\,I_2- I_3^2} \,.
\end{split}
\end{equation}

The calculation of the thermal and thermoelectric conjugate conductivities, $\bar{\kappa}$, and $\alpha$ proceeds in a very similar fashion. In this case however our ansatz for the form of the perturbations contains two time-dependent components ${b}_{x}=b_{x}(z,x)-\epsilon\, t \,\tau \, A(z,x)$ and 
$h_{tx}=h_{tx}(z,x)-\epsilon \,t \,Q_{tt}(z,x) \frac{f(z)}{z^2}$.\footnote{ The form of the coupled metric and gauge field perturbation is chosen such that the equations of motion remain time-independent. In writing the form of the metric perturbation as above we have tacitly modified our background metric ansatz such that $Q_{xz}=0$. This is done to avoid the more complicated form of the metric perturbation necessary if $Q_{xz} \neq 0$. It has been checked that the response function results derived from the near horizon behaviour do not change as a result of this modification.} The restriction of regularity in ingoing coordinates again determines the leading scalings for the fields. The  result for the  $h_{tx}$ and $b_x$ fields is found to be:
 \begin{align}\label{eq:therm_pert_exp}
h_{tx}(z,x) & \simeq \sqrt{\frac{Q_{xx}}{Q_{yy}}} \left(h_{tx}^{0}(x) +4 \pi T h_{tx}^{l}(x)(1-z)\log{\left(4 \pi T (1-z)\right)}\right)+\mathcal{O}(1-z) \\ \nonumber
b_{x}(z,x) & \simeq b_{x}^{0}(x)+\mathcal{O}(z-1) \\ \nonumber
 \end{align}
 while all other fields exhibit the same scalings and constraints as displayed in equations \eqref{eq:elec_pert_exp}. The constraints displayed in \eqref{eq:elec_ingoing_res} are modified by the presence of a  logarithmic term in the $h_{tx}$ expansion and the lack of an external electric field. The resulting constraints are given by:
 \begin{align} \label{eq:therm_ingoing_res} 
 h_{zx}^{0}(x)= h_{tx}^{0}(x), \quad h_{zz}^{0}(x)= 2 h_{tz}^{0}(x)- h_{tz}^{0}(x), \quad b_{z}^{0}(x)= b_{t}^{0}(x), \quad h_{tx}^{l} &=-\frac{Q_{tt} \tau}{4 \pi T} \sqrt{\frac{Q_{zz}}{Q_{xx}}}
 \end{align}

In order to obtain expressions for the response functions which do not involve the background fields we expand the conserved quantities of, ${\cal J}$, and ${\cal Q}$ to leading and subleading order, respectively, near the black hole horizon.
 The leading order term in the expansion of ${\cal Q}$ again indicates that $h_{tx}^0(x)$ is:
 \begin{align}\label{eq:Q_form_therm}
 {\cal Q}&=-4 \pi \,T\, h_{tx}^0(x)
 \end{align}
 While the remaining two constraints displayed in equation \eqref{eq:JQhor_therm} may be integrated over a single period such that they become linear equations in ${\cal J} h_{tx}^0$ and $\tau$.
 \begin{align}\label{eq:JQhor_therm}
&\sqrt{\frac{Q_{xx}}{Q_{yy}}} \left(\frac{{\cal J}}{Z(\phi )}-A_0 \frac{h_{tx}^0}{Q_{zz}}\right)+b_t^{0 \prime}(x)=0 \nonumber \\
& \sqrt{\frac{Q_{xx}}{Q_{yy}}} \left(\frac{h_{tx}^0}{2 \pi  T Q_{zz}} \left(\pi  T \partial_z\log \left(\frac{Q_{xx} Q_{zz}}{Q_{tt}^3 Q_{yy}}\right)-8 \pi  T+6\right)-\frac{{\cal J}\, A_0}{4  \pi  T Q_{zz}}\right)+\frac{\partial }{\partial x}\frac{h_{tz}^0}{Q_{zz}}+\tau=0
\end{align}
 This information is sufficient to calculate the thermal and conjugate thermo-electric conductivities:
  \begin{equation}\label{eq:thermal_anal_cond}
\begin{split}
 \bar{\kappa}& =\frac{{\cal Q}}{T \,\tau}=\frac{16 \pi^2 T \, I_2}{2\,I_1\, I_2 - I_3^2}
  \\
  \alpha&=\frac{{\cal J}}{T\, \tau}  =- \frac{4 \pi\, I_3}{2\,I_1\, I_2 - I_3^2}
\end{split}
  \end{equation}

 Once we have an expression for $\bar{\kappa}$ we may easily calculate the conjugate thermal conductivity, $\kappa$ which corresponds to the the thermal conductivity at zero electric current. This is done via the relation  
 \begin{align}\label{eq:conj_thermal_anal_cond}
 \kappa &= \bar{\kappa} - \frac{\alpha^2 T}{\sigma}= \frac{8 \pi ^2 T}{I_1}
 \end{align}

It is reassuring to note that the $\alpha=\bar{\alpha}$ as it should since our system does not break time reversal invariance. We may also check that the high temperature limit of the response functions behaves in the expected fashion. This limit may be extracted by expanding around $\frac{T}{\mu}=\frac{T}{\sqrt{6-8 \pi T}} \rightarrow \infty$. In this limit the solution resembles that of a Schwarzschild-\AdS{} black hole and so appropriate field substitutions are:
\begin{align}\label{eq:highT_subs}
& A_0 \rightarrow \delta_\epsilon\,  \mu, \quad Q_{zz} \rightarrow 1+ \delta_\epsilon\,  \Delta Q_{zz}, \quad Q_{tt} \rightarrow 1+ \delta_\epsilon\,  \Delta Q_{tt}, \quad Q_{xx} \rightarrow 1+ \delta _\epsilon\, \Delta Q_{xx}\,, \\ 
\nonumber
& Q_{yy} \rightarrow 1+ \delta_\epsilon\,  \Delta Q_{yy}\, \quad 
Q_{xz} \rightarrow \delta_\epsilon\, \Delta Q_{zz}, \quad  \phi \rightarrow \delta _\epsilon\, \Delta \phi, \quad \mu \rightarrow \delta_\epsilon\, \mu 
 \end{align}
 where we may now expand around $\delta_\epsilon=0$.\footnote{ The high temperature limit may also be accessed by undoing the scalings of the action and equations of motion which are used to keep the location of the horizon fixed at $z=1$ as the temperature is changed. As the location of the black hole horizon, $z_p$, then approaches the conformal boundary as $\frac{T}{\mu} \rightarrow \infty$ we may find the high temperature behaviour of the response functions by expanding around $z_p=0$. The appropriate expansions of the fields in this limit is determined by their known conformal behaviour. As expected the results from this approach agree with those presented in the text.} We find in this limit that $\alpha= \bar{\alpha} \rightarrow 0$ and $\bar{\kappa} \rightarrow \infty$ diverges as the effects of momentum dissipation are removed from the system. The electrical conductivity, $\sigma$, asymptotes to $Z(0)=1$, the known result for Schwarzschild-\AdS{4}.  Furthermore, we can confirm that in the hydrodynamic regime the relation between the conductivities \eqref{eq:hydroaks} is satisfied. In fact, from \eqref{eq:elec_anal_cond} and \eqref{eq:thermal_anal_cond} we learn that in the hydrodynamic limit the background geometry should satisfy  $ 2\,I_1\,I_2- I_3^2 = 0$.\footnote{ Strictly speaking this relation is valid in the translationally invariant case, where the DC conductivity contribution of the Drude peak diverges.}

 It was shown in \cite{Horowitz:2012ky} and \cite{Donos:2014cya} that the breaking of translational invariance in holographic models produces a low frequency AC conductivity well fit by the Drude model of conductivity. We confirm that this is true in our model in the appendix. It will therefore be interesting to test if our model obeys the Wiedemann-Franz law at any point in its phase diagram. We therefore calculate the Lorenz factors for both $\kappa$ and $\bar{\kappa}$ as follows:
\begin{subequations}
\begin{align}
\bar{L} &= \frac{8 \pi ^2 \, I_2}{I_1}
\label{eq:L_bar_eqn} \\ 
L &= \frac{4 \pi ^2\, (2\,I_1\, I_2 -\, I_3^2)}{I_1^2}
\label{eq:L_eqn}
\end{align}
\end{subequations}
In  \S\ref{subsec:num_results} we will see that the that these expressions are not constant in either the metallic or insulating phases of our model, confirming departures from the Wiedemann-Franz law.

\section{Transport results for holographic systems}
\label{subsec:num_results}

 We regard the present effort as a first step of exploring the vast phase diagram of the the effective holographic theories, identifying interesting corners for further study.  In this section we discuss features of the DC conductivities and the optical conductivity, and qualitative changes in the physics as we vary parameters. We argue that these changes indicate the existence of metal-insulator quantum phase transitions at various loci in the phase diagram.\footnote{ The phrase, phase diagram, here refers to changing both the sources in a given theory (by tuning the period of the lattice set by $k$) as well as explorations across theories (by changing Lagrangian parameter $\upsilon$).}
           
Let us start with a brief  description of the numerical techniques we employ, before describing our main results. For those interested, much more detail regarding the development and testing of our numerical methods is provided in Appendix \ref{sec:Appendix1}. In order to solve our PDEs numerically we discretize them using spectral methods \cite{trefethen2000spectral,boyd2001chebyshev}. A Chebyshev grid was employed in the radial direction and a Fourier grid in the spatial direction, which imposes spatial periodicity, as discussed above. The solution of the non-linear background equations uses both Newton and quasi-Newton methods, whereas the solution of the perturbation equation only necessitates the inversion of a matrix, for which we use direct methods.

\subsection{DC conductivities}
\label{subsubsec:phas_trans}

We start by exploring the direct electric and thermoelectric conductivities. These are calculated using horizon data, as explained in \S\ref{subsec:anal_cond}. The electric conductivity is also calculated as the zero frequency limit of the optical conductivity. Besides providing a check of the numerics, the low frequency behaviour of the optical conductivity helps in understanding and elucidating the IR physics.

Since we are working at finite temperature, we cannot probe the metal-insulator quantum phase transition directly. Nevertheless, by interpolation of our knowledge of both the temperature dependence of the DC and optical conductivities to sufficiently low temperatures, we can diagnose the presence of such transition as function of parameters. In our exploration we fix $C=1.5,$ and discuss the phase diagram as function of $\upsilon$ as well as the perturbation wavenumber $k$. Varying these two parameters we find both metallic and insulating regimes, and transitions between them.

 \begin{figure}
\center
      \includegraphics[width=8cm]{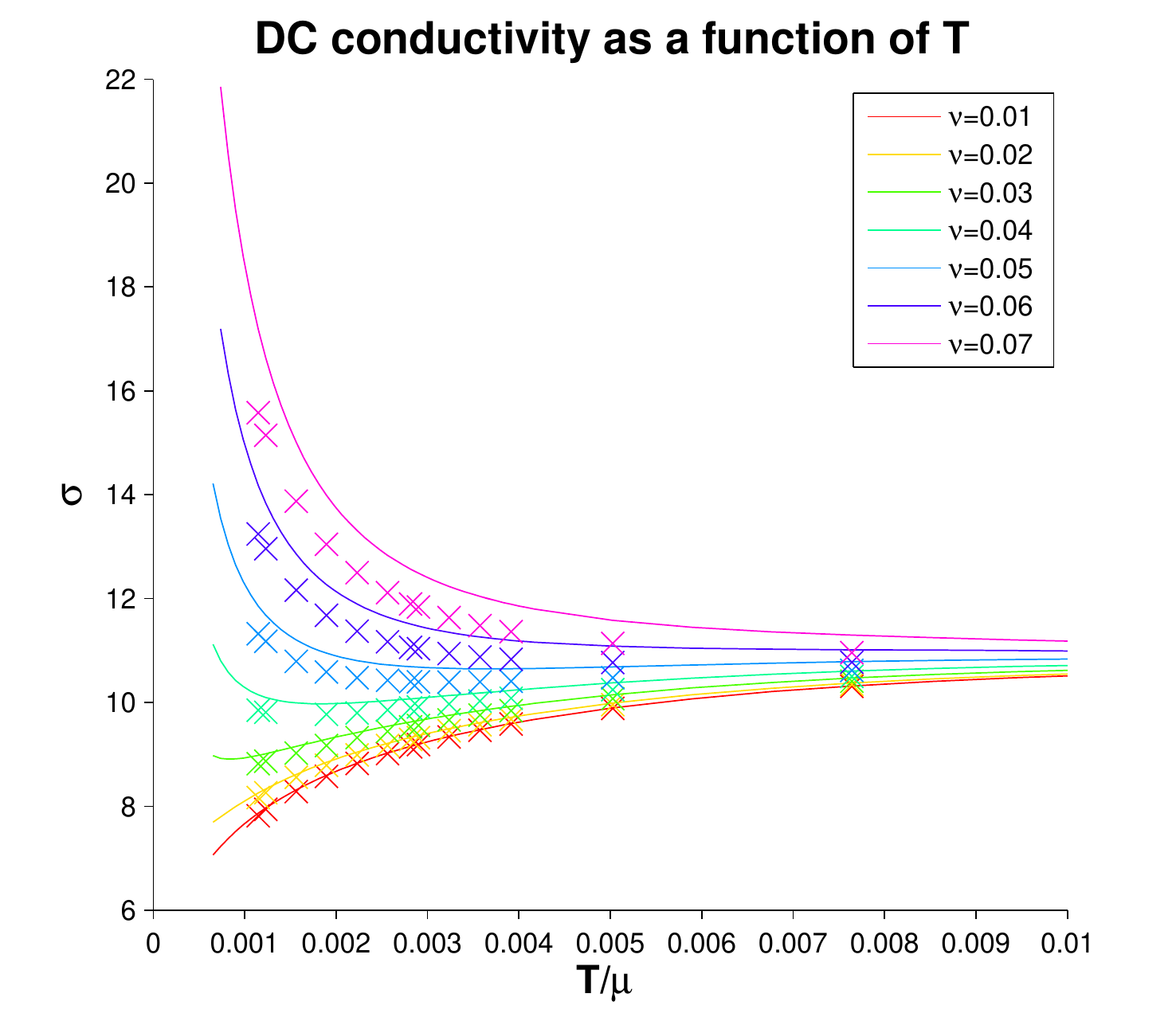}
      \includegraphics[width=8cm]{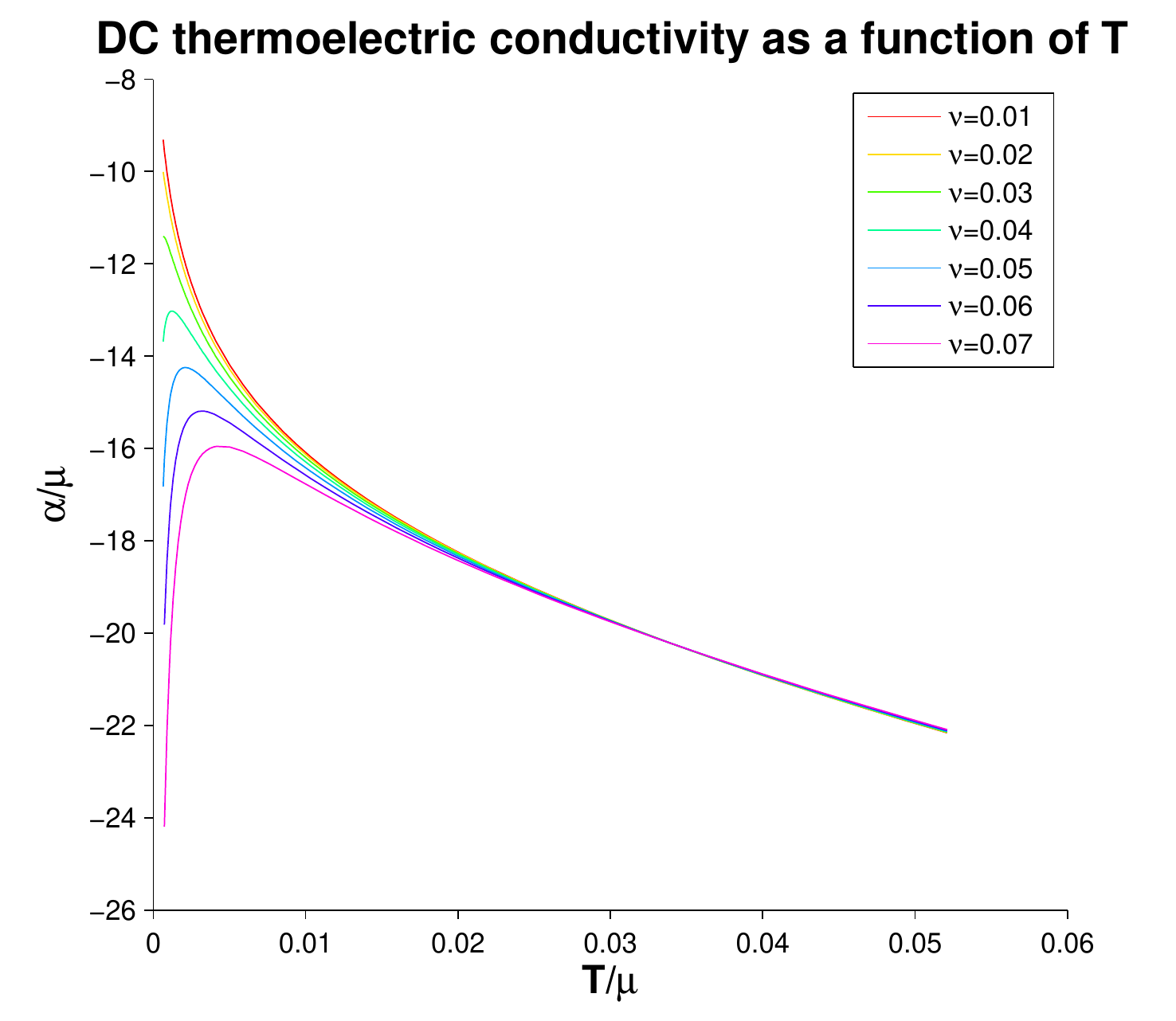}
    \caption[]{Plots of  the DC electrical and thermoelectric conductivities for a range of models against  T for various theories parameterized by $\upsilon$, with $C =1.5, k=1$ held fixed. We clearly see the existence of the metallic and insulating regimes separated by an intermediate region.}
\label{fig:elec_thermoelec_cond}
\end{figure}

In Fig.~\ref{fig:elec_thermoelec_cond} we show a representative sample of the DC conductivities with $k=1$ and varying $\upsilon$. We clearly observe the transition from an insulating to metallic behaviour as the value of $\upsilon$ is increased. For low value of $\upsilon$, which includes the Einstein-Maxwell model of \cite{Horowitz:2012ky}, we find a distinct insulating behaviour: the conductivity decreases at low temperatures and seems to vanish at zero temperature. On the other hand, for sufficiently large $\upsilon$ the opposite behaviour is manifest: the conductivity is increasing with temperature and seems to diverge at zero temperature.  In the transition region, the conductivity shows no distinct trend -- this is the region which is a bad or incoherent metal at low temperatures. 

A similar trend can be seen in the DC thermoelectric conductivity, which is also monotonically increasing as we lower the temperature, for small values of $\upsilon$. As we increase $\upsilon$ the curve begins to kink downwards at low temperatures until eventually a well defined turning point is formed. This turning point migrates towards larger temperatures as we continue to increase our control parameter. We conclude that the transition between metallic and insulating behaviour exists in this observable as well.

\begin{figure}
\center
    \hspace*{1cm}
     \includegraphics[width=8cm]{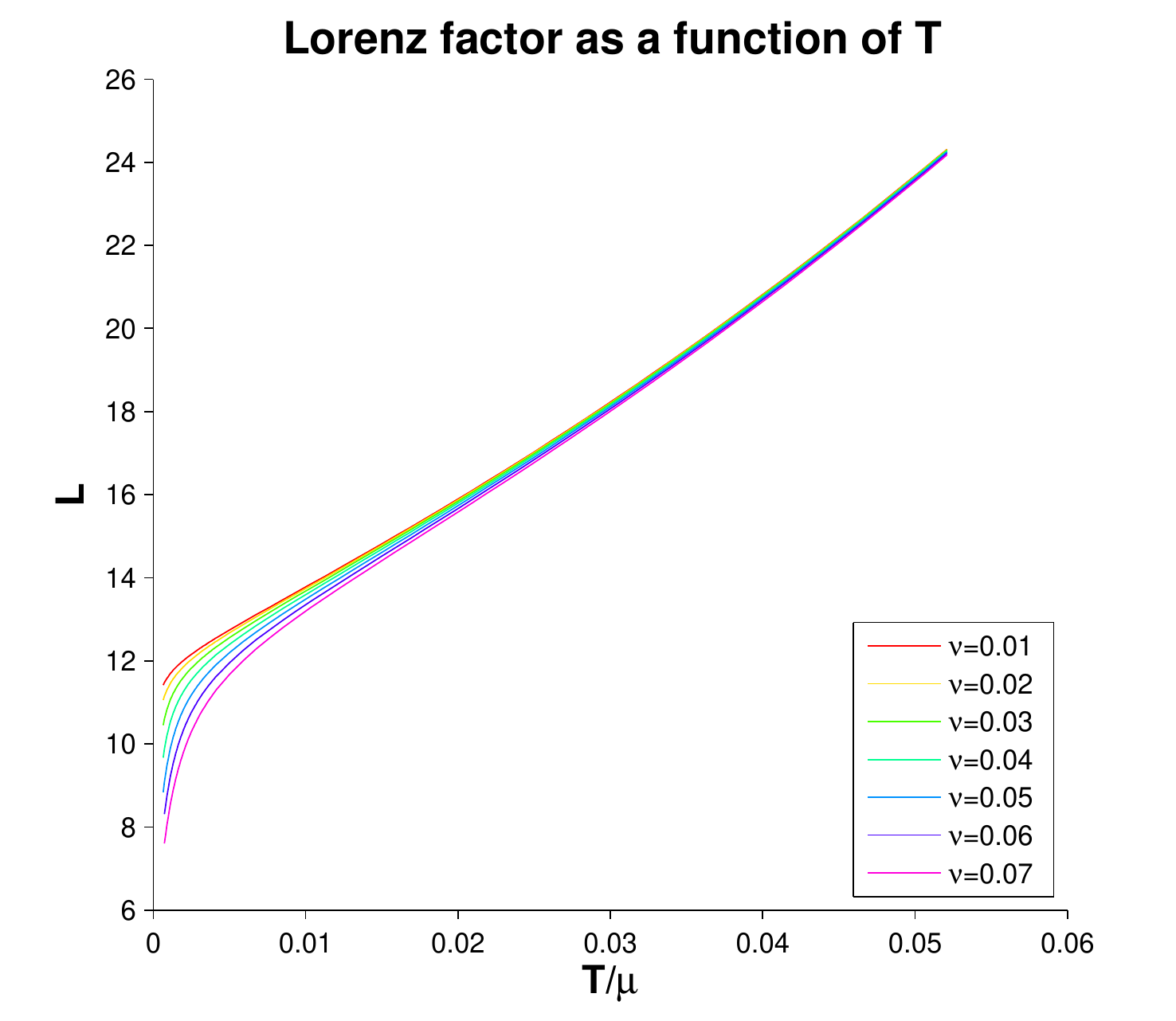}
    \includegraphics[width=8cm]{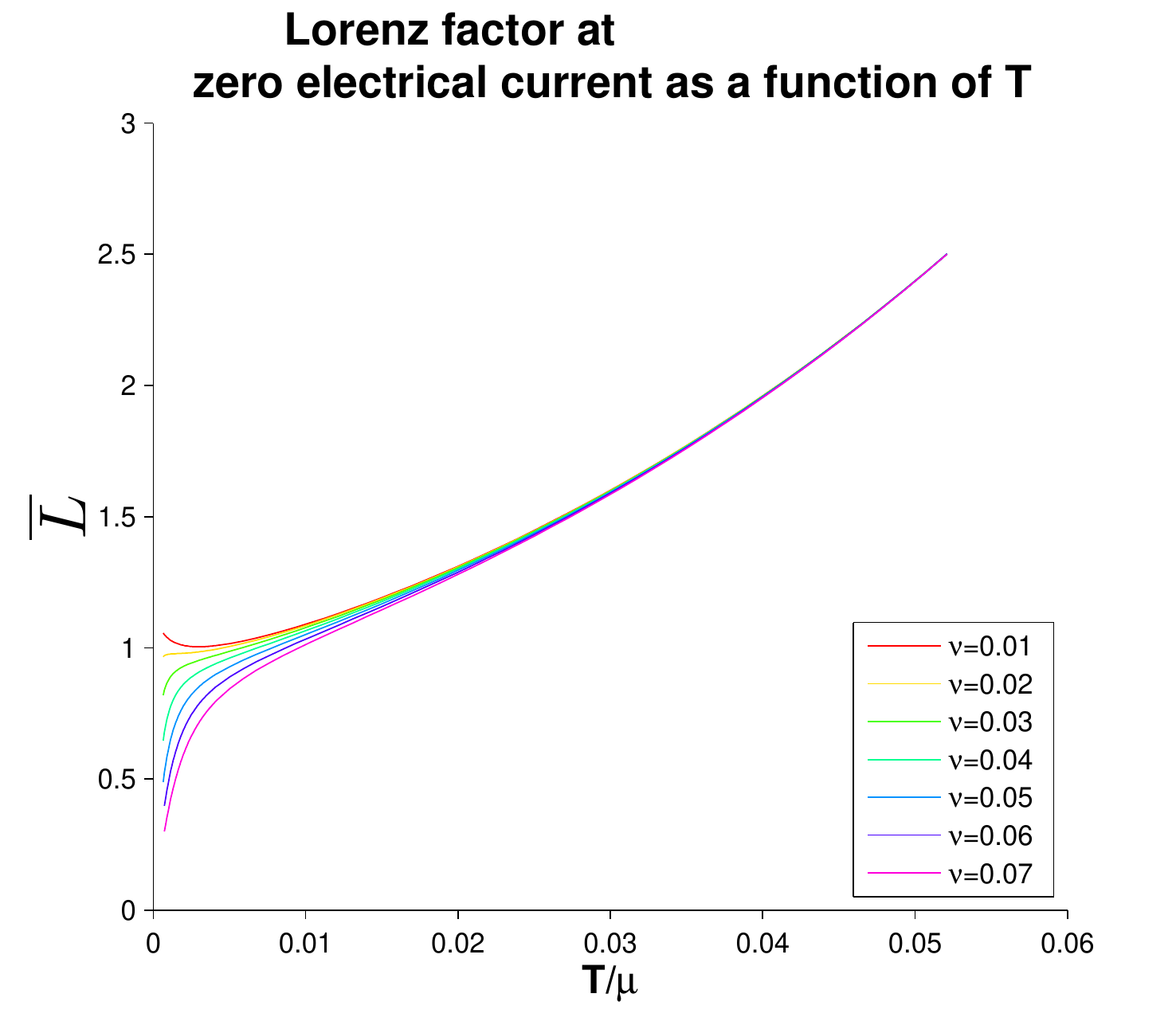}
    \caption[]{Plots of the Lorenz factors associated with $\kappa$ and $\bar{\kappa}$. The fact that these factors are neither constant as a function of temperature nor $\upsilon$ indicates that the Wiedemann-Franz law is violated. This is in accord with the lack of a phase transition in the thermal conductivity.}
\label{fig:Lorenz_factors}
\end{figure}

It is interesting to note that no such transitions occur in the thermal conductivity, which displays a simple monotonic increase as a function of temperature for all values of parameters we examined. Thus our theories are all good thermal conductors. In some sense this is not surprising as the Wiedemann-Franz law is explicitly violated in our expressions for Lorenz factors in equations \eqref{eq:L_bar_eqn} and \eqref{eq:L_eqn}. This is confirmed by the numeric results presented in Fig.~\ref{fig:Lorenz_factors}.  

\begin{figure}
\center
    \hspace*{1cm}
     \includegraphics[width=8cm]{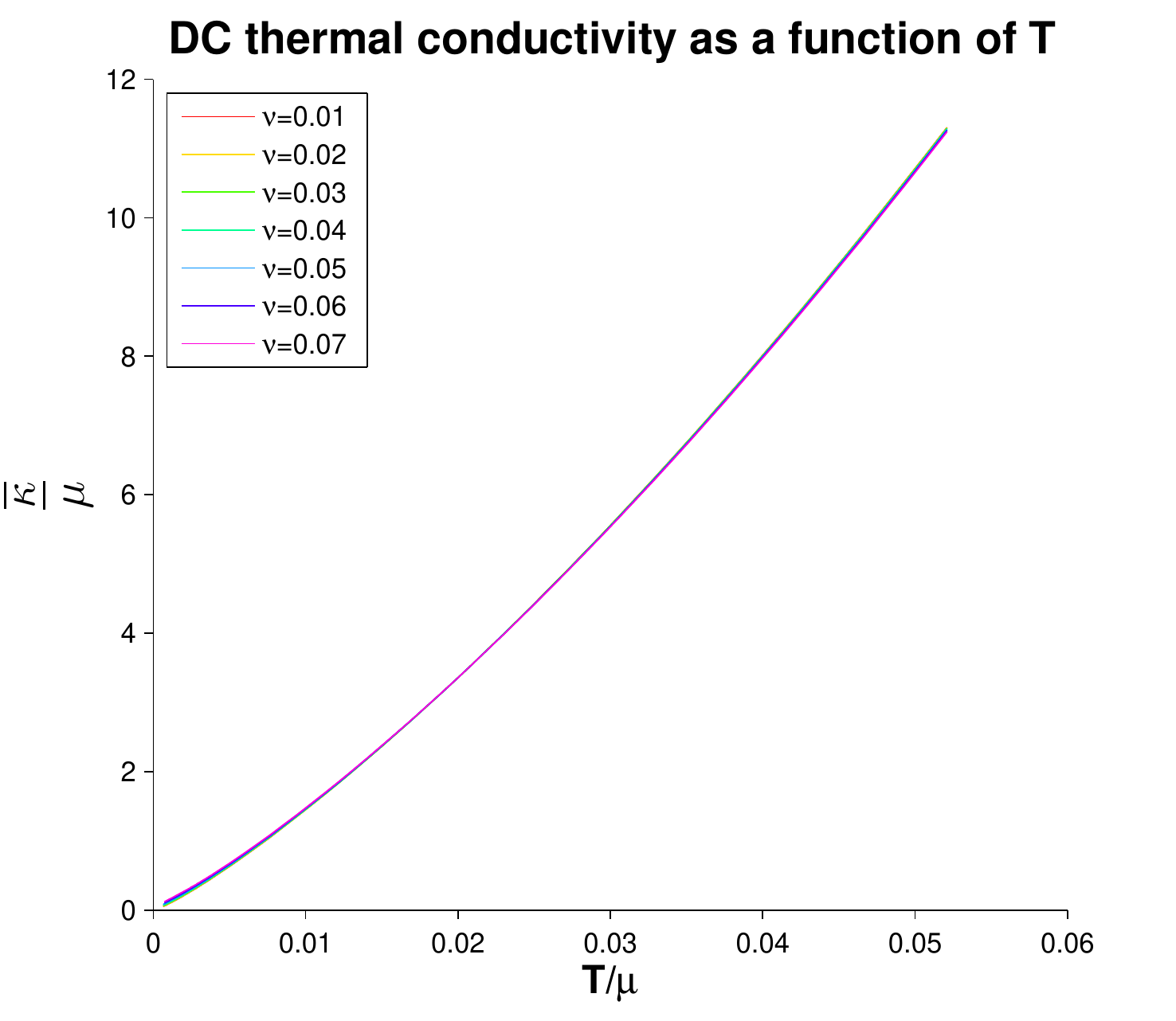}
     \includegraphics[width=8cm]{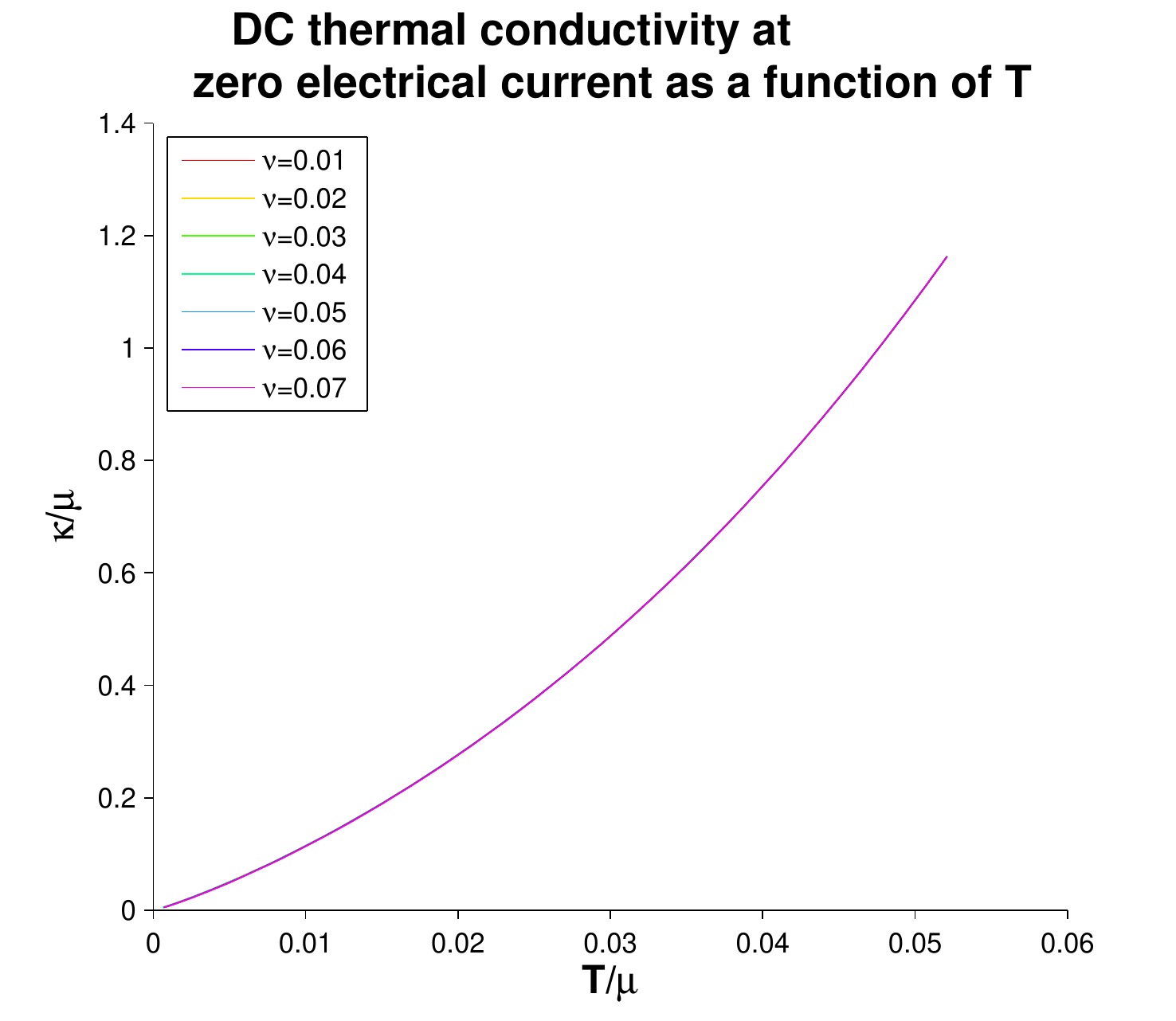}
    \caption[]{Plots of  both forms of the thermal conductivity, $\kappa$ and $\bar{\kappa}$. We note both the qualitative similarity of the two quantities and the insensitivity to the variation in the $\upsilon$ parameter. In all cases that we have examined, including those associated with the data used to construct Fig.~\ref{fig:met_insul_trans_plane}, the thermal conductivities were seen to increase monotonically with temperature.}
\label{fig:thermal_cond}
\end{figure}
%

\subsection{Metal-insulator transitions}
\label{sec:mit}

\begin{figure}
\center
    \hspace*{1cm}
    \includegraphics[width=8cm]{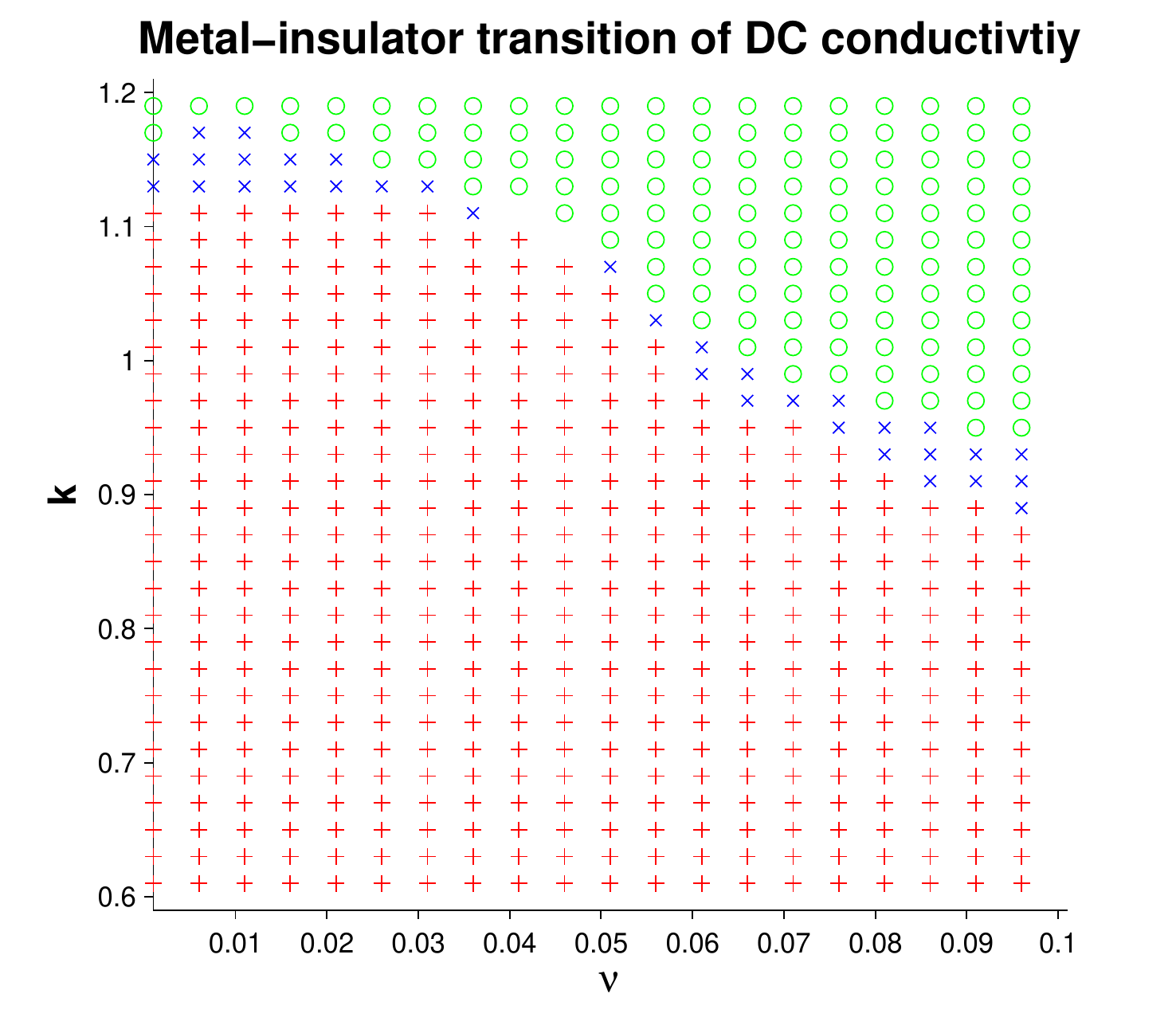}
    \caption[]{The phase plot in the $(\upsilon,k)$ plane illustrating the metal insulator transition. The green data points correspond to regimes which are clearly metallic in character with monotonically increasing DC electrical conductivity at low temperatures. Likewise the red data points correspond to insulating phases characterized by a monotonically decreasing DC electrical conductivity at low temperatures. The intermediate region between these corresponds to the transition region where a turning point is still evident in the profile at low temperatures.} 
\label{fig:met_insul_trans_plane}
\end{figure}

We have seen that the finite temperature results lend themselves to interpretation as indicative of a zero temperature metal-insulator transition. We characterize such a phase transition by a qualitative change in the low temperature behaviour of the response functions.
 One can delineate regions of the phase diagram where the holographic theories we consider describes good metals, bad metals and insulators.  We now search for such transitions as function of $\upsilon$ and $k$. 
 
 The result, displayed in Fig.~\ref{fig:met_insul_trans_plane}, separates the parameter space into the regimes of clear metallic and insulator phases, separated by intermediate regimes in which the conductivity is neither monotonically increasing or decreasing.   We see that increasing $\upsilon$ reduces the relevance of the sourced inhomogeneity, such that the transition to a metallic phase occurs at lower values of $k$.
 
The intermediate phase which straddles the phase transition region can be viewed as one where the competition between metallic and insulating orders is strong. It is interesting to speculate in analogy with the domain model for magnetic phase transitions, that one is encountering pockets of the competing phases. This leads to incoherence in the transport, reducing for instance the conductivity from its metallic value.

We note that the phase separation between metallic and insulating behaviour occurs along a locus 
$k_\star(\upsilon)$ which is monotone decreasing -- as we increase the lattice wavelength, we encounter a transition at lower values of $\upsilon$. This  suggests that the efficacy in translating the lattice between UV and IR regions in the geometry is playing a role in the presence/absence of  ``charge carriers''.\footnote{ We use the phrase ``charge carriers'' somewhat loosely since we are talking about transport in a strongly interacting system with no obvious quasiparticles.}  The DC conductivity is effectively a proxy for the weight of the charge carrier spectral function at vanishing frequency. The support of this spectral function is localized in the vicinity of the horizon. This immediately follows from the fact that we have a membrane paradigmesque formula for the conductivity. 

\begin{figure}
\includegraphics[keepaspectratio=False,width=\textwidth,height=0.5\textwidth]{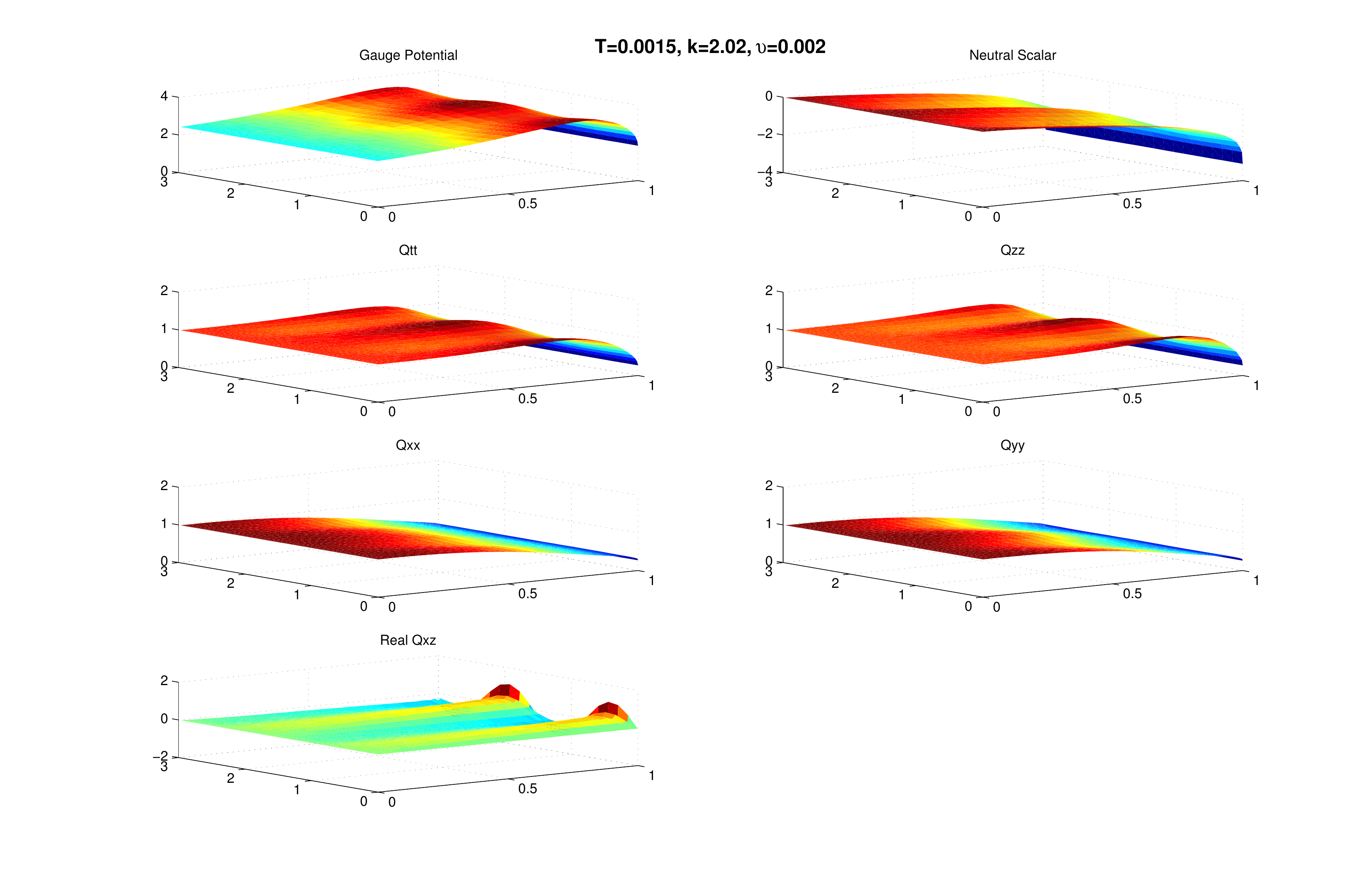}
\includegraphics[keepaspectratio=False,width=\textwidth,height=0.5\textwidth]{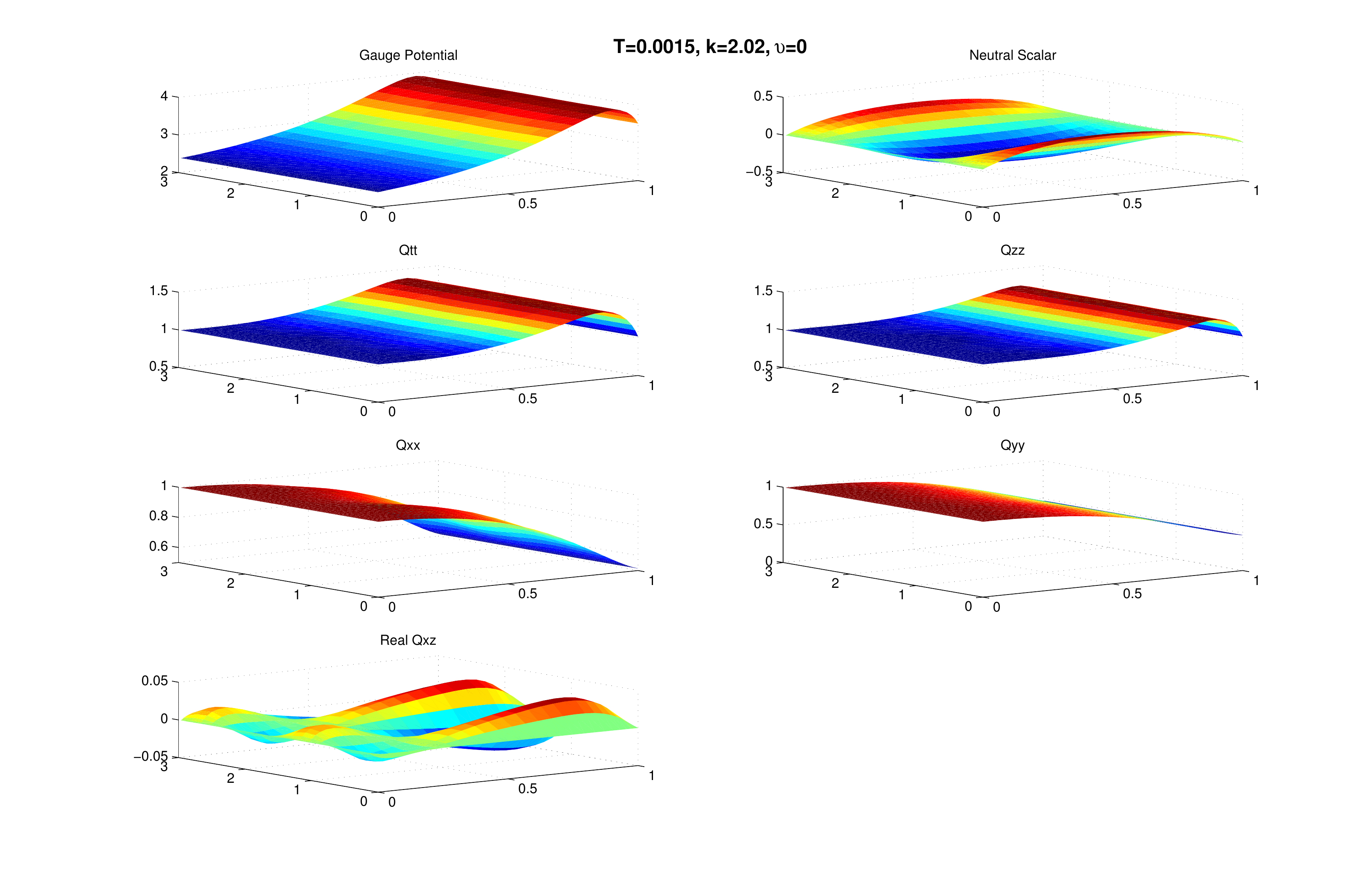}
\caption[]{Setting $\upsilon$ to be strictly zero means that the IR evolution of the scalar field flattens out. As this behaviour of the scalar controls many aspects of the IR physics qualitative changes in the response functions are to be expected.}
\label{fig:upsilon_zero}
\end{figure}

\begin{figure}
\makebox[\linewidth]{%
\subfigure{\includegraphics[width=0.7\textwidth]{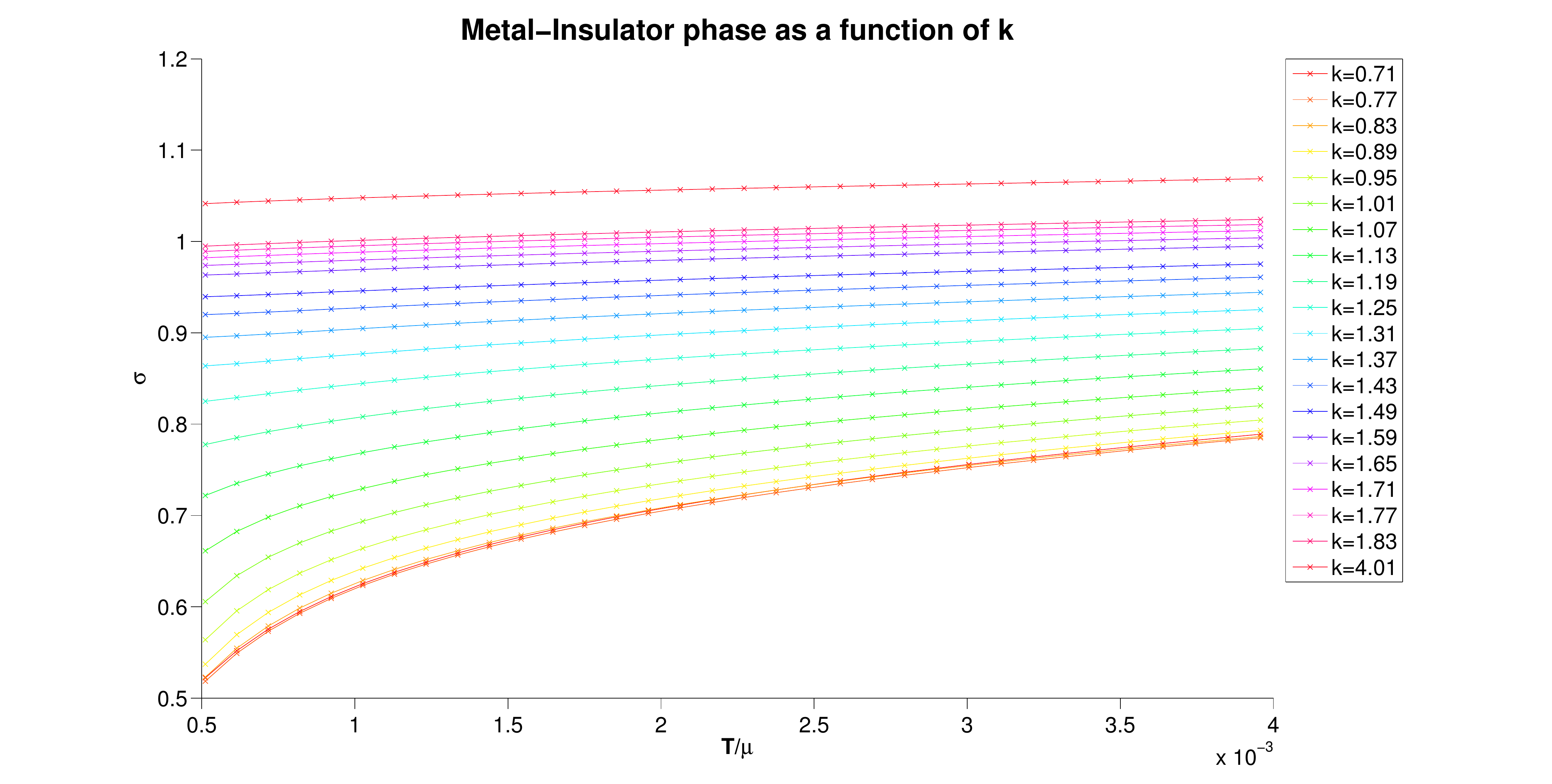}}%
}
\caption[]{The DC conductivity  in the limit  $\upsilon=0$ for different lattices. We notice that for lower values of the wavelength $k$ the conductivity is decreasing at low temperatures. As we increase $k$ the DC conductivity begins to flattens at lower temperatures though for this choice of amplitude, $C=1.5$, for the sourced inhomogeneity the conductivity never transitions into the metallic phase for the temperatures which we can reliably access.}
\label{fig:deltagamma_trans}
\end{figure}

Finally, we specialize to $\upsilon=0$ which describes the Einstein-Maxwell model of \cite{Horowitz:2012ky}. This allows us to probe the existence of a phase transition as a function of the wavenumber $k$ alone.\footnote{ Metal-Insulator transitions in Einstein-Maxwell theory deformed by helical lattices were discussed in \cite{Donos:2012js}.} The results for the DC electrical conductivity are displayed in Fig.~\ref{fig:deltagamma_trans} and clearly show the existence of a quantitative change in the temperature dependence as $k$ is varied. However at temperatures which we may reliable access the phase is always insulating for our choice of parameters (with $C=1.5$ held fixed). This is in contrast to the metallic phases observed for the parameters chosen in \cite{Horowitz:2012ky}.\footnote{ We have checked that for the parameters $C,k$ chosen in \cite{Horowitz:2012ky} we also encounter metallic phases.} We have seen evidence that the transition may occur at lower temperatures however these are difficult to probe reliably. The qualitative shift in behaviour when moving from finite $\upsilon$  to $\upsilon=0$  can be illustrated by direct examination of the background field solutions as seen in Fig.~\ref{fig:upsilon_zero}. 

 \subsection{Optical conductivity}
 \label{subsubsec:AC_analysis}      

\begin{figure}
\makebox[\linewidth]{%
\subfigure{\includegraphics[width=0.6\textwidth]{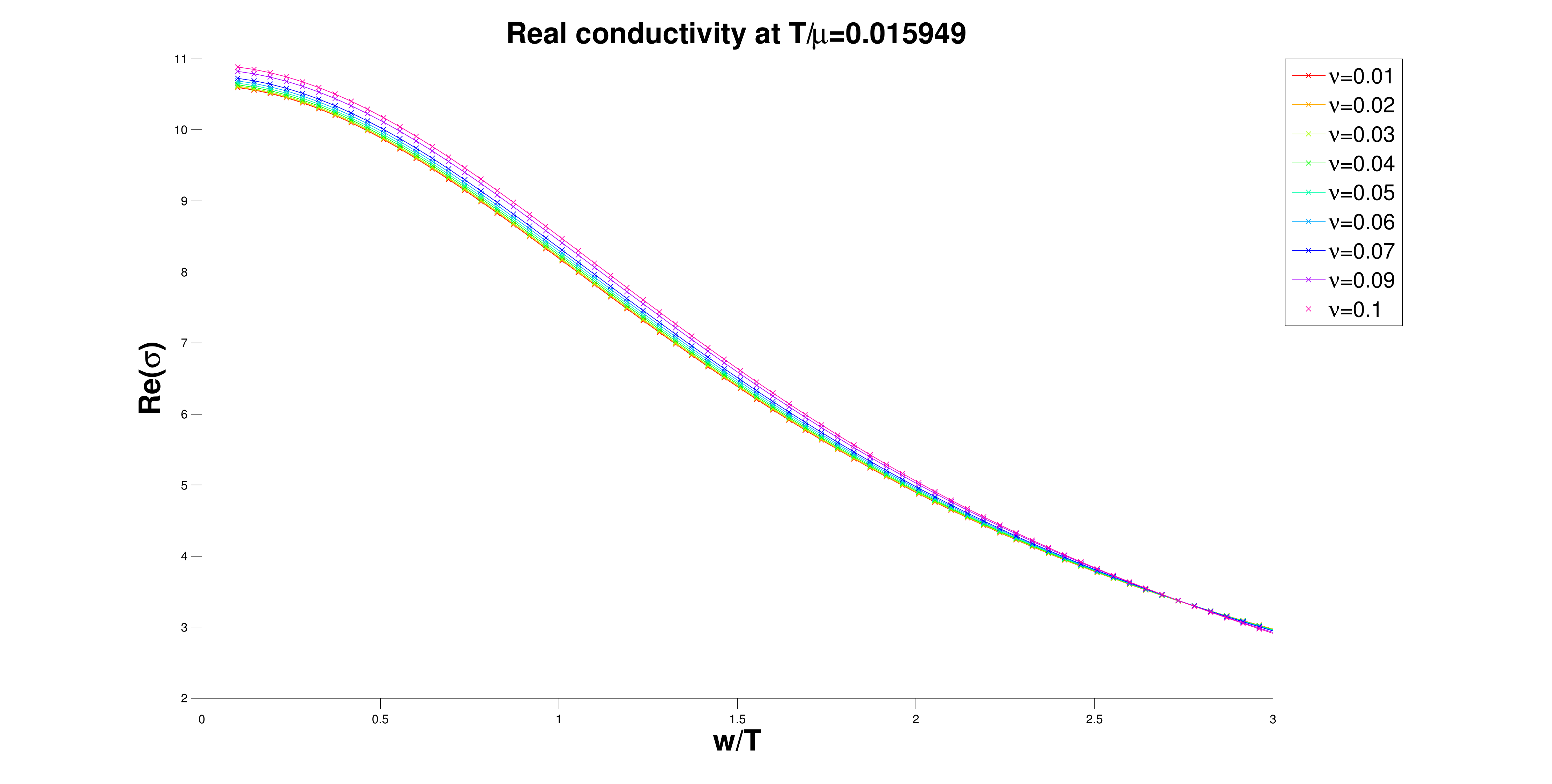}}%
\subfigure{\includegraphics[width=0.6\textwidth]{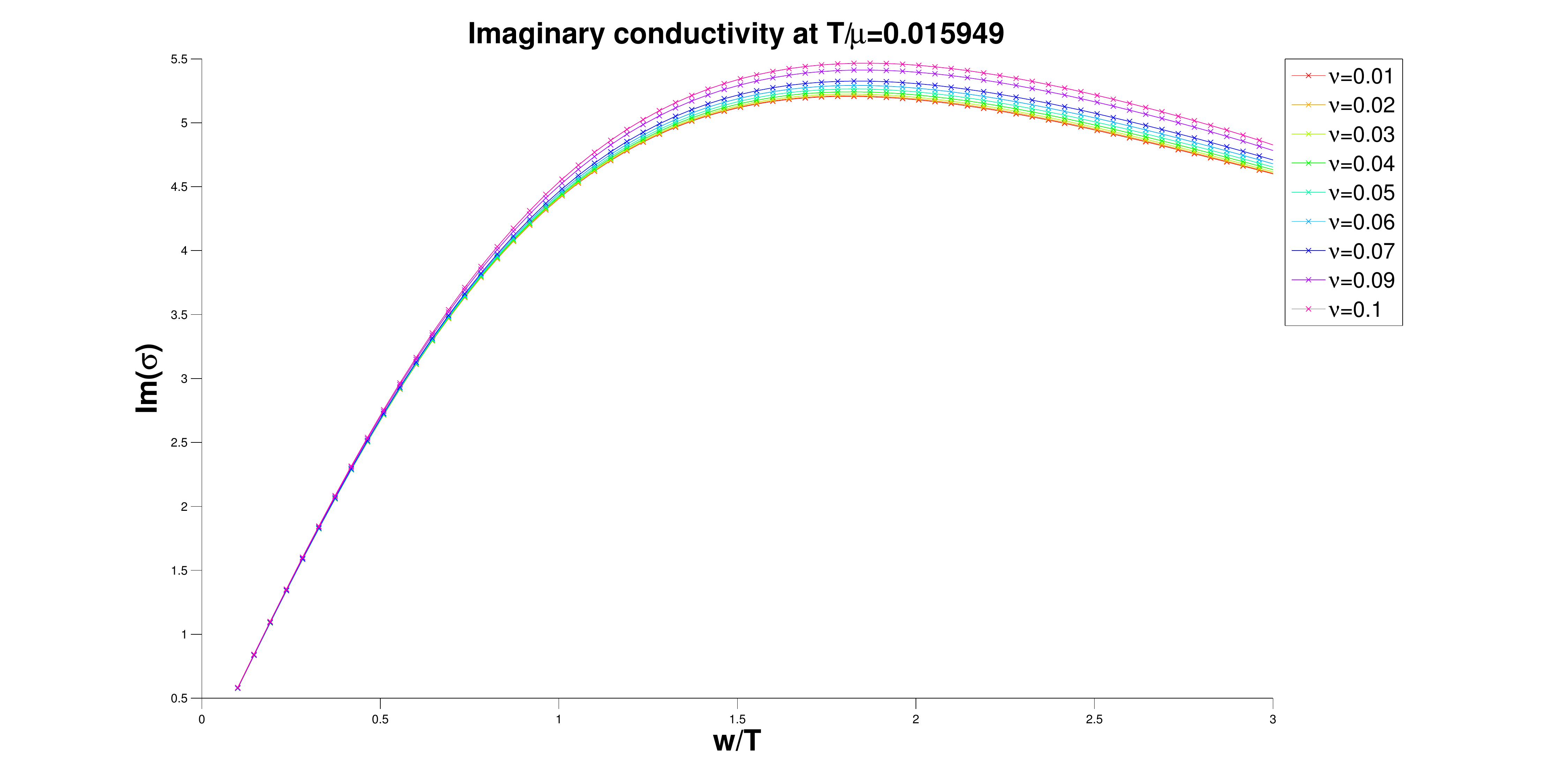}}%
}
\makebox[\linewidth]{%
\subfigure{\includegraphics[width=0.6\textwidth]{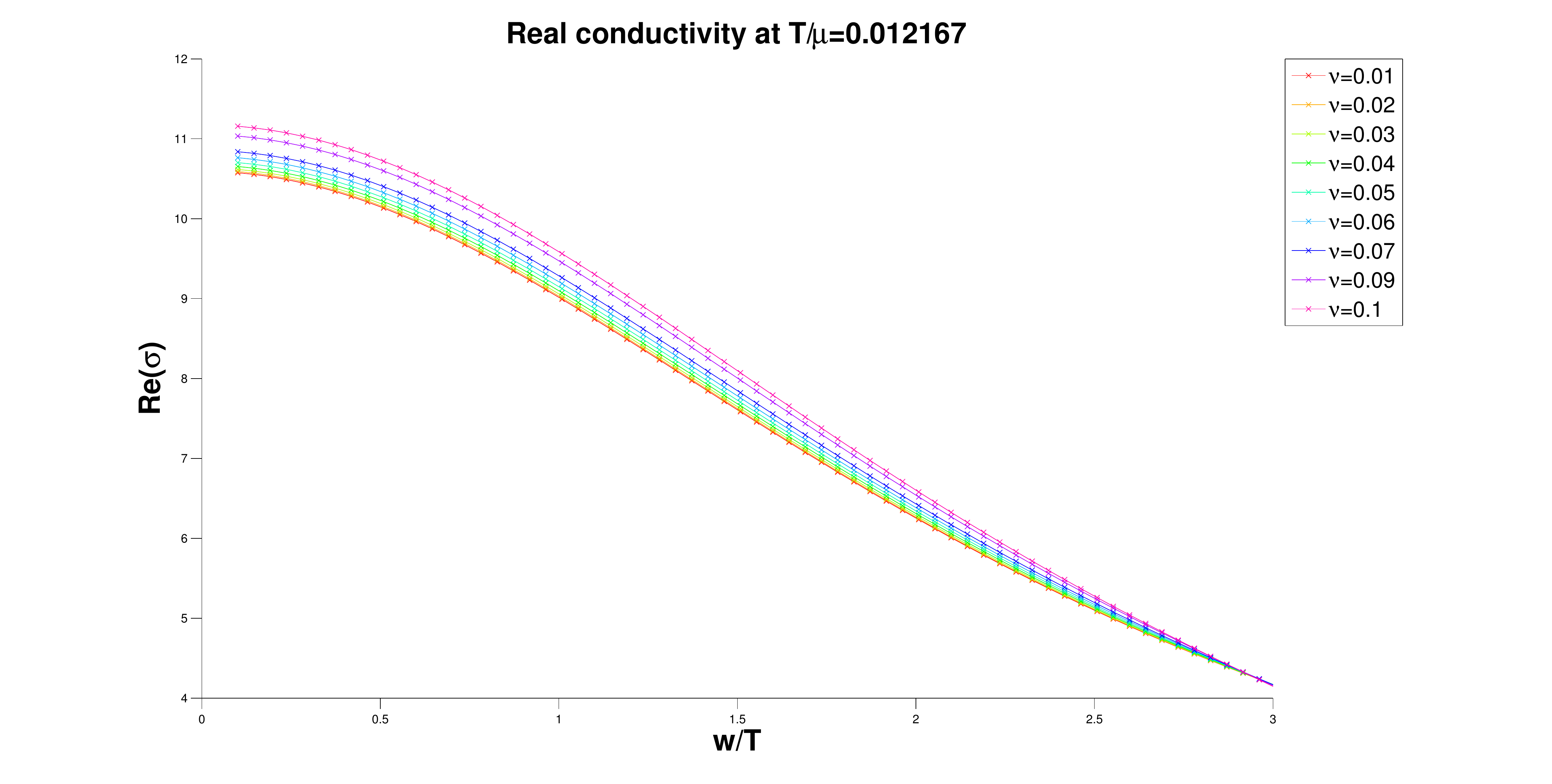}}%
\subfigure{\includegraphics[width=0.6\textwidth]{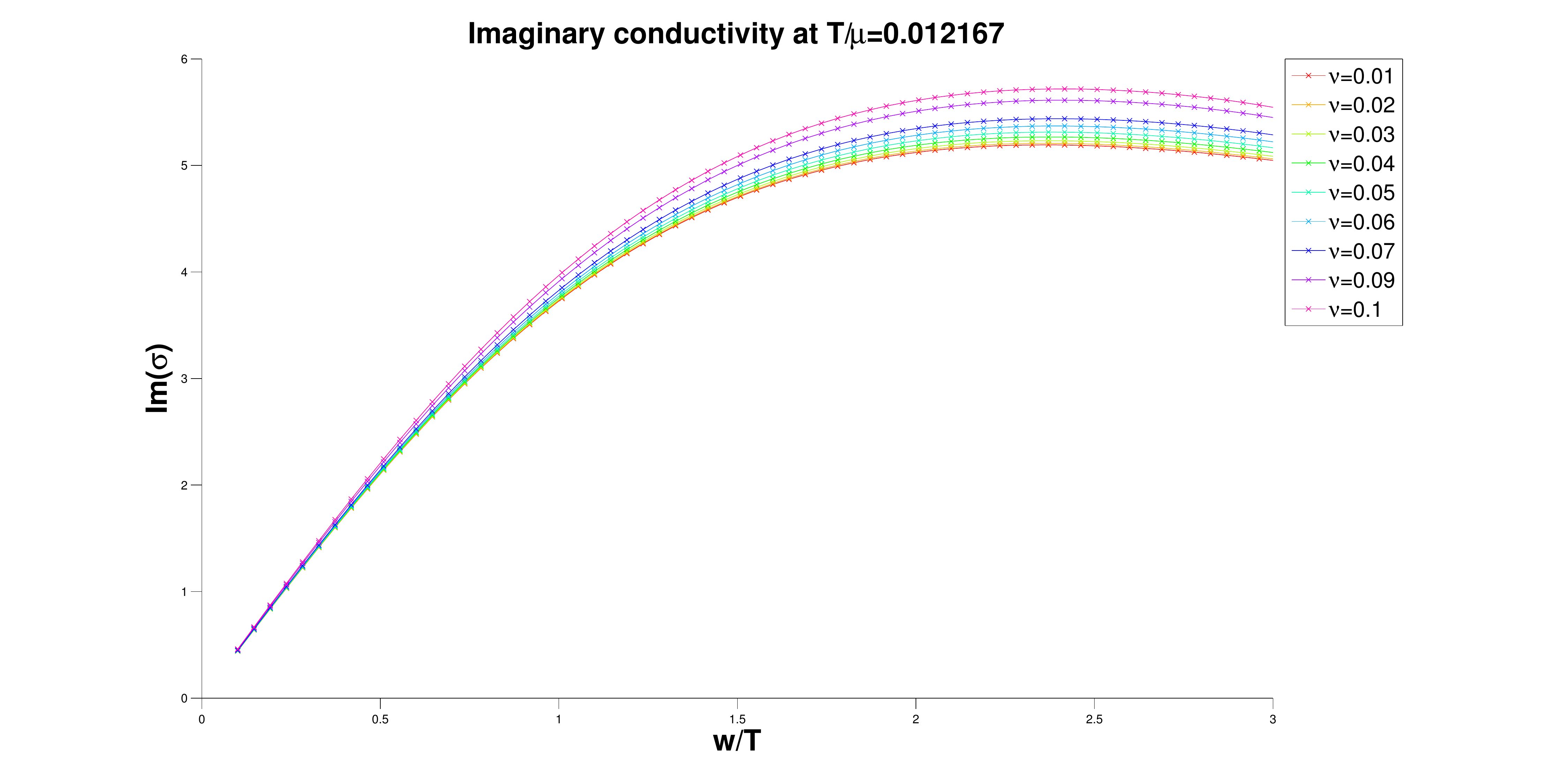}}%
}
\makebox[\linewidth]{%
\subfigure{\includegraphics[width=0.6\textwidth]{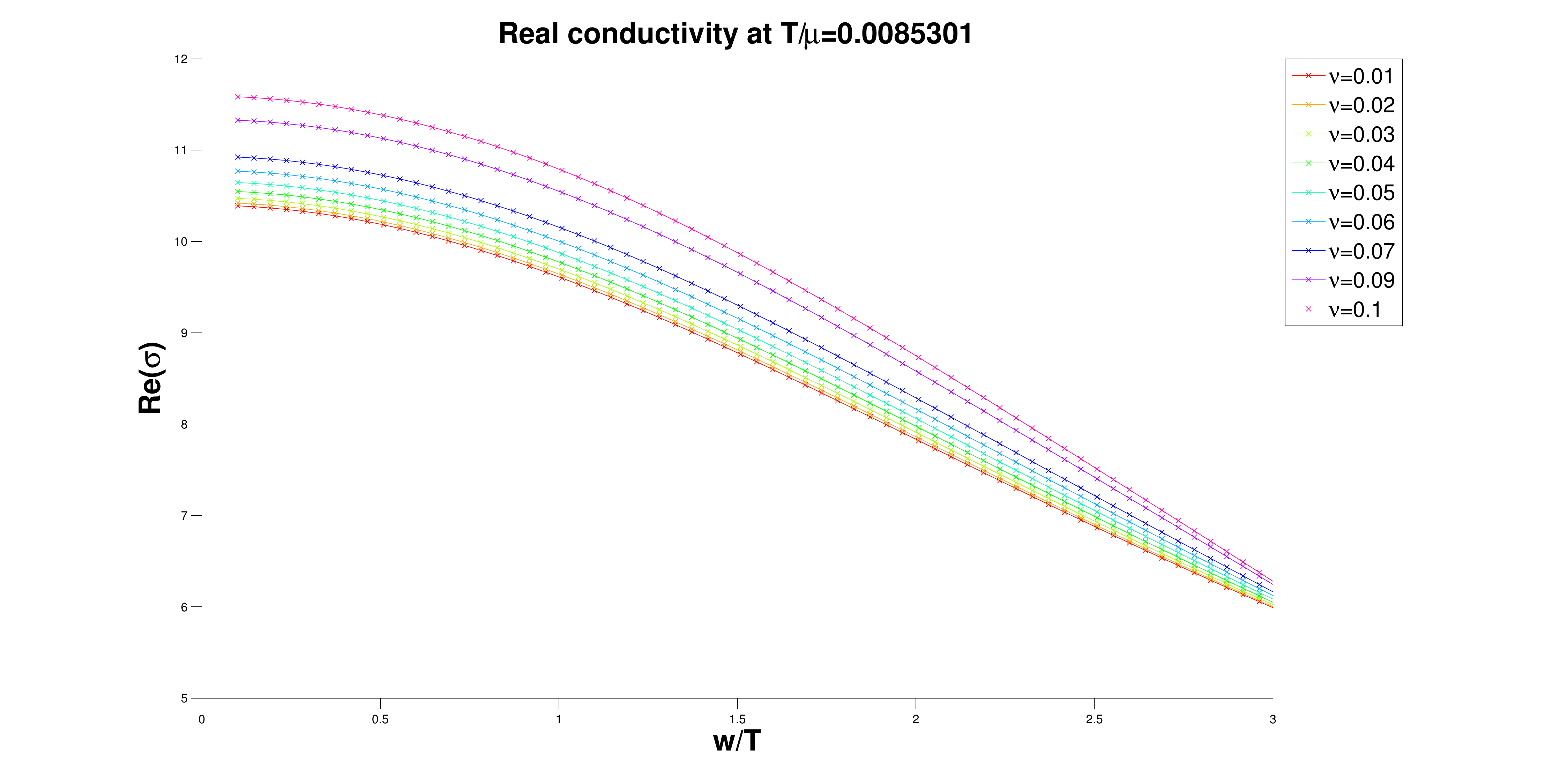}}%
\subfigure{\includegraphics[width=0.6\textwidth]{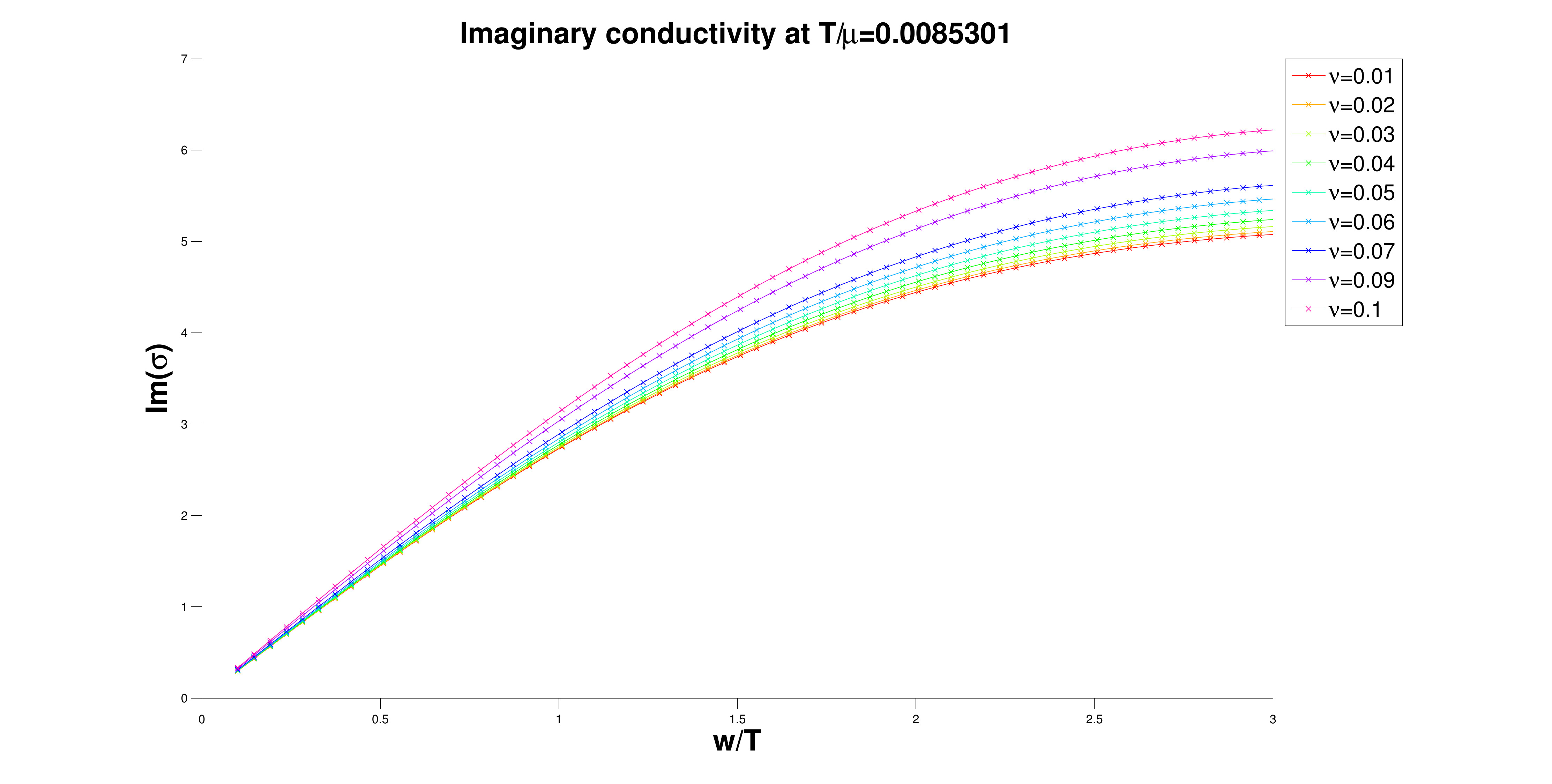}}%
}
    \caption[]{We examine the effect of the control parameter $\upsilon$ on the AC conductivity as $\frac{T}{\mu}$ is lowered. As the value of $\frac{T}{\mu}$ is lowered the influence of $\upsilon$ on the form of the real conductivity curves becomes more apparent. In this limit the curves for different values of $\upsilon$ can be seen to clearly differentiate at lower values of ${\mathfrak w}$.}
\label{fig:AC_analysis_upsilon}
\end{figure}

 It is interesting to examine the AC electrical conductivity in the vicinity of the transition between good metals and insulators. We observe that the influence of the $\upsilon$ parameter on the profile of the real and imaginary parts of the AC conductivity is minimal until $\frac{T}{\mu}$ is sufficiently small. This is in keeping with the profiles of the DC electric conductivity shown in Fig.~\ref{fig:elec_thermoelec_cond}, where the profiles do not begin to strongly differentiate until values of $\frac{T}{\mu} \simeq 0.06$ are reached. At lower values of $\frac{T}{\mu}$ the characteristic profiles associated with Drude behaviour begin to acquire a distinct spread as a function of $\upsilon$. This AC conductivity of manifestation of the metal-insulator transition, cf., Fig.~\ref{fig:AC_analysis_upsilon}. 
             
Our testing the zero frequency limit of the optical conductivity, agrees cleanly with the explicit evaluation of the DC conductivity using the membrane paradigm formulae. In the real part of $\sigma({\mathfrak w})$ we observe the characteristic Drude peak. We can confirm that the intercept $\sigma({\mathfrak w}=0)$ agrees with the DC result obtained from the membrane paradigm, providing us with a nice consistency check of the numerics.

\begin{figure}
\makebox[\linewidth]{%
\subfigure{\includegraphics[width=0.6\textwidth]{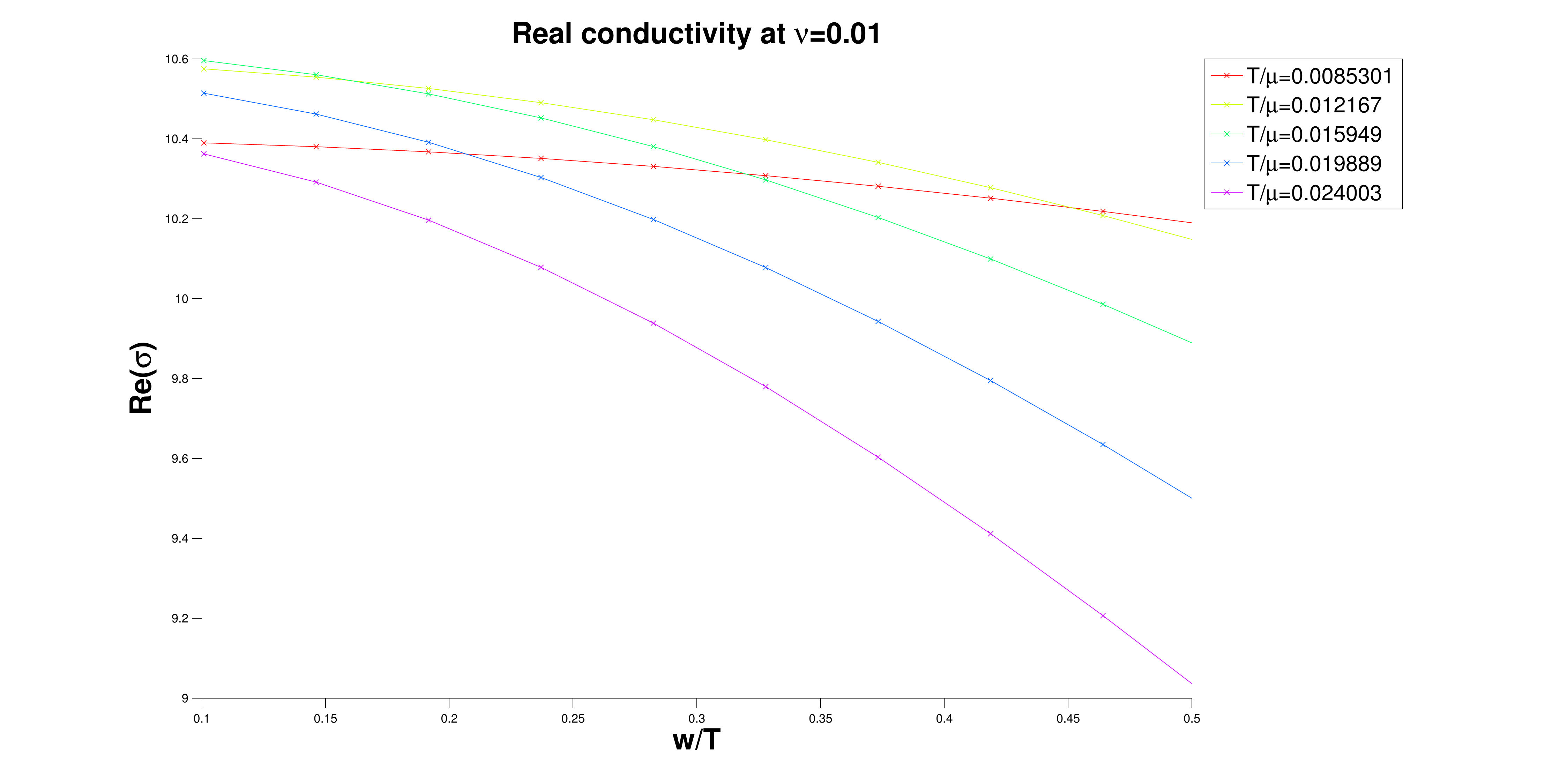}}%
\subfigure{\includegraphics[width=0.6\textwidth]{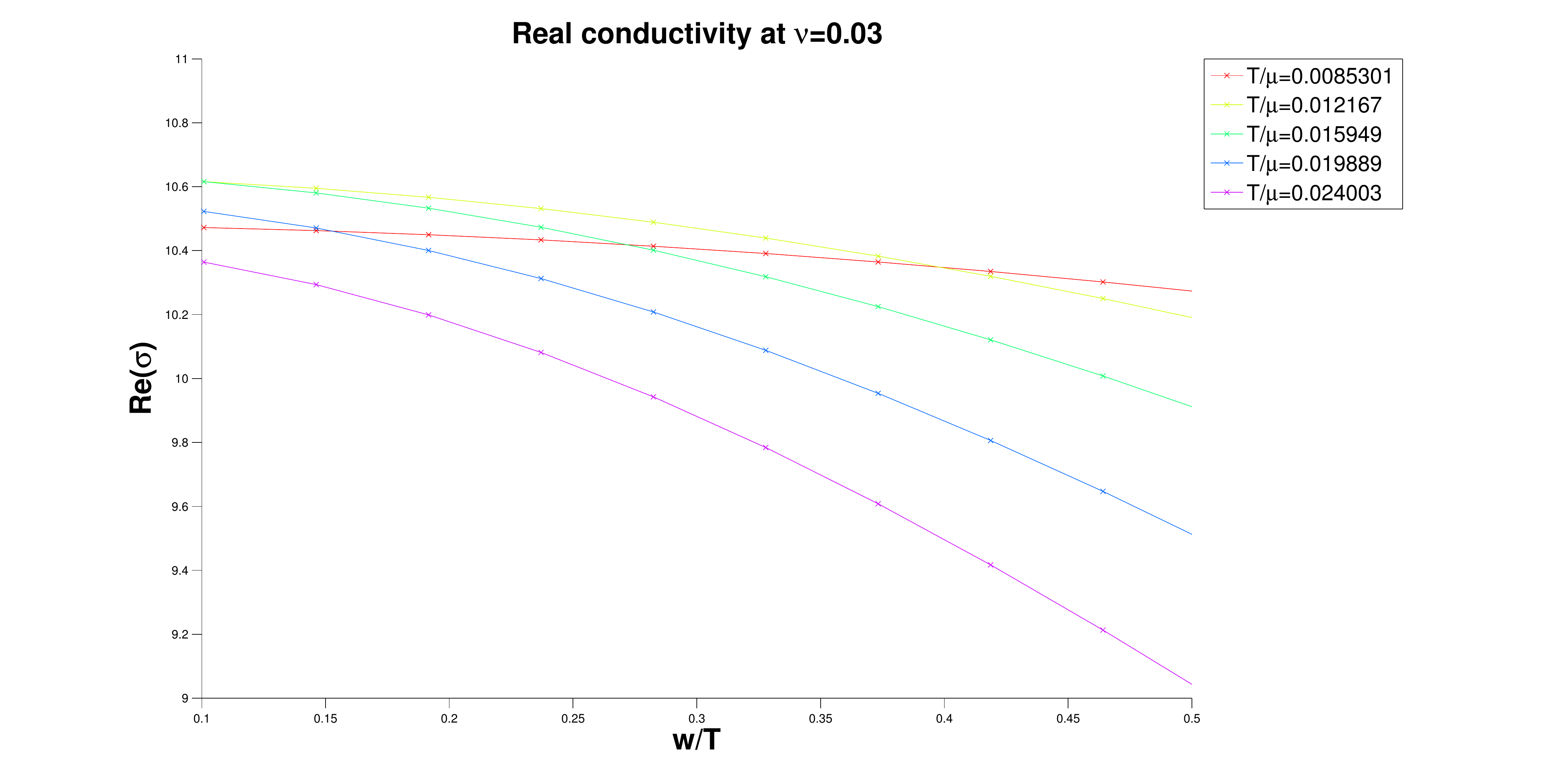}}%
}
\makebox[\linewidth]{%
\subfigure{\includegraphics[width=0.6\textwidth]{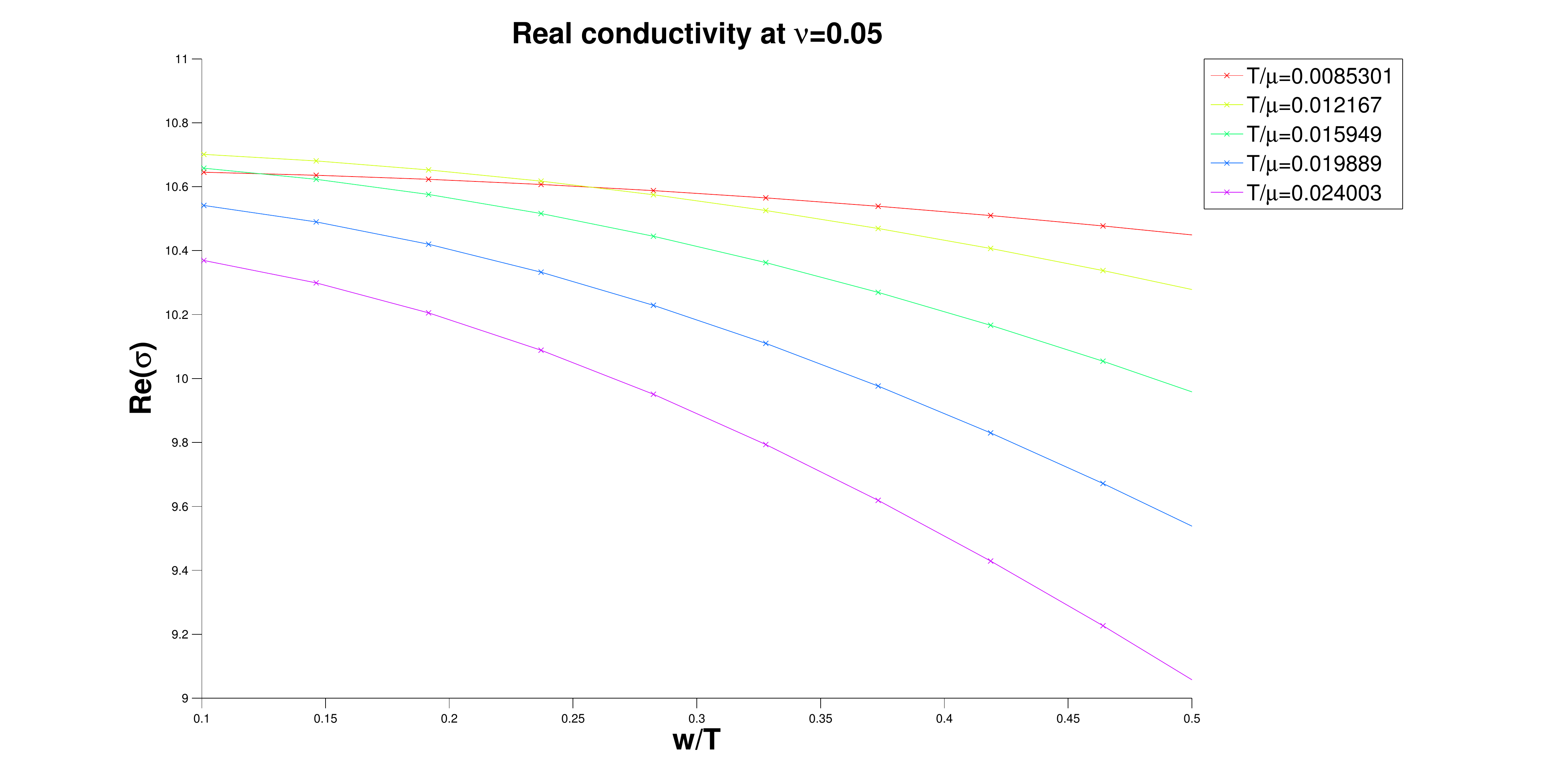}}%
\subfigure{\includegraphics[width=0.6\textwidth]{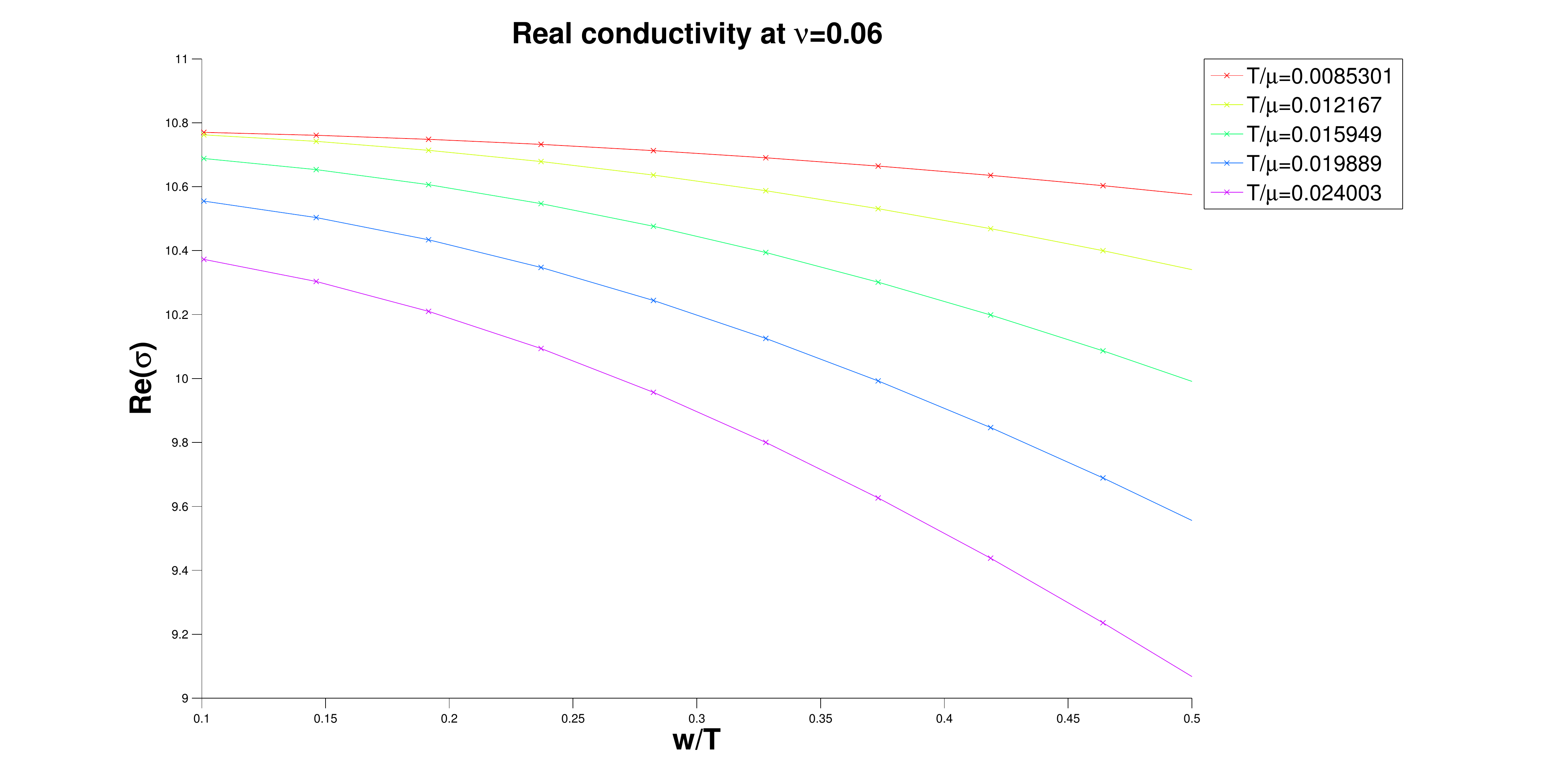}}%
}
\makebox[\linewidth]{%
\subfigure{\includegraphics[width=0.6\textwidth]{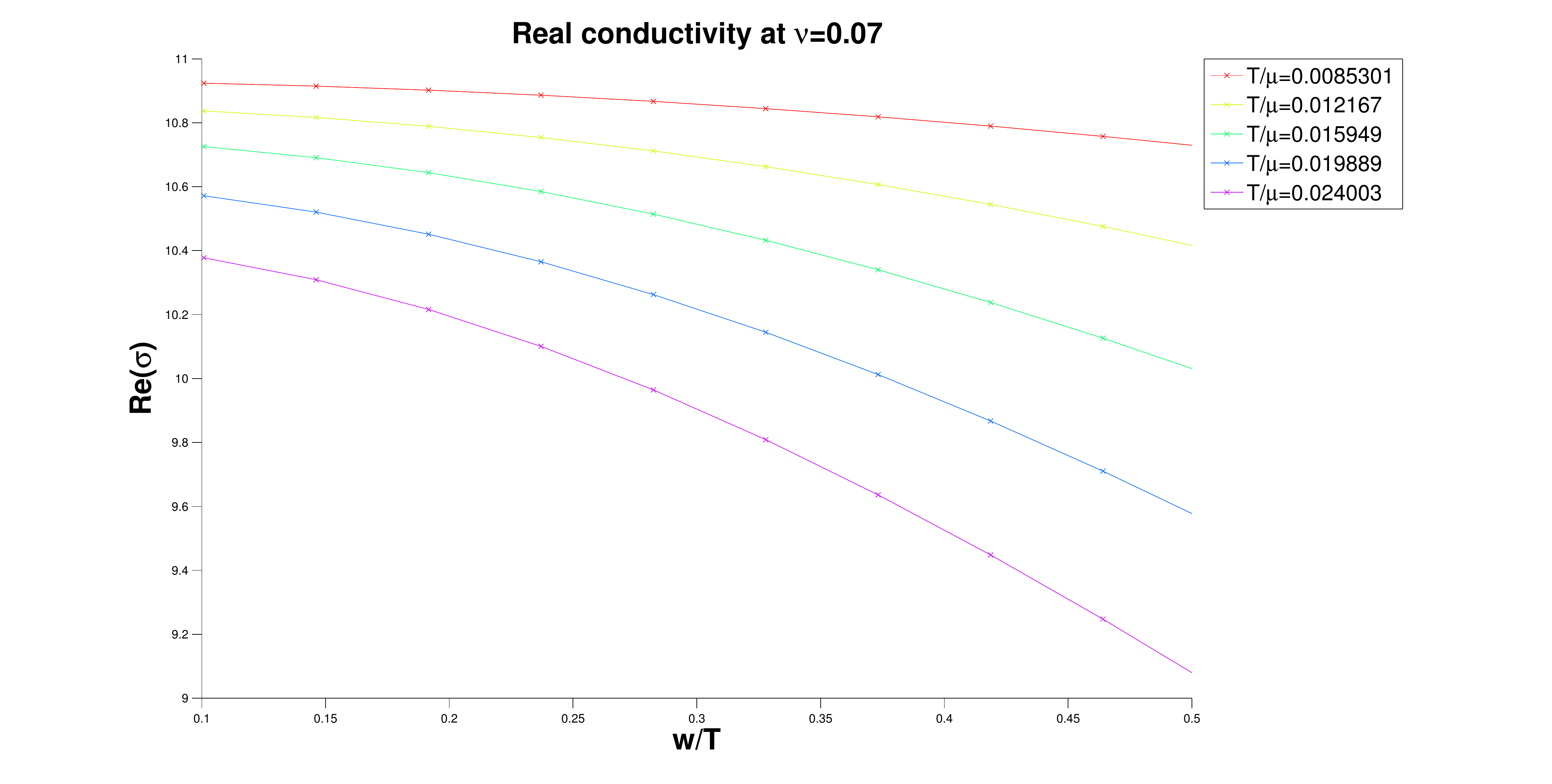}}%
\subfigure{\includegraphics[width=0.6\textwidth]{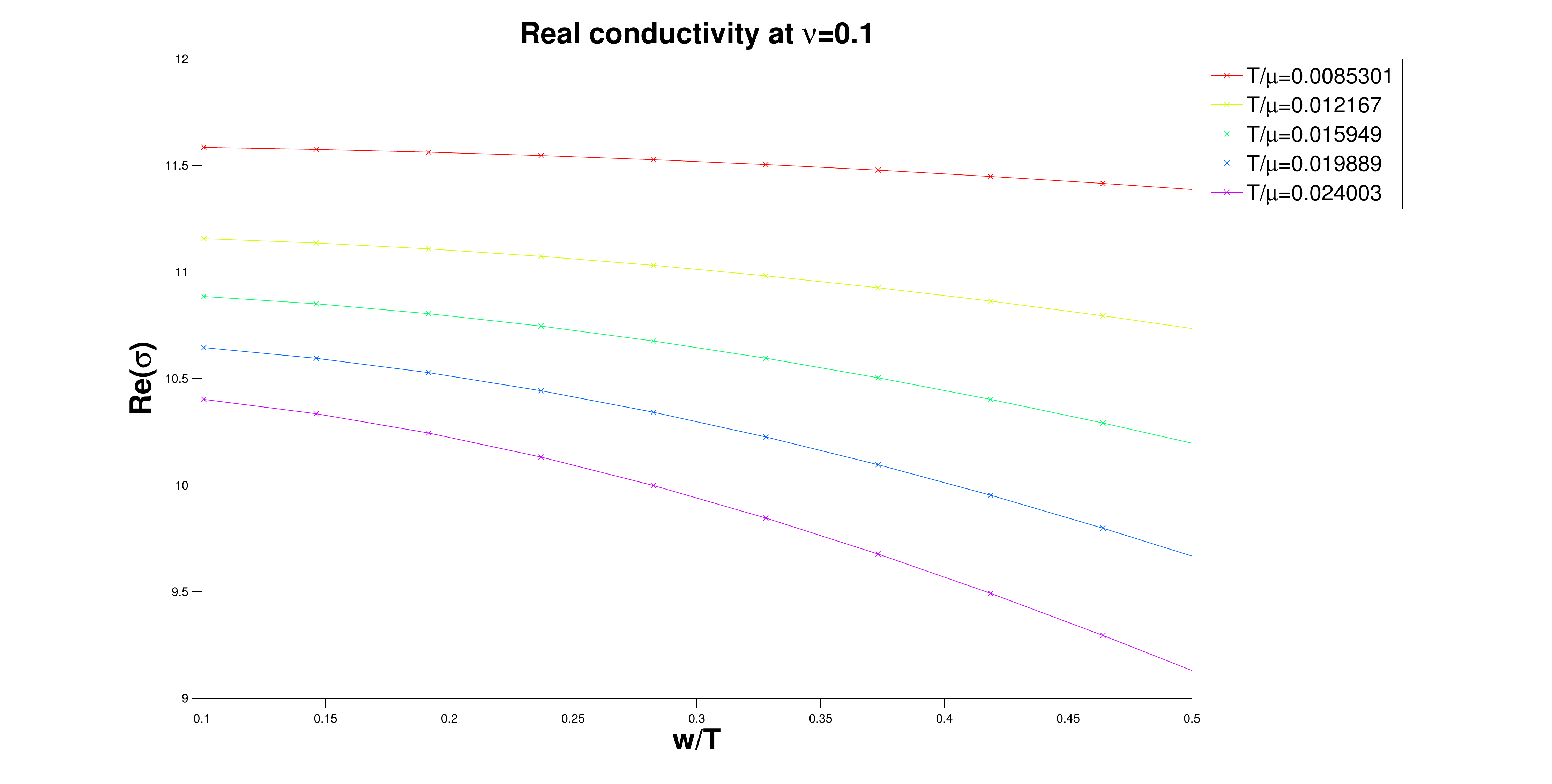}}%
}
\caption[]{Examining the real part of the AC conductivity as a function of  $\frac{T}{\mu}$ for a variety of $\upsilon$. At small $\upsilon$ the constant $\frac{T}{\mu}$ curves intercept and overlap for lower values of  ${\mathfrak w}$. This can be interpreted as the AC representation of the incoherent phase. As $\upsilon$ is increased the constant $\frac{T}{\mu}$ curves separate and differentiate as we transition into a conducting phase.}
\label{fig:AC_analysis_temperature}
\end{figure}

Inspection of the low frequency optical conductivity lends further evidence to our physical picture of the low temperature transport. In Fig.~\ref{fig:AC_analysis_temperature} we confirm that as the temperature is lowered for parameter choice in the insulating phase spectral weight is shifted from lower to higher frequencies, as expected.

We also check for potential mid-IR scaling regimes. A convenient way to identify this behaviour is using the diagnostic quantity which we label $F({\mathfrak w})$ (cf., \cite{Donos:2014yya}):
\begin{align}\label{eq:IR_scal_diag}
F({\mathfrak w})&=1+{\mathfrak w} \,\frac{|\sigma({\mathfrak w})|''}{|\sigma({\mathfrak w})|'}
\end{align}
If $|\sigma({\mathfrak w})|$ develops a scaling regime such that it behaves as $|\sigma({\mathfrak w})|\sim C_{1} + C_{2} \, {\mathfrak w}^{\nu}$, the quantity $F({\mathfrak w})$ will be equal to a constant given by the scaling exponent $\nu$.

\begin{figure}
\makebox[\linewidth]{%
\subfigure{\includegraphics[width=0.6\textwidth]{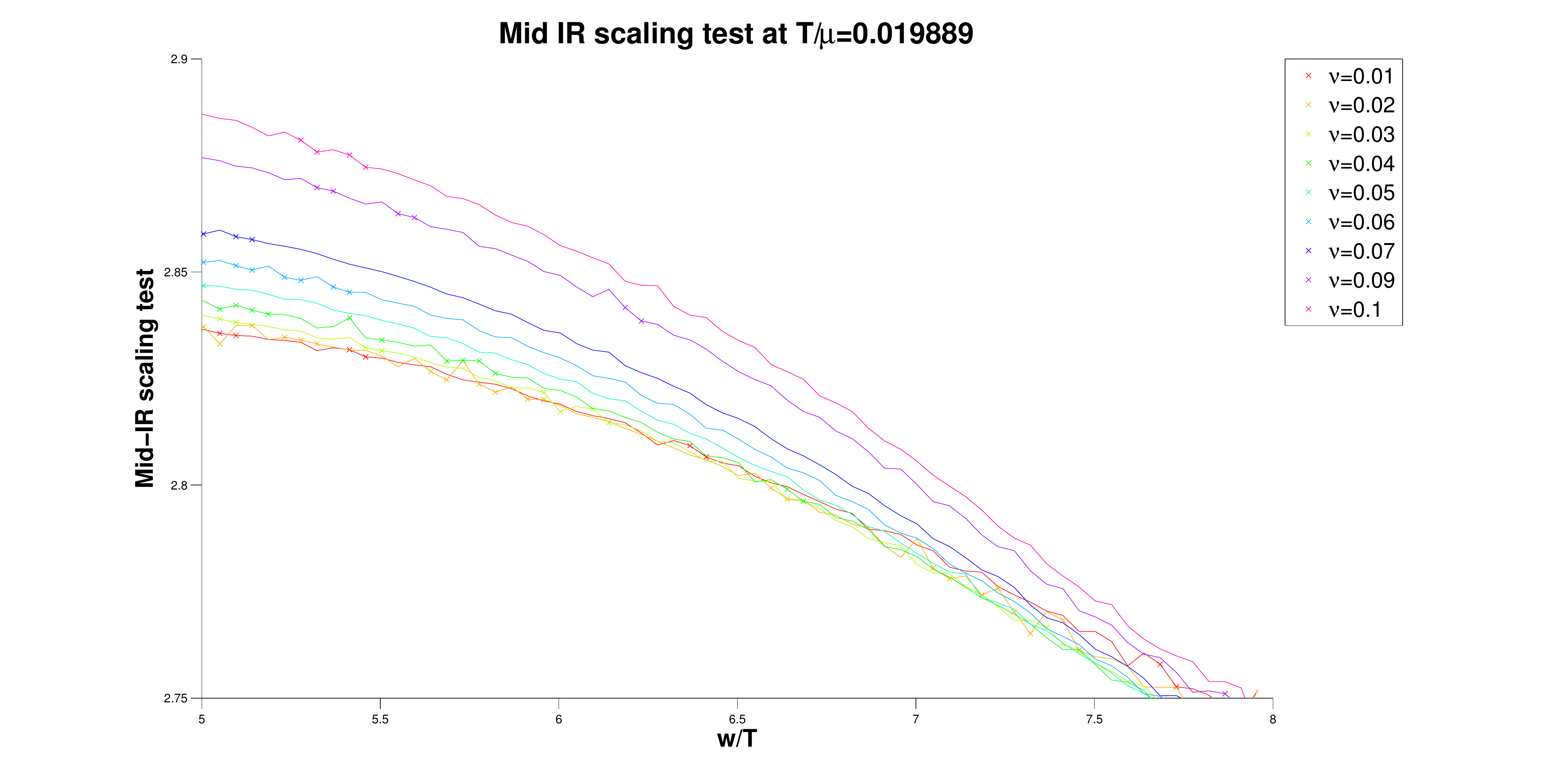}}
\subfigure{\includegraphics[width=0.6\textwidth]{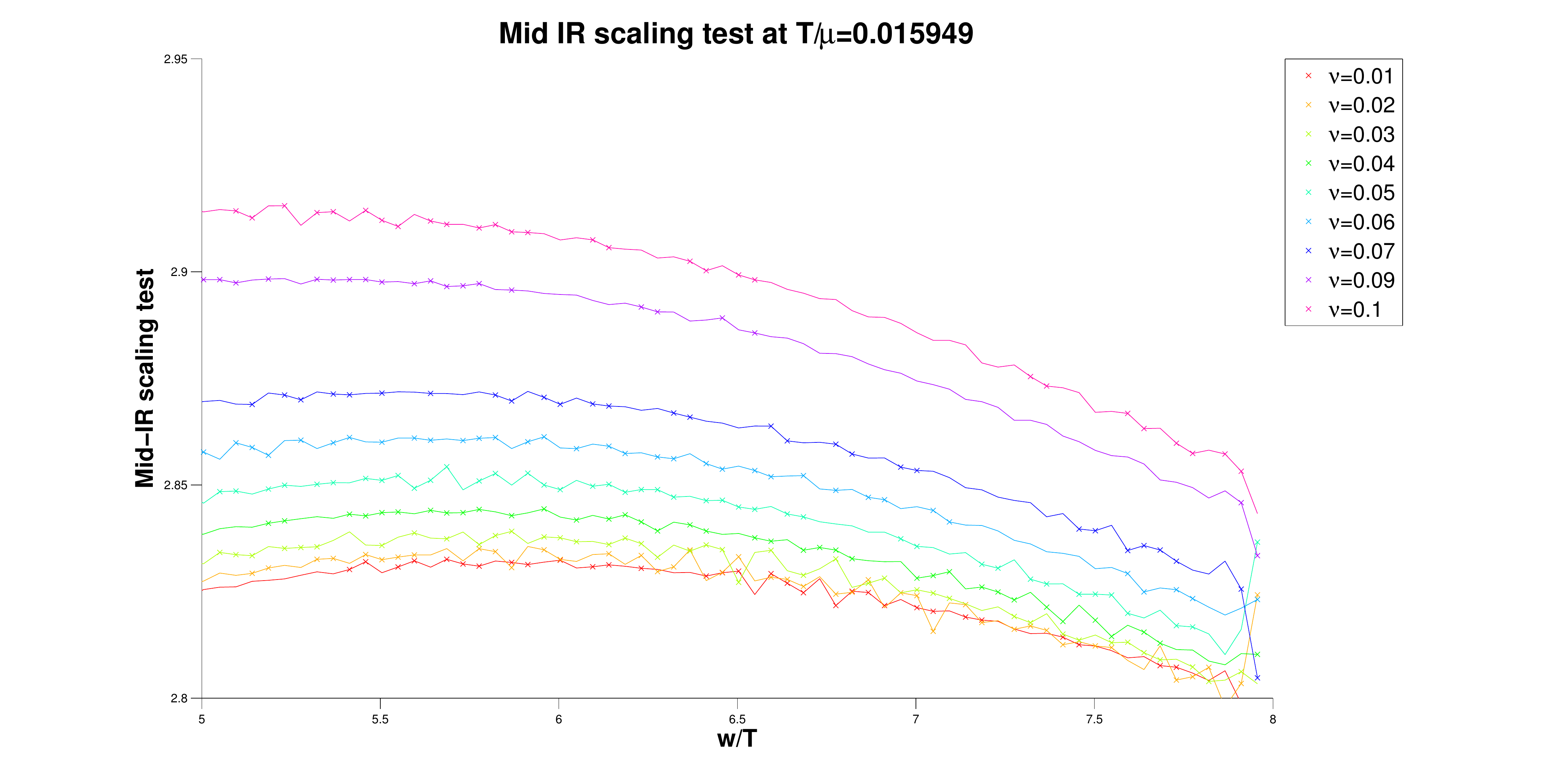}}
}
\makebox[\linewidth]{%
\subfigure{\includegraphics[width=0.6\textwidth]{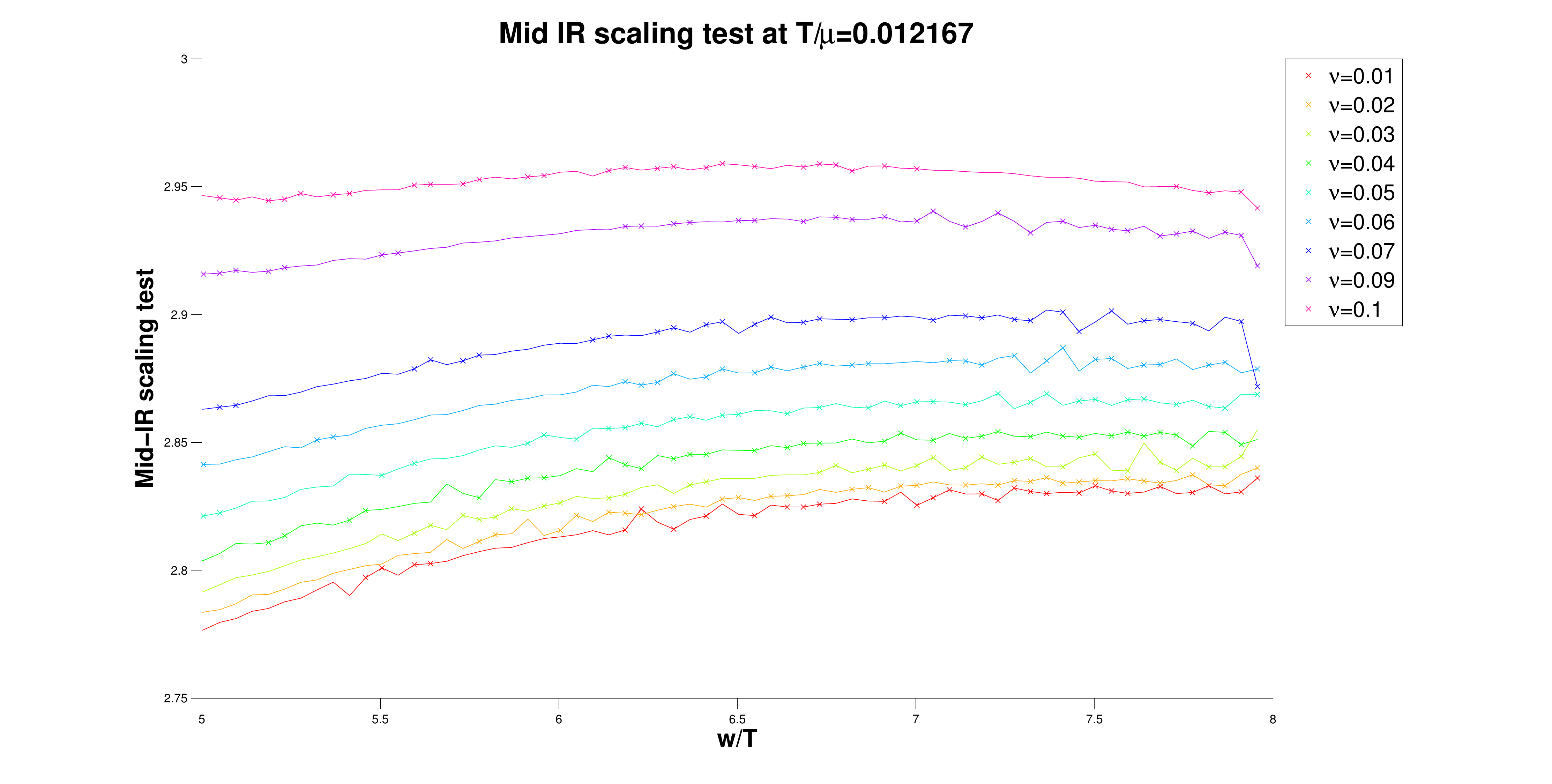}}
\subfigure{\includegraphics[width=0.6\textwidth]{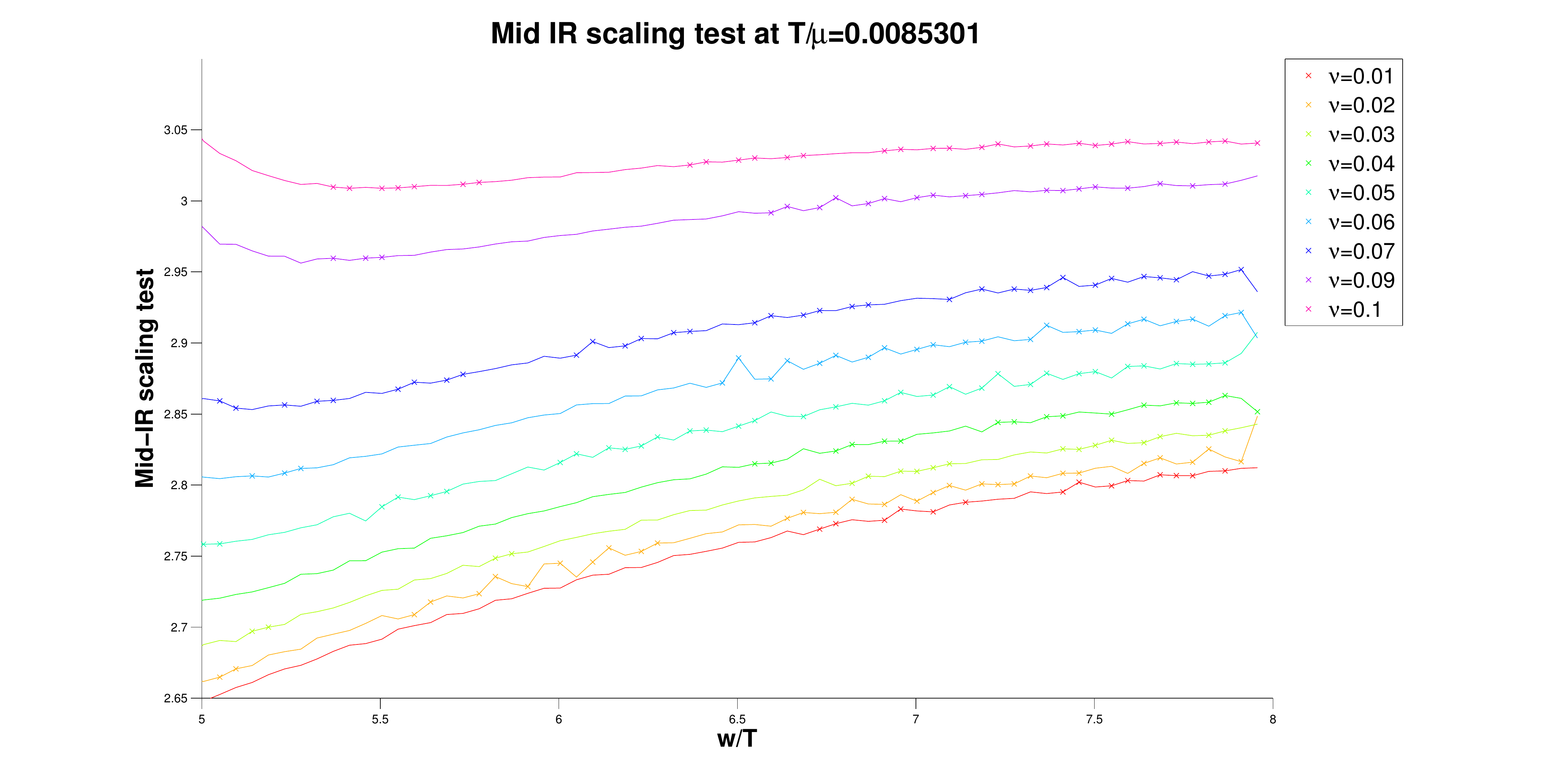}}
}
    \caption[]{The behaviour of the diagnostic function $F({\mathfrak w})$ as we scan for mid-range scaling behaviour as a function of $\upsilon$. At low temperatures, and for appropriately chosen values of $\upsilon$,  the existence of a scaling regime is possible. For example, at temperature of $\frac{T}{\mu}=0.01267$ we observe that the our test function flattens out at  $\upsilon \simeq 0.09$.}
\label{fig:IR_scaling_upsilon}
\end{figure}

\begin{figure}
\makebox[\linewidth]{%
\subfigure{\includegraphics[width=0.6\textwidth]{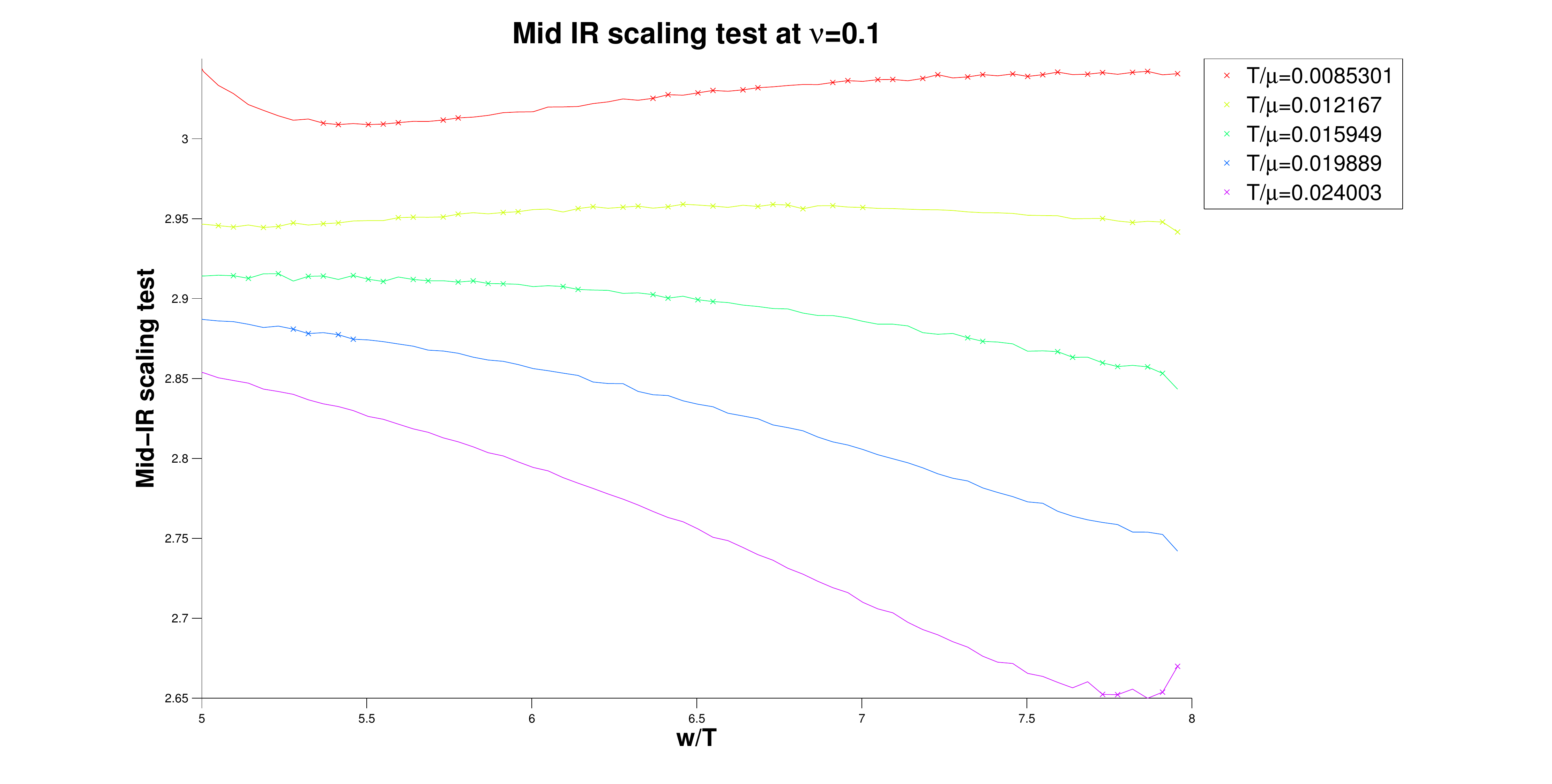}}
\subfigure{\includegraphics[width=0.6\textwidth]{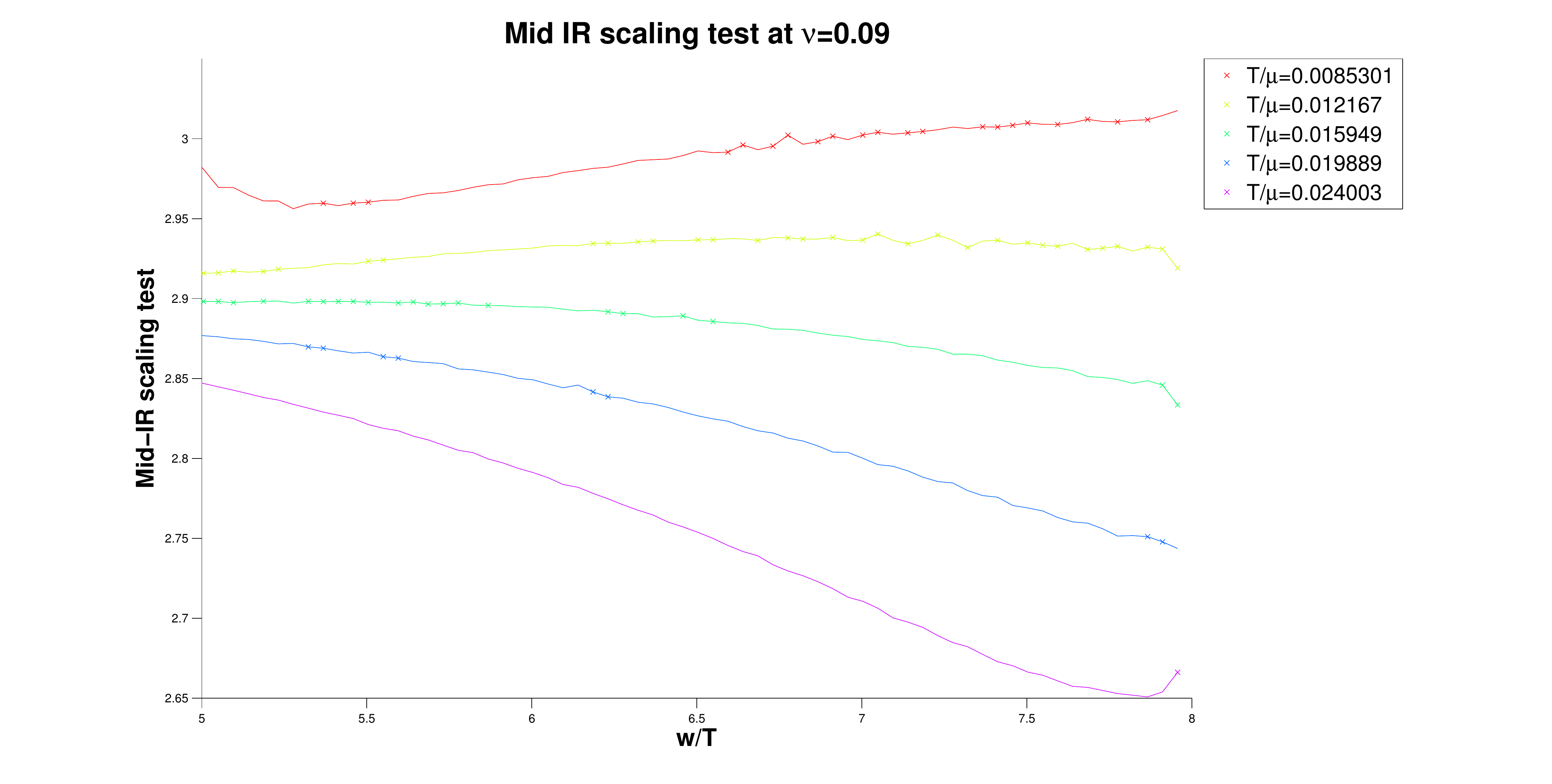}}
}
\makebox[\linewidth]{%
\subfigure{\includegraphics[width=0.6\textwidth]{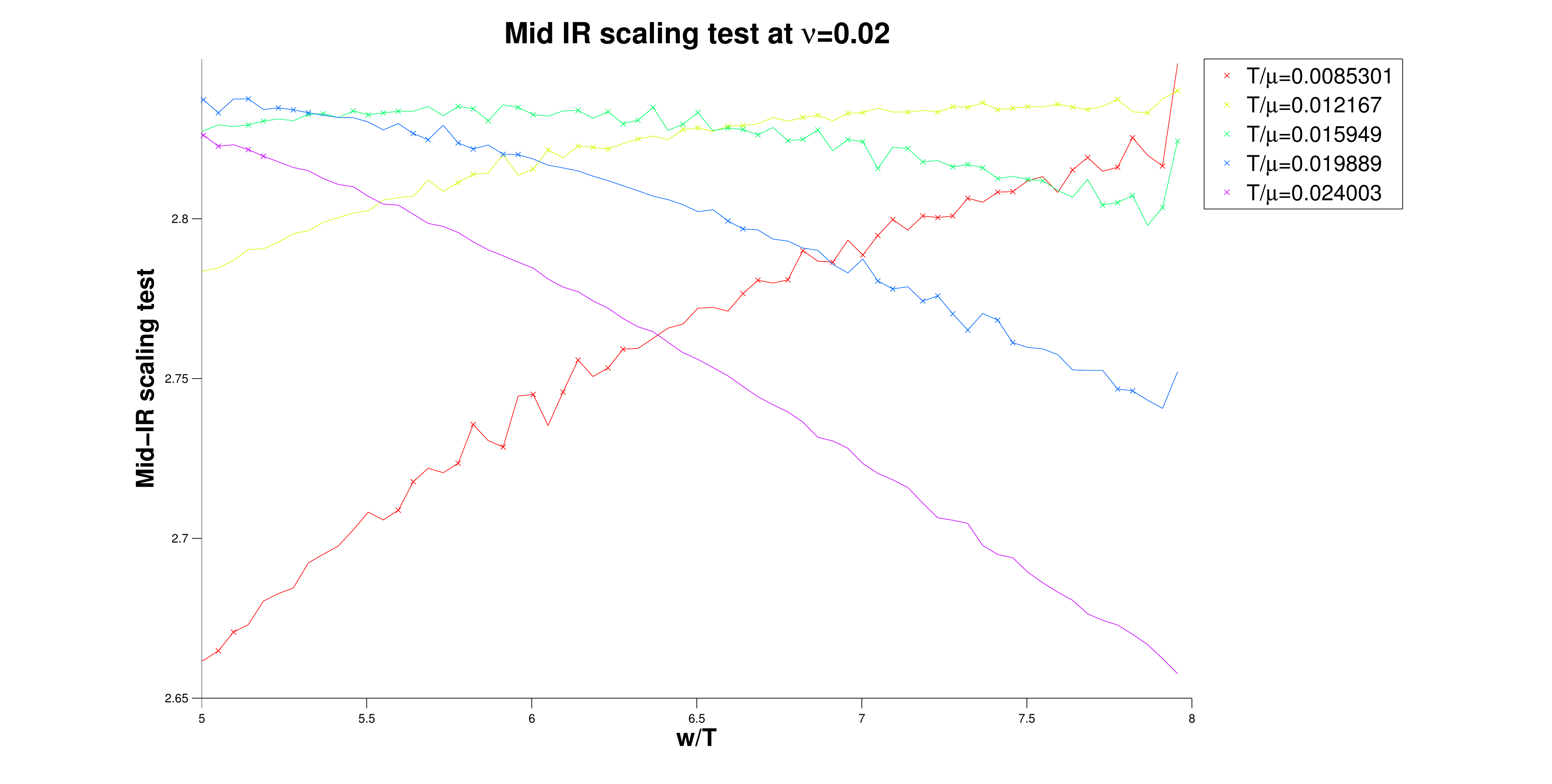}}
\subfigure{\includegraphics[width=0.6\textwidth]{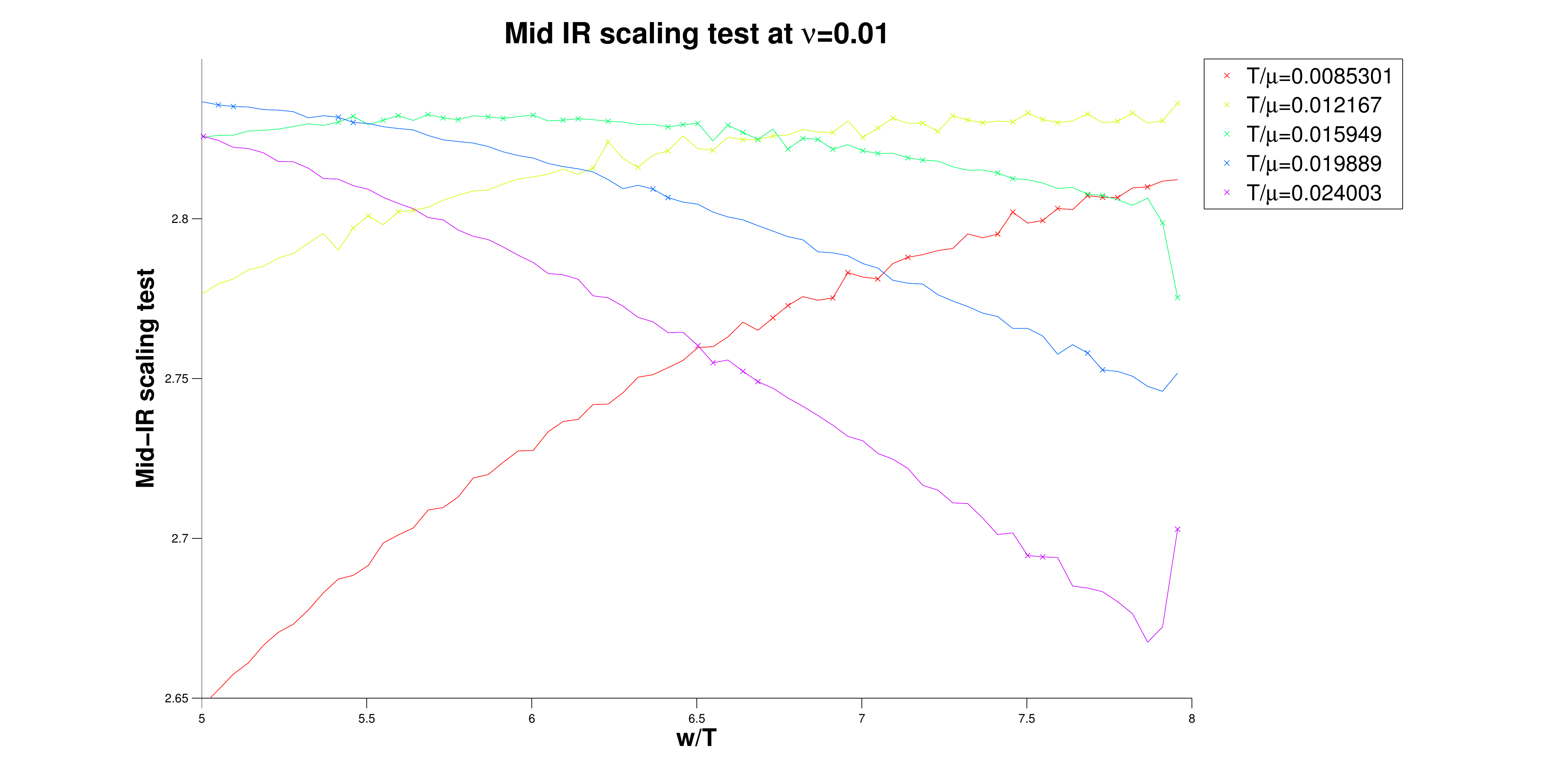}}
}
    \caption[]{The behaviour of the diagnostic function $F({\mathfrak w})$ as we scan for mid-range scaling behaviour as a function of temperature. The reversal of the slopes of $F({\mathfrak w})$ as a function of temperature, for a fixed $\upsilon$, imply that it should be possible to tune to a scaling regime by appropriately specifying the temperature.}
\label{fig:IR_scaling_temperature}
\end{figure}

In Figs.~\ref{fig:IR_scaling_upsilon} and \ref{fig:IR_scaling_temperature} we plot $F({\mathfrak w})$ for a variety of values of $\frac{T}{\mu}$ and $\upsilon$, including those displayed in Fig.~\ref{fig:AC_analysis_upsilon}. We observe that as $\frac{T}{\mu}$ is decreased the profiles of the diagnostics $F({\mathfrak w})$ begin to flatten out and the existence of a scaling regime becomes a real possibility.  However the existence of such a scaling regime requires the fine tuning of $\upsilon$ (or $\frac{T}{\mu}$) to a very narrow parameter range. This ranges questions regarding the generic nature and therefore physical significance of the behaviour. However, even when scaling regimes do exist, the scaling exponent seem to be generically  different than 2/3. Further questions regarding the significance and robustness of this scaling regime is postponed to future work.\footnote{ For a discussion of the difficulties involved in working in this frequency regime please see Appendix \ref{sec:Appendix1}.}

 \subsection{High Temperature Limit}
 \label{subsubsec:highT_limit}    

As described in \S\ref{subsec:anal_cond} the form of the response functions should exhibit high temperature behaviour consistent with that of the  Schwarzschild-\AdS{4} black hole. In this limit $\sigma \rightarrow 1$, $\alpha=\bar{\alpha} \rightarrow 0$ and $\bar{\kappa} \rightarrow \infty$. 
Therefore both the metallic and insulating low temperature phases must transition to that generic behaviour, determined by the conformal invariance of the UV theory, as the temperature is increased.  
        
An interesting point to note is that, as seen in Fig.~\ref{fig:elec_thermoelec_cond}, the DC electrical conductivity increases to values well above unity in the insulating phase. This can be understood as consequence of the sum rule obeyed by the optical conductivity -- the suppressed low frequency spectral weight in the insulating phase has to transfer to high frequencies, which are still low compared to the scale of the chemical potential.  Therefore for those insulating phases, a turning point must exist when the sign of the slope of the DC electric conductivity reverses, as the low temperature physics begins to transition to the generic high temperature behaviour. Conversely, the existence of such turning point could be another indication of the transition to an insulating phase.
     
\begin{figure}
\makebox[\linewidth]{%
\subfigure{\includegraphics[width=0.5\textwidth]{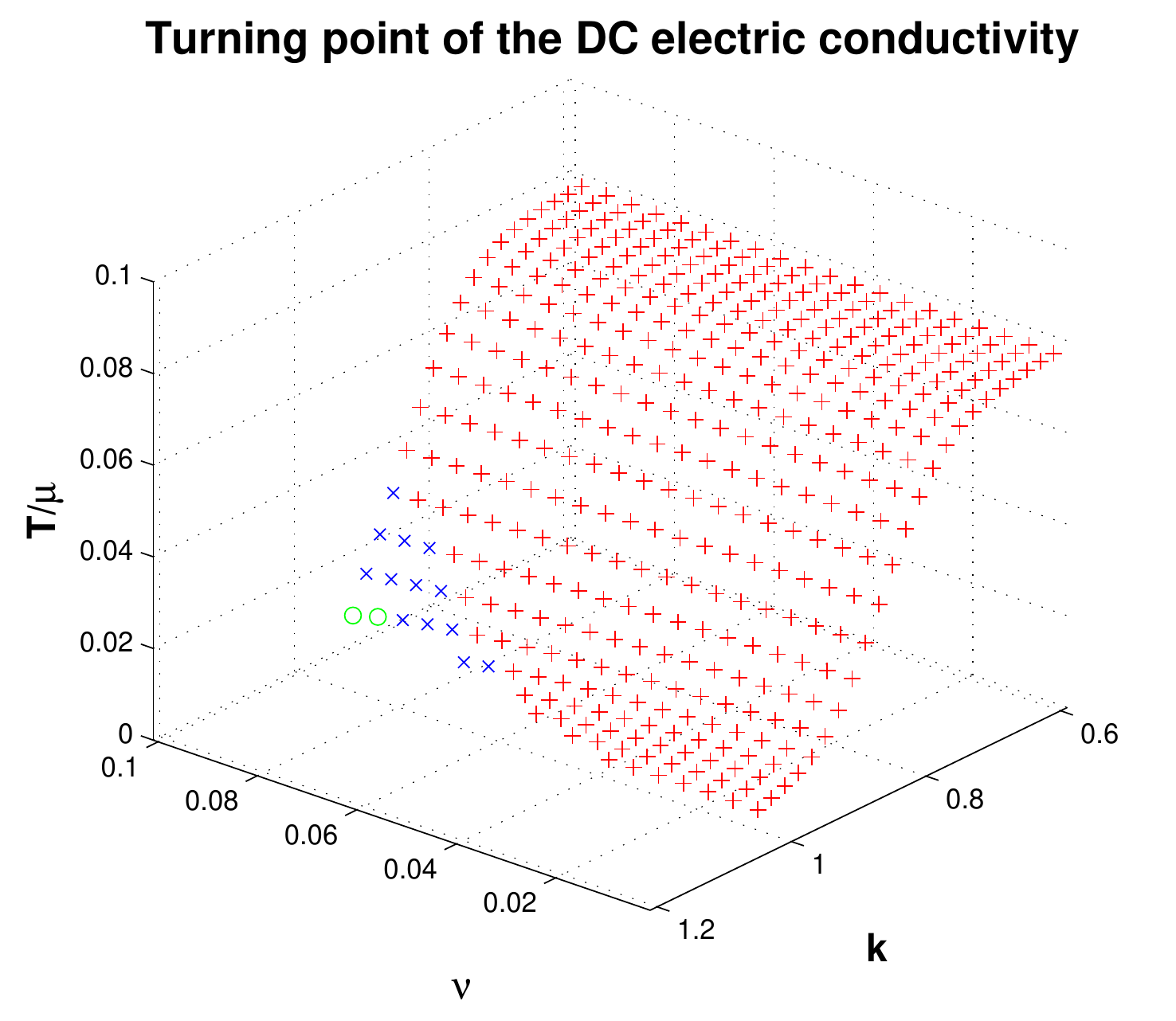}}
\subfigure{\includegraphics[width=0.5\textwidth]{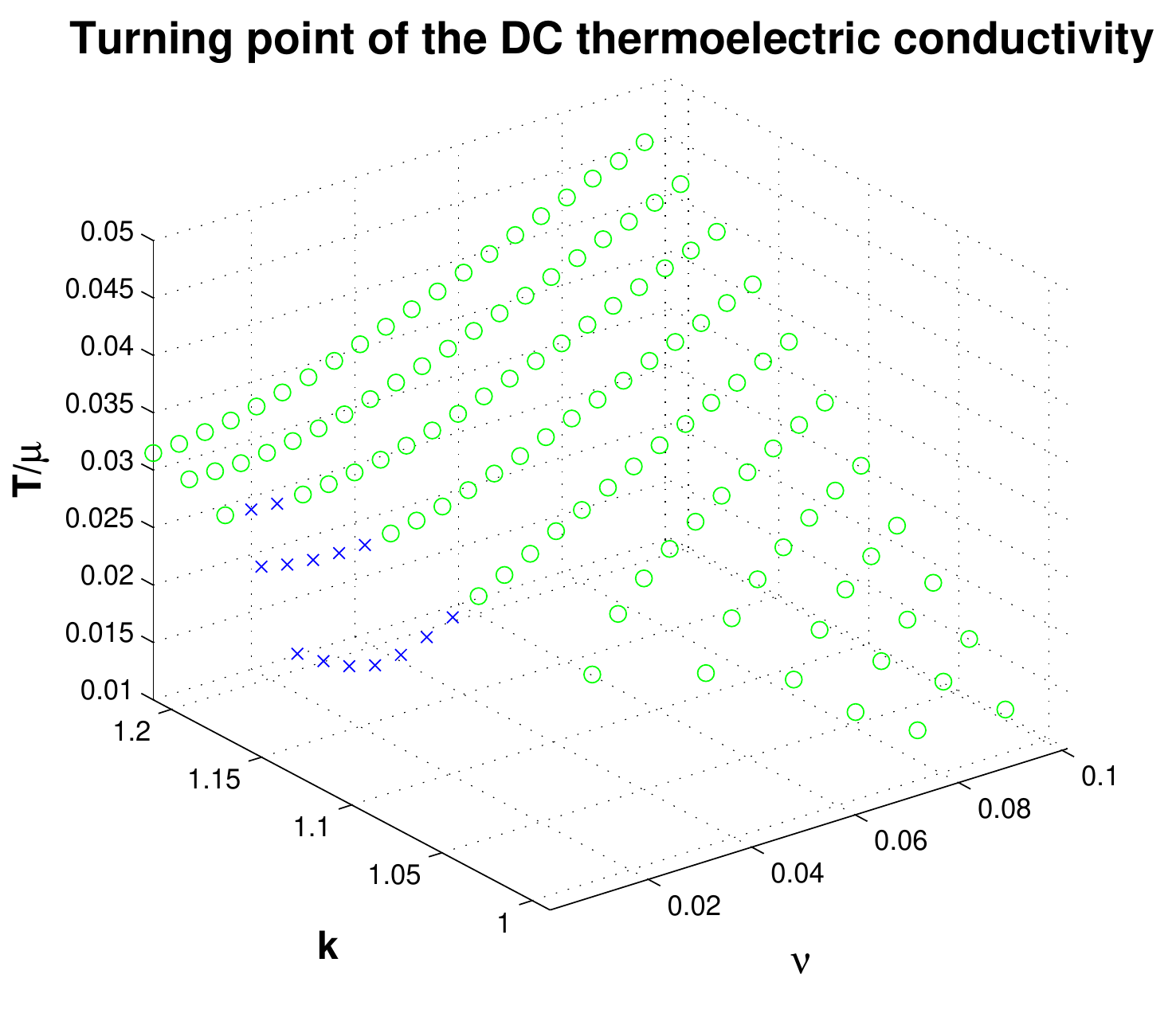}}
}
    \caption[]{The DC electric conductivity starts off at a finite value of  $\Gamma_\text{crit} = \frac{T}{\mu}$ in the insulating phase and decreases rapidly as we approach the metal-insulator transition. On the other hand, in the thermoelectric conductivity, the existence of a turning point sets in as we enter the metallic phase and steadily increases in $\frac{T}{\mu}$ as we tune $\upsilon$ and $k$ to move deeper into the metallic phase. }
\label{fig:turning_point}
\end{figure}

In Fig.~\ref{fig:turning_point} we plot the values of $\Gamma_\text{crit}=\frac{T}{\mu}$ at the turning point versus $\upsilon$ and $k$. For the DC electrical conductivity it can be seen that the position of this turning point decreases steadily as one moves from the insulating regime towards the metallic transition. We also plot the corresponding turning point in the DC thermoelectric conductivity (right panel). It should be noted however that a further, higher temperature, change of slope must exist for the thermoelectric conductivity, as it is expected to go to zero in the high temperature limit.

%

 \section{Conclusions and outlook}
 \label{sec:discuss} 

The main goal of this article has been a preliminary exploration of the large phase space of effective holographic theories, with a particular focus on understanding the efficiency of low energy transport in these models.  More specifically we focused on the low frequency conductivity in a phenomenologically motivated AdS/CFT set-up and examined the predilection of the system towards  metal-insulator phase transitions. We view our investigation as  offering a large set of toy models, with features that can be dramatically different from those of the much studied Einstein-Maxwell theory. Our analysis has identified interesting loci in the phase diagram where qualitative changes of the IR physics take place. The  natural next step is taking a closer look at these locations to get a better understanding of the physics that drives these transitions in holographic systems.

One of the interesting regimes is that of incoherent, or bad metals which lie at the interface of the metal-insulator transition. A conjecture was put forward in \cite{Hartnoll:2014lpa}, that transport in such incoherent metals was to be governed by diffusion processes. The models studied here provide a natural testing ground for this conjecture.

Additionally, it is interesting to study the low lying quasinormal modes in the incoherent and insulating phases. The motion of those quasinormal modes in the complex plane can often elucidate the dominant physics governing the transition or cross-overs in the qualitative behaviour of the conductivity. In this context, it would also be interesting to quantify the scaling of the spectral weight and DC conductivity at low frequencies. In particular, examination of the residue of the lightest quasinormal mode (the hydrodynamic mode that contributes to charge diffusion),  should illuminate whether the reduction in the conductivity as we enter the regime of incoherent metals (from the metallic side) is caused by the drop in spectral weight or if some other physics is responsible. 

From a  gravitational viewpoint it would be useful to know if there is a characteristic feature of the near-horizon geometry which results in this phenomenon. Our preliminary investigations were inconclusive in ascertaining a sharp feature of the near-horizon geometry, which could be held responsible for the incoherence in charge transport. Ideally one would conjure up a geometric observable that is sensitive to transport. What we can definitely conclude is that for fixed $\upsilon$  varying $k$ demonstrates clear changes in the relevance of the inhomogeneity in all the metric components. Since decreasing $k$ increases the relevance of the inhomogeneous scalar source, this makes it clear that smaller values of $k$ will have a more stronger impact in the IR transport. Comparing across values of $\upsilon$ is of course more complicated, since we exploring behaviour in the space of theories. Indeed even in the homogeneous case we see differing IR behaviour as we tune $\upsilon$. 

We have further seen that in certain small regimes of our parameter space intermediate frequency scaling may be possible. While our analysis of this effect has not been comprehensive, it appears that we require a certain amount of fine-tuning in order to achieve scaling behaviour. Moreover, in most of the cases we looked at the exponent $\nu$ was not $2/3$, which was the value seen for the cuprates \cite{van2006scaling} and in $\upsilon =0$ models \cite{Horowitz:2012gs,Horowitz:2012ky}. While this is per se not surprising, since our explorations with $\upsilon \neq 0$ move us in the space of UV field theories, it is curious that the effect is rather dramatic in the low frequency conductivity.

In conclusion, we hope that the results presented in this work illustrate that the class of models considered provides an interesting phenomenological environment for exploring detailed features of charge transport in holographic systems. We hope to continue this exploration, and address some of the interesting questions raised in the course of this work, in the not too distant future.

\acknowledgments 

We thank Aristos Donos, Jerome Gauntlett, and Ben Withers for interesting discussions.
M.~Rangamani would also like to thank KITP, Santa Barbara for hospitality during the final stages of this project.
He was  supported in part by the STFC Consolidated Grant ST/L000407/1, by the European Research Council under the European Union's Seventh Framework Programme (FP7/2007-2013), ERC Consolidator Grant Agreement ERC-2013-CoG-615443: SPiN (Symmetry Principles in Nature),  and by the National Science Foundation under Grant No. NSF PHY11-25915 (KITP). 
M.~Rozali and D.~Smyth are supported by a discovery grant from NSERC of Canada.

\appendix
 \section{Code Analysis}
 \label{sec:Appendix1} 
 
 In this appendix we provide further information on our numerical methods and exhibit some checks on the results presented in this paper.  We first briefly describe our approach to solving the background and perturbation equations. We then present our convergence results and other checks of the numerics.

 \subsection{Numerical Procedure and Implementation Details}

The derivation of the equations, boundary conditions and gauge conditions were done in Mathematica. These expressions, in appropriately discretized form, were then exported to Matlab which served as the principal platform for equation solving and post-processing work. Significant portions of the code were transferred to  C\verb!++! code, which utilized the Armadillo and BLAS linear algebra libraries. 

The non-linear nature of the background equations of motion requires an iterative process for the solution. This was accomplished using a combination of Newton method with line search and the quasi-Newton, Broyden method algorithms. The Broyden method was found to be the most efficient approach, as it did not necessitate the computationally expensive process of assembling the Jacobian matrix. As the results presented in this paper rely on knowledge of the temperature dependence of the conductivity at each point in the $\upsilon, k$ parameter space, it was necessary to numerically solve the PDEs many times. As such the utilization of optimized code and efficient solver algorithms was important in maintaining a manageable computational load. An example of a useful solver strategy for the background was:
\begin{itemize}
\item First attempt to use the Broyden method. If the initial guess is sufficiently close to the sought after solution, convergence should be reached rapidly.
\item  If the Broyden method fails to converge the solver switches to Newton method augmented with a three-point safeguarded parabolic line search. Once the norm of the residual has reduced below a safe tolerance, the Broyden method can once again be utilized to quickly bring about convergence to the required accuracy.
\item  Once an initial solution is obtained suitably sized (adiabatic) variations in parameters allow for the mapping of parameter space, without having to resort to the Newton method. 
\end{itemize}
        The solution of the perturbation equations is in some sense easier as the linear nature of the equations means that they can be inverted in one step without need for iteration. An additional complication is introduced by the fact that knowledge of the temperature dependence of the AC conductivity requires us to scan over both temperature and frequency at each value of the parameters $\upsilon,  k$. As the quality of the numerical convergence is not uniform with temperature or frequency, we must be careful to establish which regions of the $(T, \omega, \upsilon, k)$ space we can reliably probe and bear this in mind when analyzing our results. We also must be careful to ensure that we have sufficiently resolved the background solution relative to the desired resolution for the perturbation solution.
        
\subsection{Convergence Tests}
\label{sec:conv}

        We now present the following convergence results for the background and perturbation solutions:      
\begin{itemize}
\item The background solution:
 \begin{itemize}
 \item Convergence of the solutions to equations \eqref{eq:background_eqns} as a function of the grid size $N$.
\item Convergence of the DeTurck gauge condition to zero as a function of $N$.
 \end{itemize}
\item The linear perturbation solutions:
 \begin{itemize}
\item  Convergence as  function of the grid size $N_p$ for the linearized equations.
\item  Convergence of the deDonder and Lorentz gauge conditions \eqref{eq:deDonder_Lorentz} to zero as a function of $N_p$.
\item  Convergence of the auxiliary (unimposed) horizon boundary conditions, as described below equation \eqref{eq:form_pert}, to zero as function of $N_p$.
 \end{itemize}  
\end{itemize}  

\begin{figure}
\makebox[\linewidth]{%
\subfigure{\includegraphics[width=0.6\textwidth]{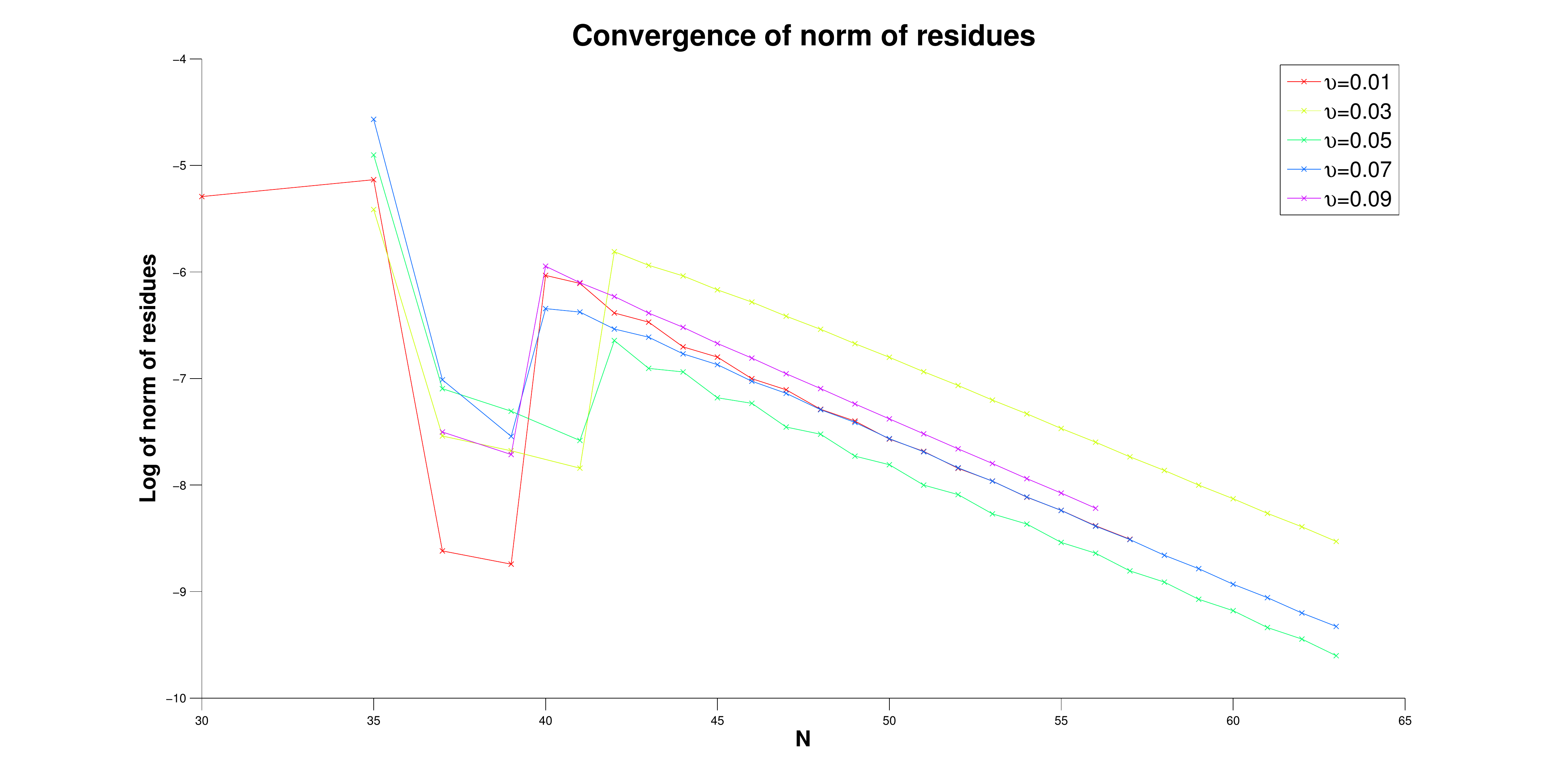}}
\subfigure{\includegraphics[width=0.6\textwidth]{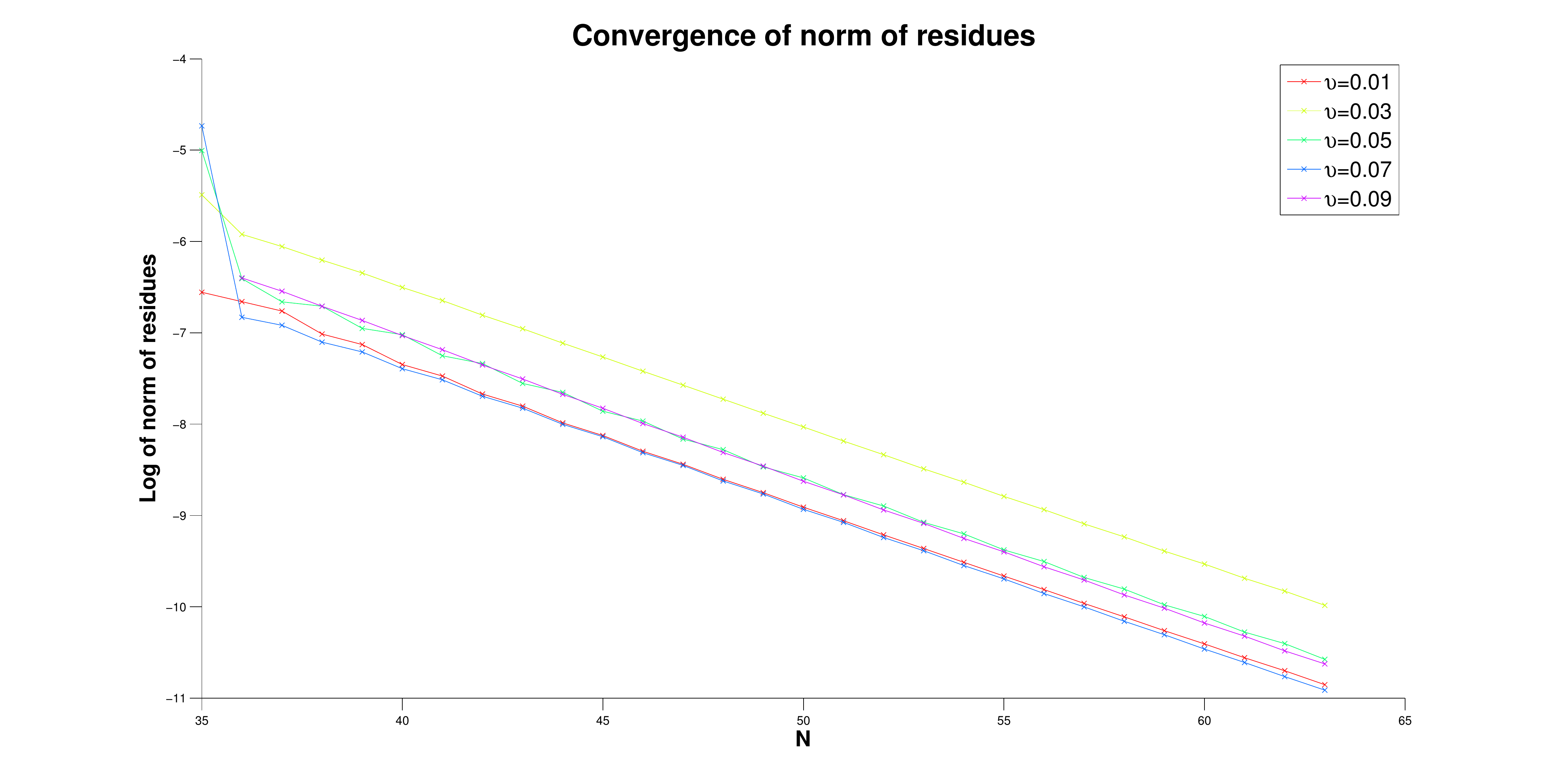}}
}
    \caption[]{Convergence of the log of the norm of the difference of the solutions as a function of $N_z$ for temperatures of $(0.0016, 0.0021)$ for a variety of $\upsilon$ with $N_x=45$. These temperatures correspond to $\frac{T}{\mu}$ values of approximately $(6.6e^{-4}, 8.6e^{-4})$ and therefore can be considered to be small on the scale set by the chemical potential. We see that we obtain exponential convergence as a function of $N_z$, and that, while these plots become noisier for smaller temperatures, the exponential convergence reasserts itself as $N_z$ increases.}
\label{fig:Nxfix_bk_res_converg}
\end{figure}

\begin{figure}
\makebox[\linewidth]{%
\subfigure{\includegraphics[width=0.6\textwidth]{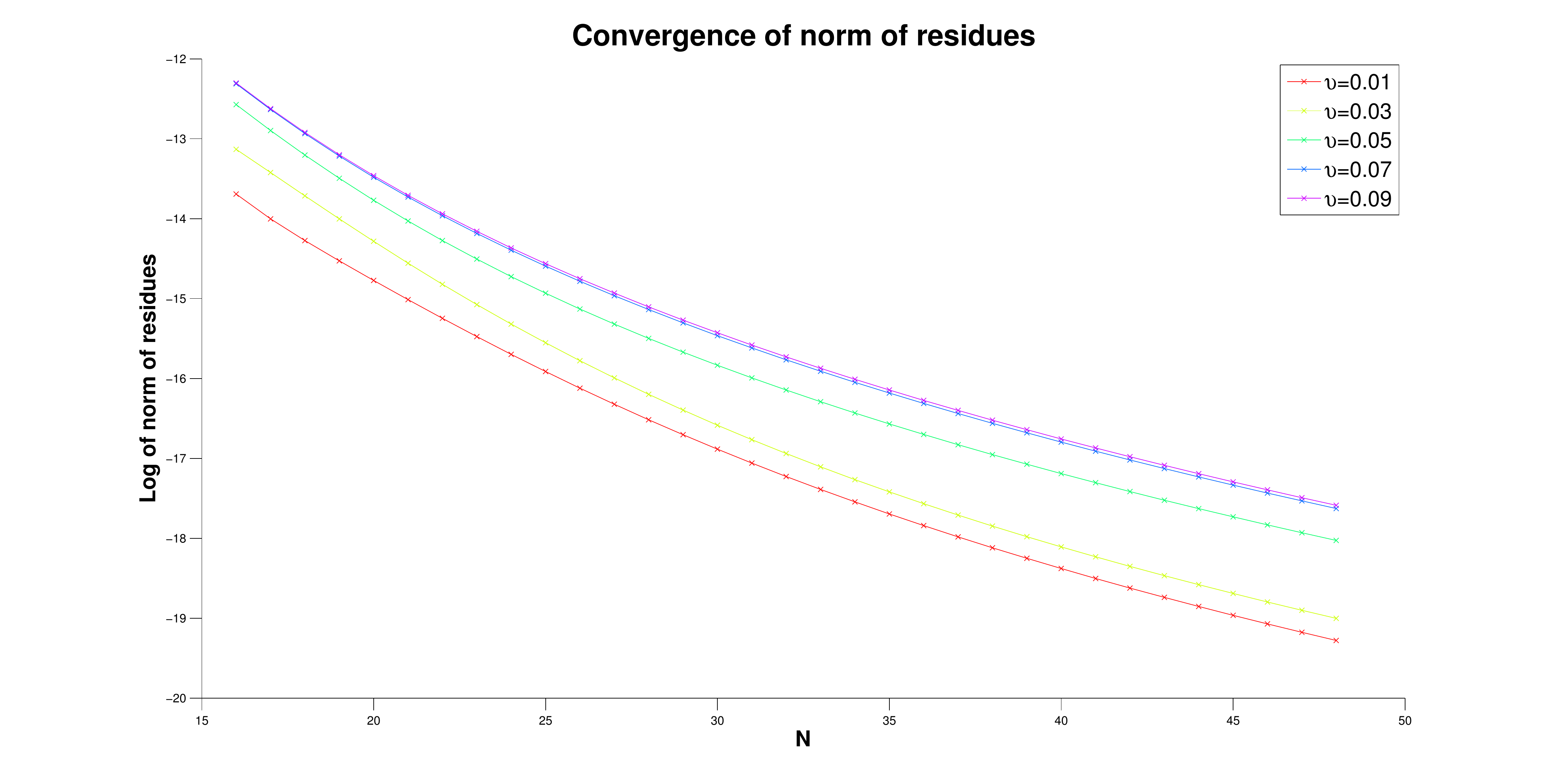}}
\subfigure{\includegraphics[width=0.6\textwidth]{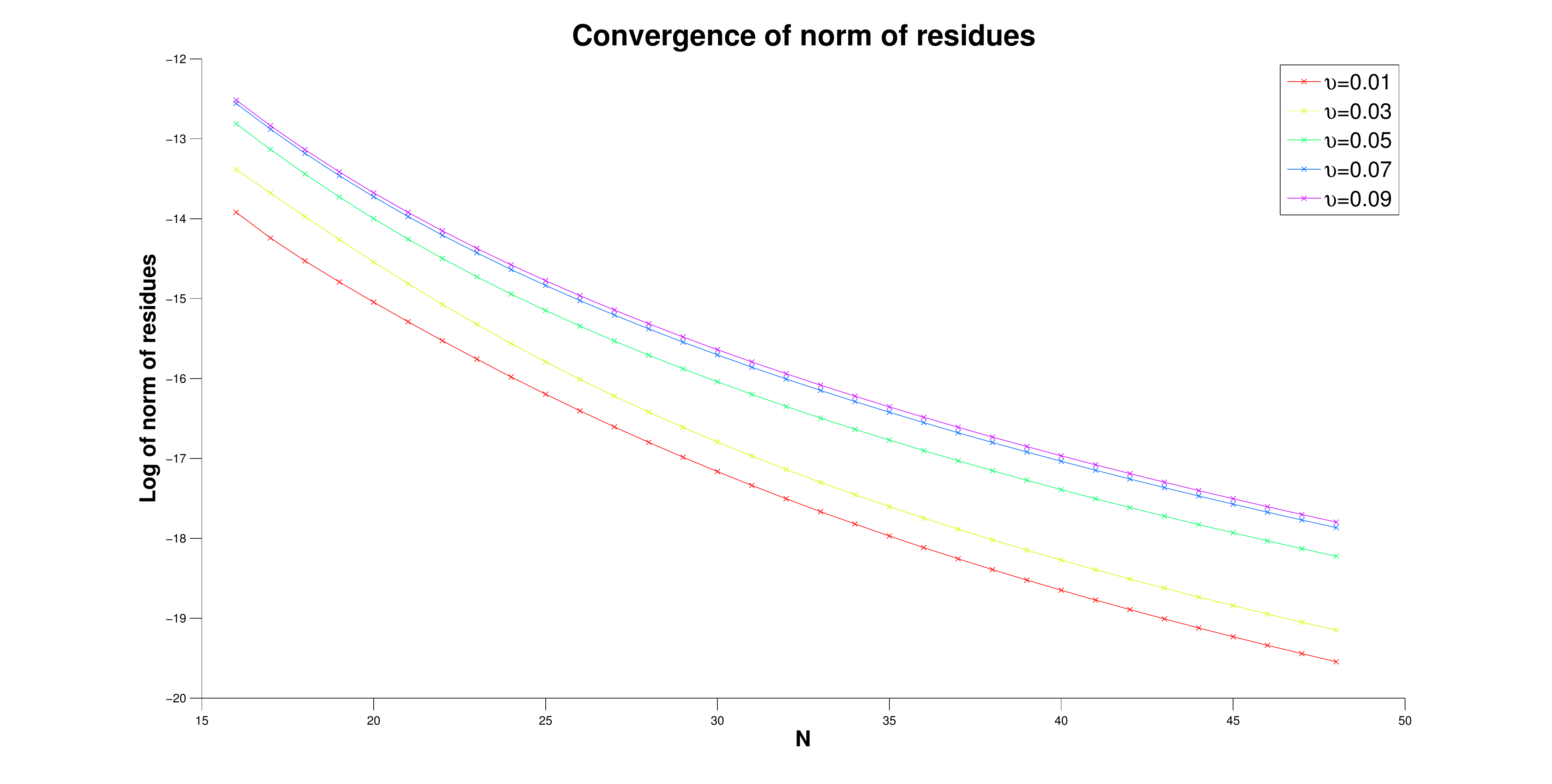}}
}
    \caption[]{Convergence of the log of the norm of the difference of the solutions as a function of $N_x$ for the same values of $\upsilon$ and temperature as the previous plot and with  with $N_z=45$. We note that the convergence deviates  significantly from exponential and that the small scale of the y axis indicates that convergence is occurring more slowly than for the case of increasing $N_z$.}
\label{fig:Nzfix_bk_res_converg}
\end{figure}

  In  Figs.~\ref{fig:Nxfix_bk_res_converg} and \ref{fig:Nzfix_bk_res_converg}, we consider the convergence of the solutions as a function of $N$, for various low temperatures and $\upsilon$. We consider the convergence properties of the solutions as a function of both the transverse grid size, $N_x$, and radial grid size, $N_z$, separately. This may be done by fixing one of the grid sizes and running convergence tests on the other. We display the results below for the convergence of the log of the norm of the difference in solutions for three separate, low temperatures. In the first set we fix $N_x=45$ and vary $N_z$ while in the second we do the reverse. From these tests we may draw the (perhaps expected) conclusion that convergence of the solutions depends more strongly on the radial grid.
  
 We note that the asymptotic expansion of the fields near the conformal boundary contain logarithmic terms at high orders of the expansion. Therefore the exponential convergence of the spectral methods we are using is expected to fail for fine enough grids. However, we find that for the range of parameters we consider here, this is not an important issue. Similarly, at sufficiently low temperatures we expect that finite differencing approximation near the horizon is more suitable. It may be that utilizing such methods together with domain decomposition approaches will allow us to reach still lower temperatures in future work.  

\begin{figure}
\makebox[\linewidth]{%
\subfigure{\includegraphics[width=0.6\textwidth]{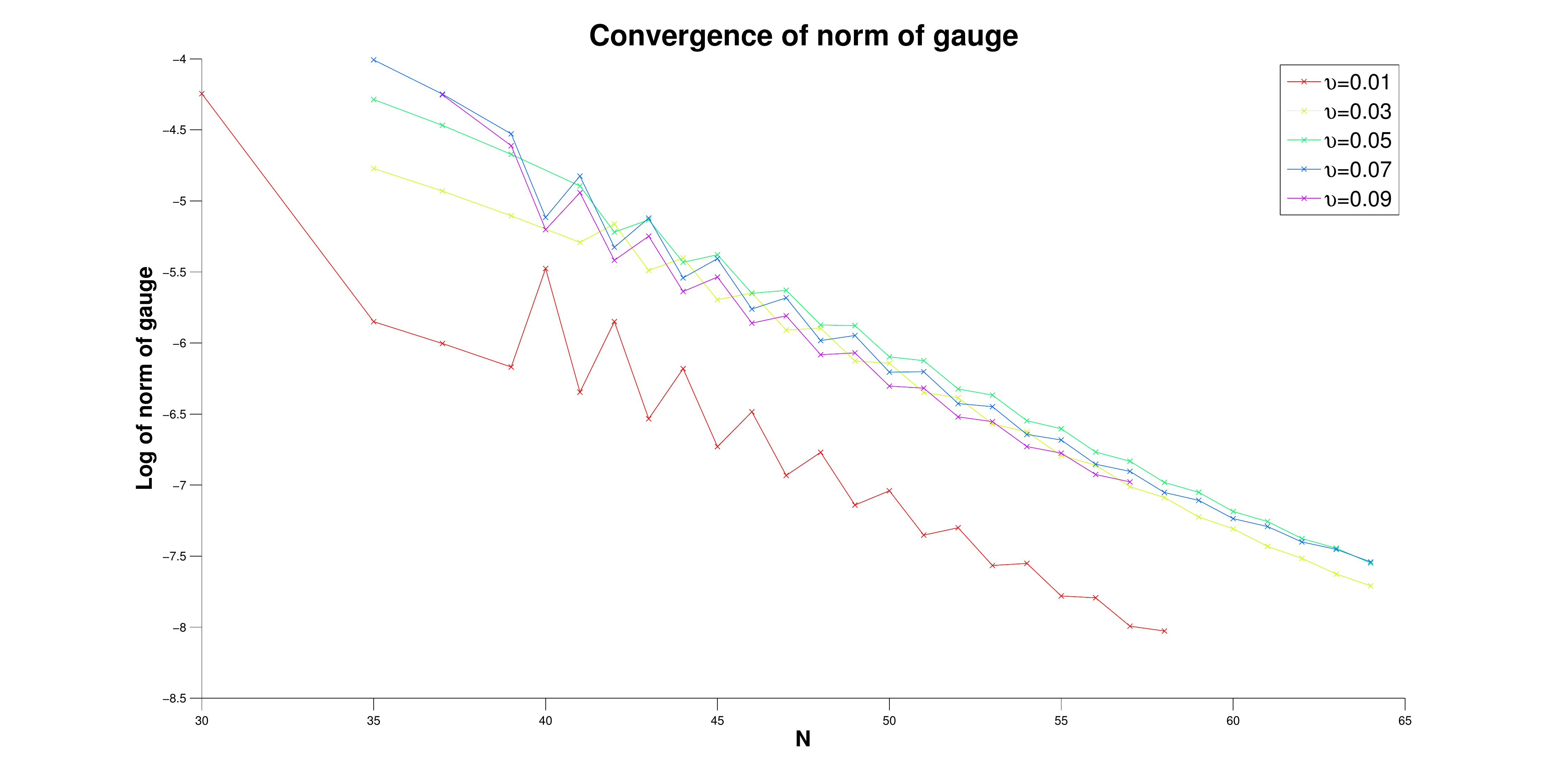}}
\subfigure{\includegraphics[width=0.6\textwidth]{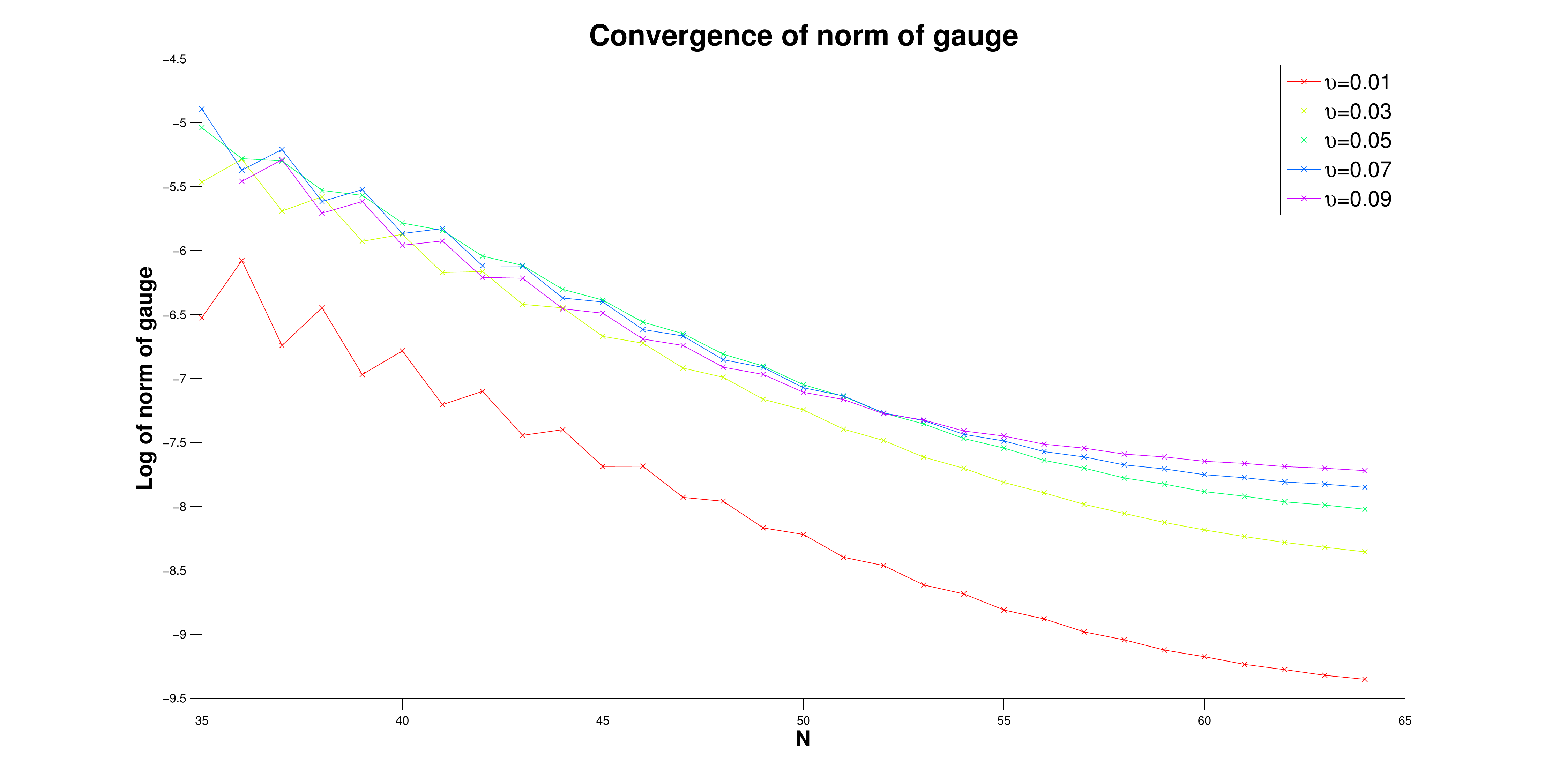}}
}
    \caption[]{Examining the convergence of the log of the norm of the gauge condition as a function of $N_z$.  For the lower temperature (leftmost graph) we observe exponential convergence with the now expected noise at lower values of $N_z$. We note that for the higher temperature (rightmost graph) larger magnitudes of $\upsilon$ exhibit two distinct scaling regimes with a crossover occurring at approximately $N_z=55$. While the convergence is exponential in both cases it is markedly faster in one case. This may mean that it is necessary to go to higher resolutions if very accurate solutions are required in this region of parameter space. This issue however was not encountered in the results presented earlier in the paper.}
\label{fig:Nxfix_bk_gaug_converg}
\end{figure}

\begin{figure}
\makebox[\linewidth]{%
\subfigure{\includegraphics[width=0.6\textwidth]{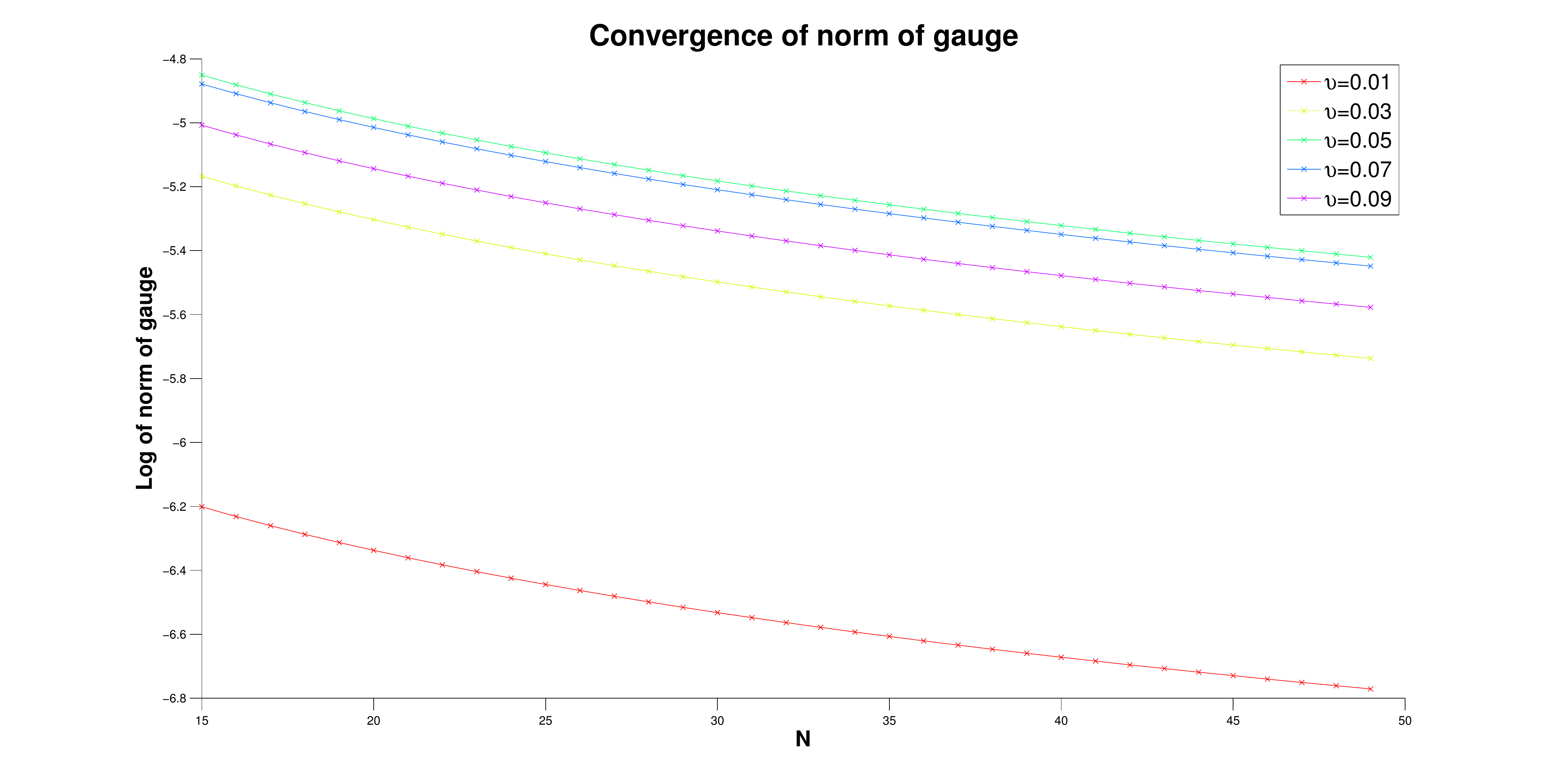}}
\subfigure{\includegraphics[width=0.6\textwidth]{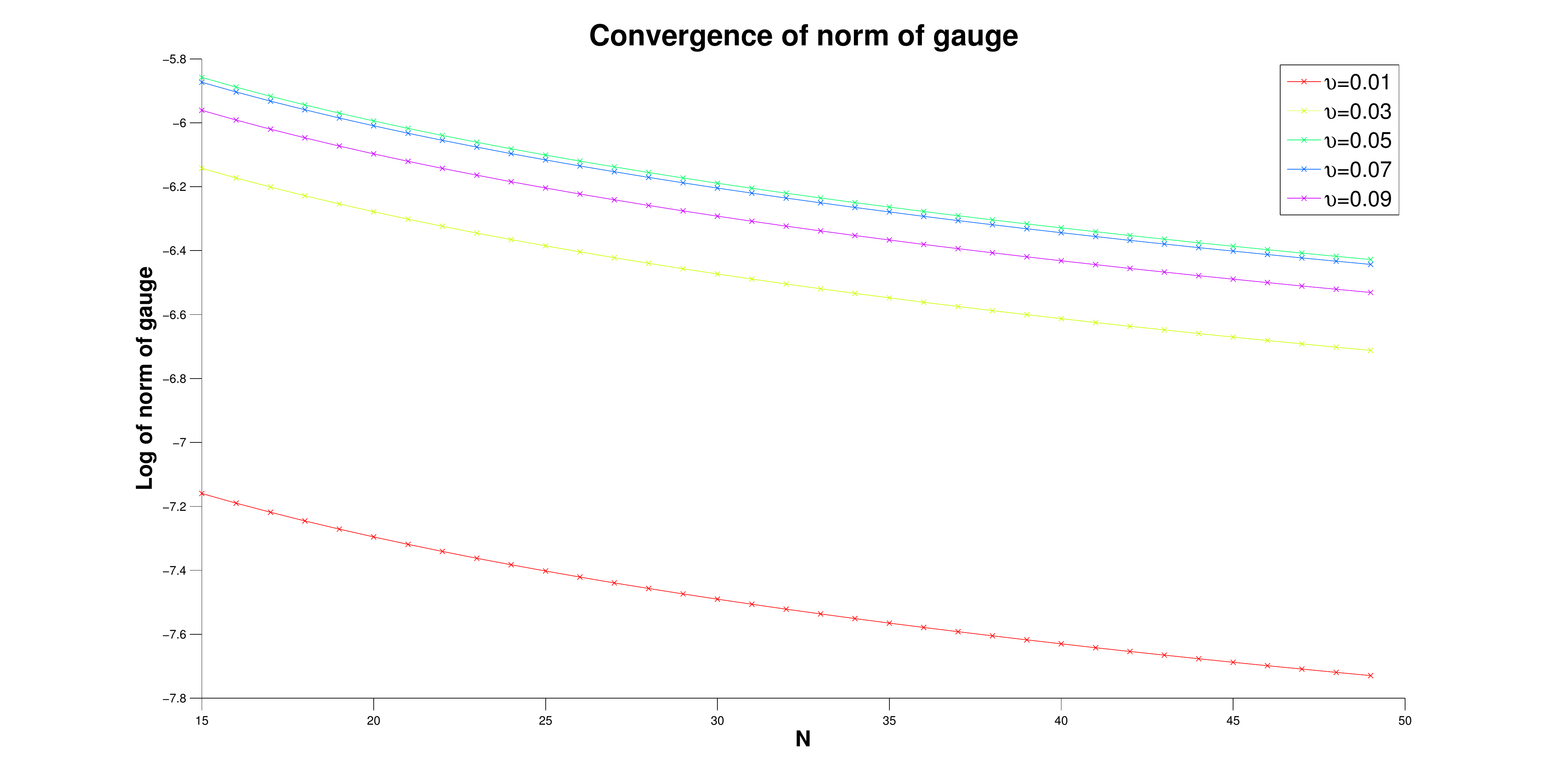}}
}
    \caption[]{For the same values of  $\upsilon$ and temperature as  Fig.~\ref{fig:Nxfix_bk_gaug_converg} we see that the convergence of the log of the norm of the gauge condition is significantly slower then exponential when $N_x$ is increased for a fixed $N_z$.}
\label{fig:Nzfix_bk_gaug_converg}
\end{figure}

In  Fig.~\ref{fig:Nxfix_bk_gaug_converg} we perform a similar series of tests for the convergence of the norm of the gauge condition towards zero. We again find that better convergence behaviour is achieved by increasing $N_z$ in preference to $N_x$. We also note that different scaling regimes may exist in the convergence of the norm of the gauge condition as a function of the resolution, cf., Fig.~\ref{fig:Nzfix_bk_gaug_converg}.

\begin{figure}
\makebox[\linewidth]{%
\subfigure{\includegraphics[width=0.6\textwidth]{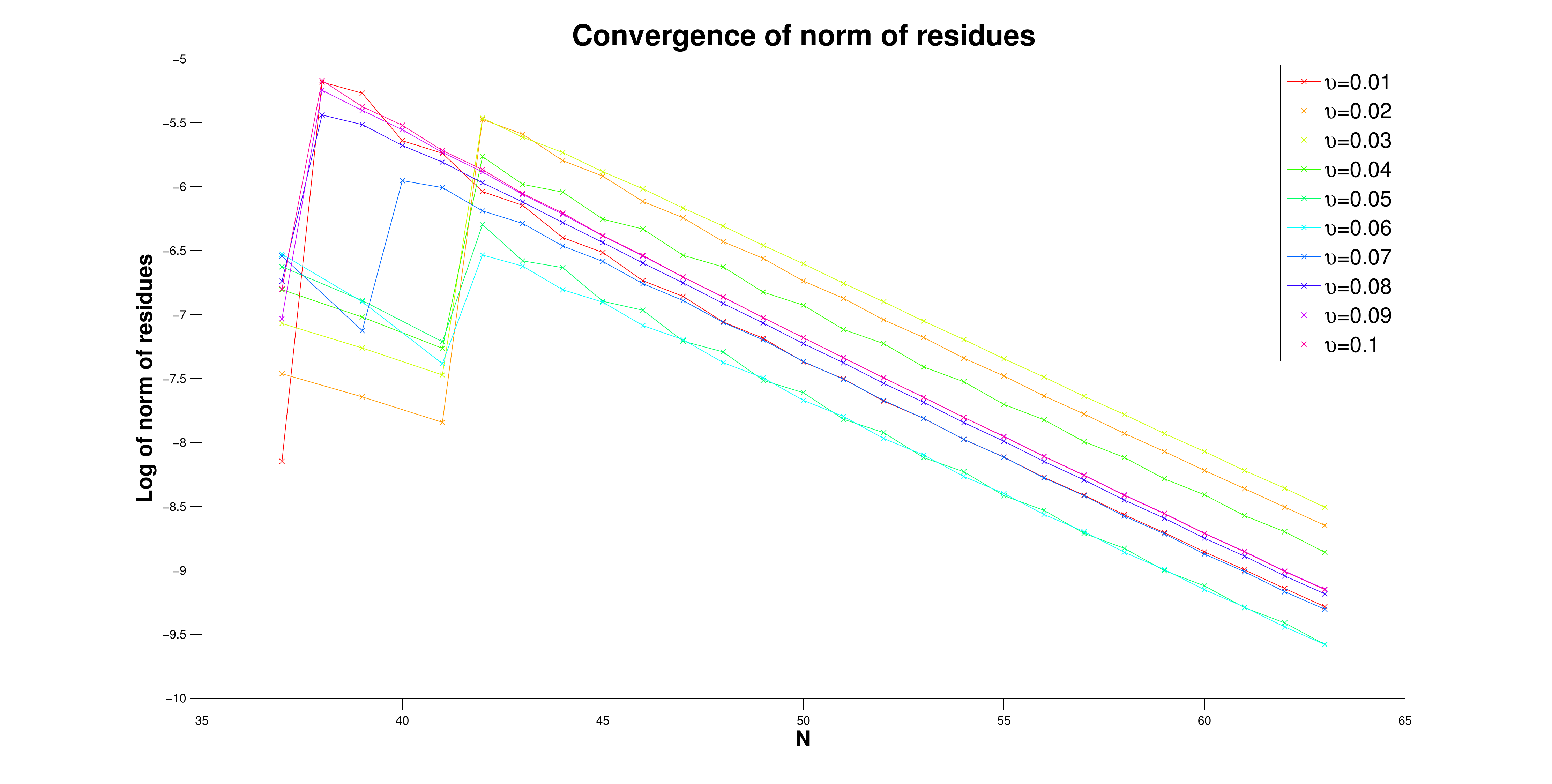}}
\subfigure{\includegraphics[width=0.6\textwidth]{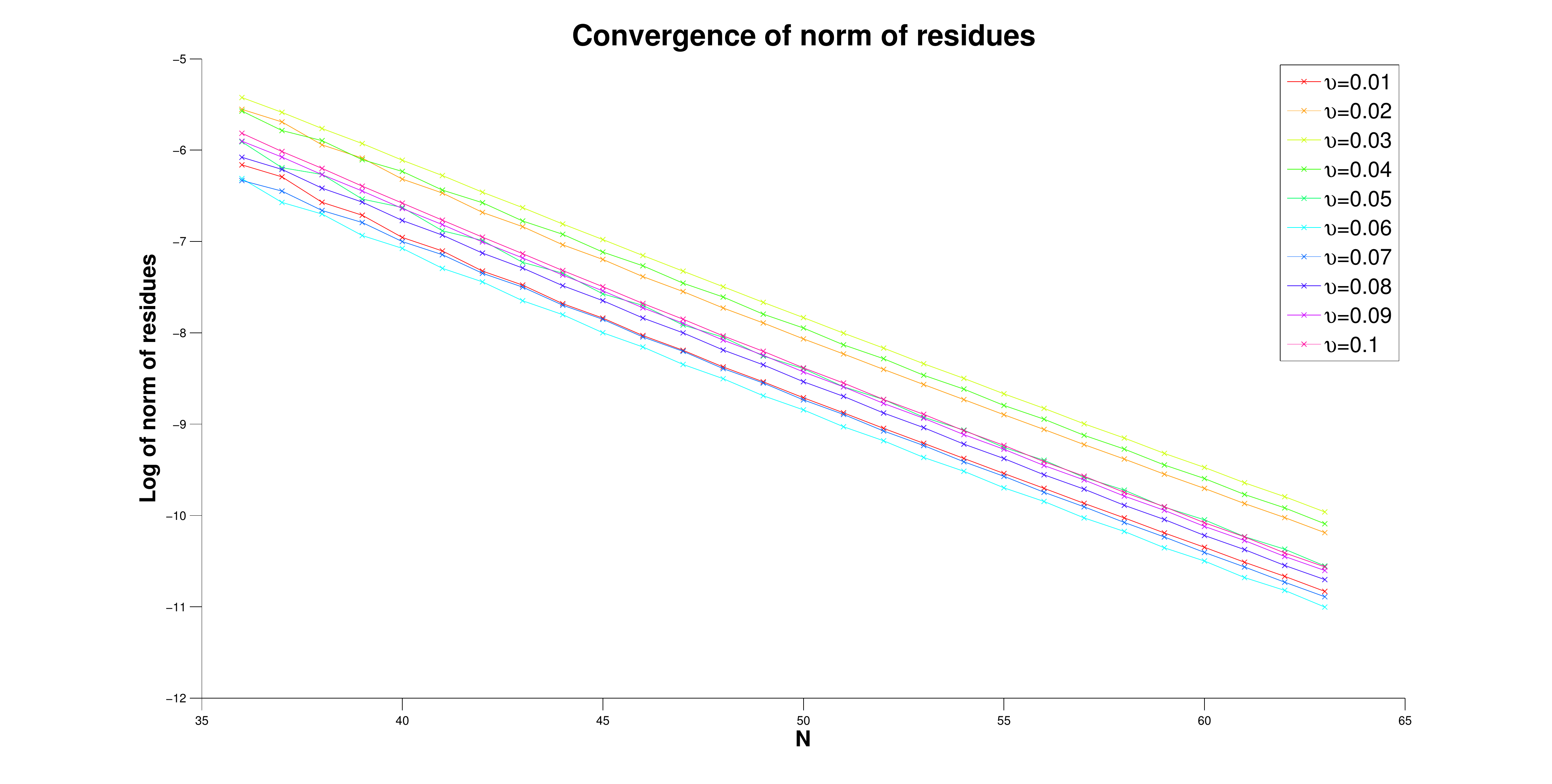}}
}
    \caption[]{Convergence of the log of the norm of the residues with both $N_z$ and $N_x$ increasing but with $N_x$ lagging $N_z$ by 20. Again the $\frac{T}{\mu}$ values are approximately $(6.6e^{-4}, 8.6e^{-4})$ moving from left to right. We see good exponential convergence even for the lower temperature case.}
\label{fig:Nxlag_bk_res_converg}
\end{figure}

 From this series of experiments we conclude that the best resolution for a fixed number of grid points is obtained when more points are allocated to the radial grid number, $N_z$. As a verification of this approach in Fig.~\ref{fig:Nxlag_bk_res_converg} we plot the log of the norm of the residues as above with $N_z$ and $N_x$ both increasing but with $N_z=N_x+20$. 
 

  We now consider convergence results for the perturbation equations. Firstly it was noted that asymmetric grids are less useful in this case and that the best results were obtained when $N_x$ and $N_z$ were increased in tandem. In addition it was found that for smooth convergence to be obtained the background should be available at a higher resolution then the perturbation resolution. 
We are now interested in the convergence as a function of $\upsilon$, $T$ and $\omega$. 
  
\begin{figure}
\makebox[\linewidth]{%
\subfigure{\includegraphics[width=0.6\textwidth]{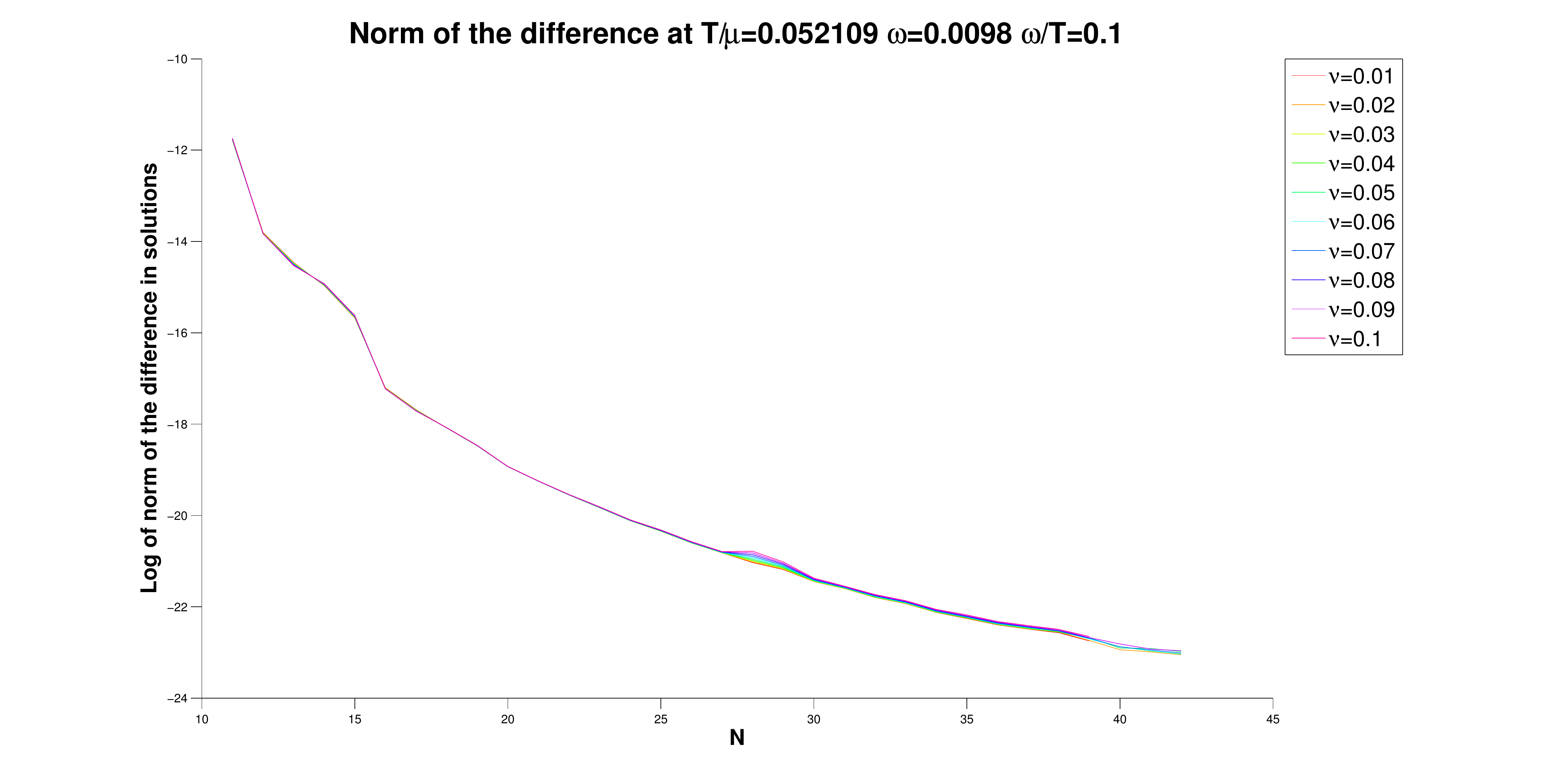}}
\subfigure{\includegraphics[width=0.6\textwidth]{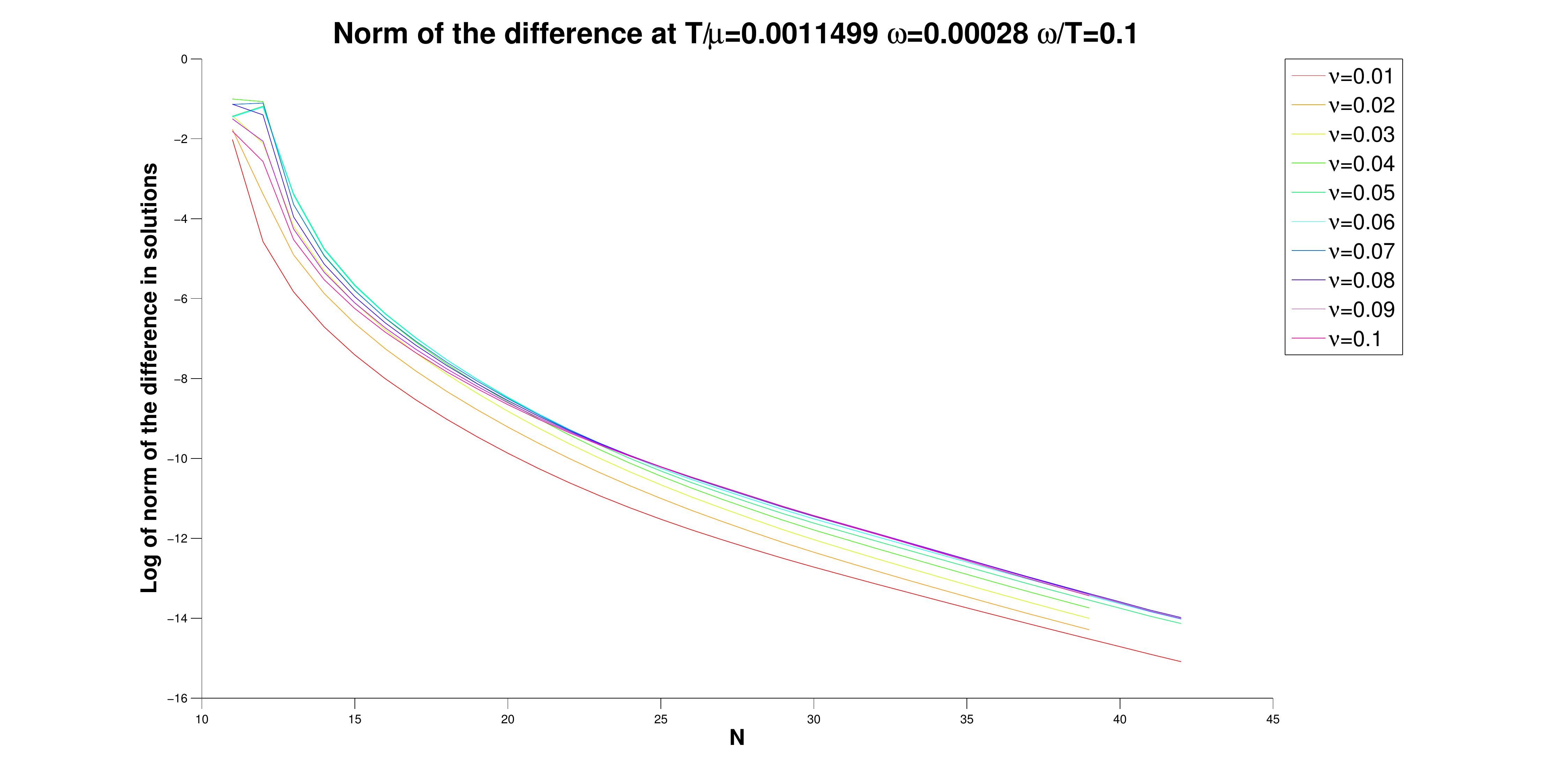}}
}
\makebox[\linewidth]{%
\subfigure{\includegraphics[width=0.6\textwidth]{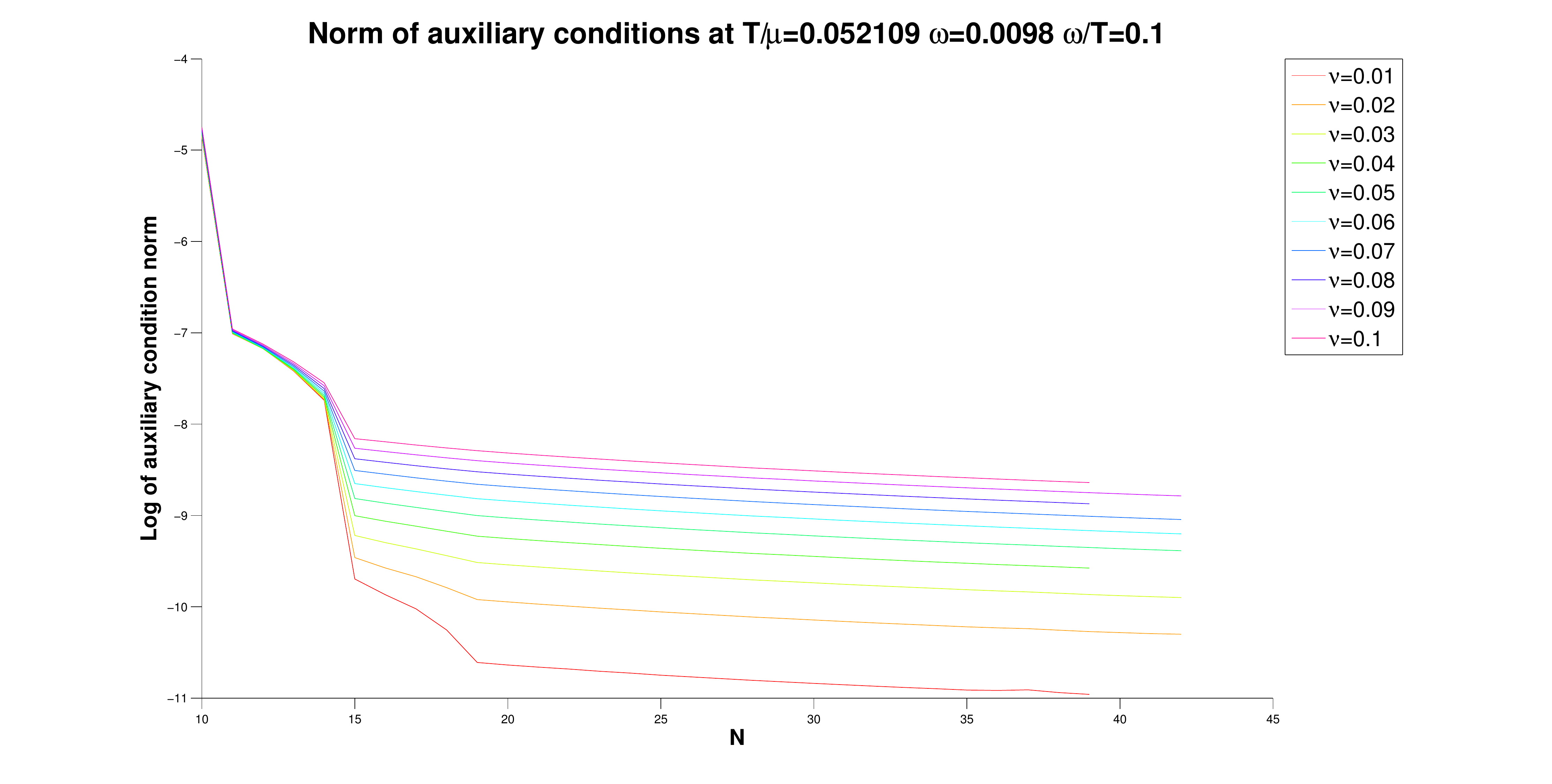}}
\subfigure{\includegraphics[width=0.6\textwidth]{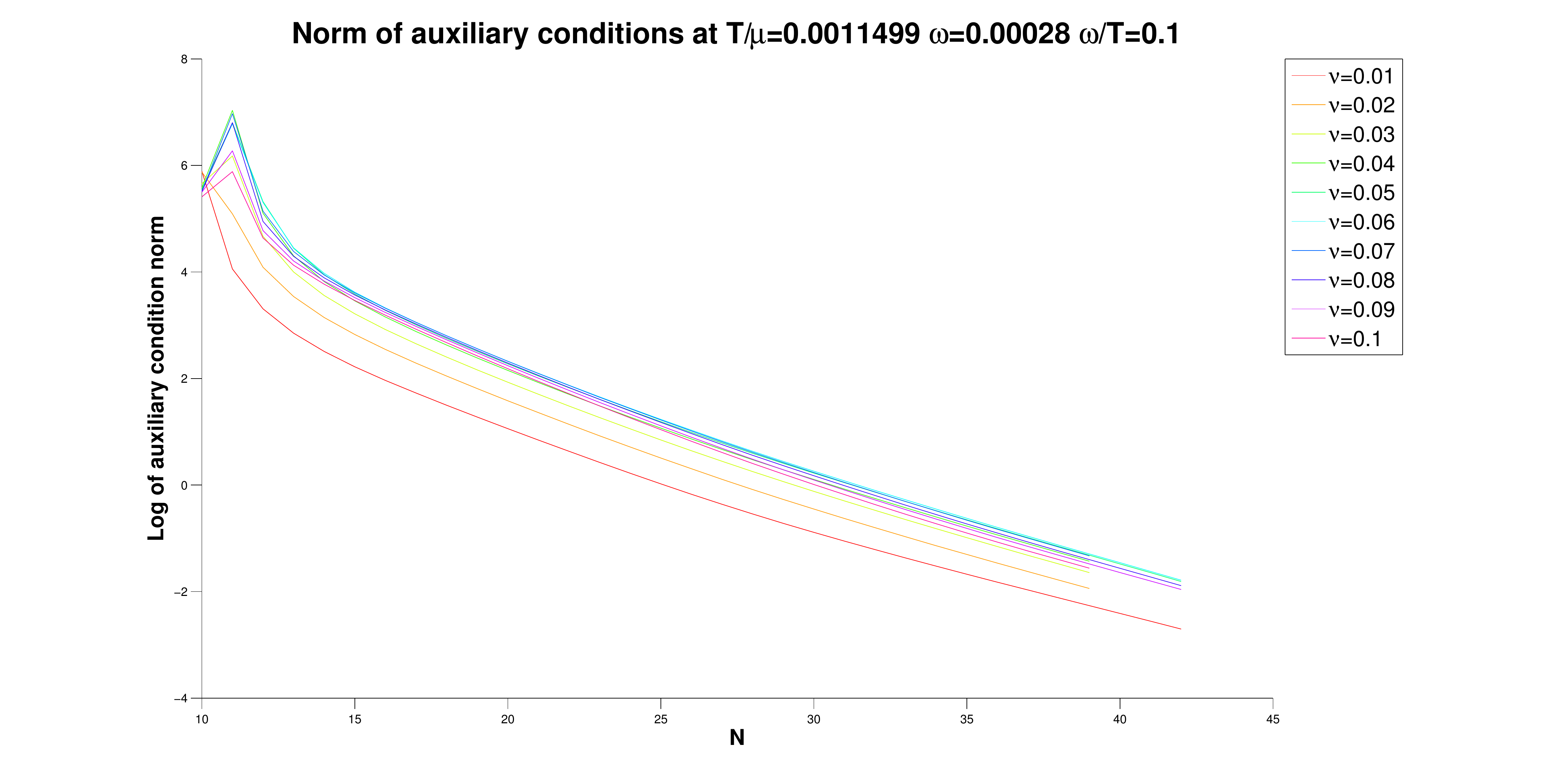}}
}
\makebox[\linewidth]{%
\subfigure{\includegraphics[width=0.6\textwidth]{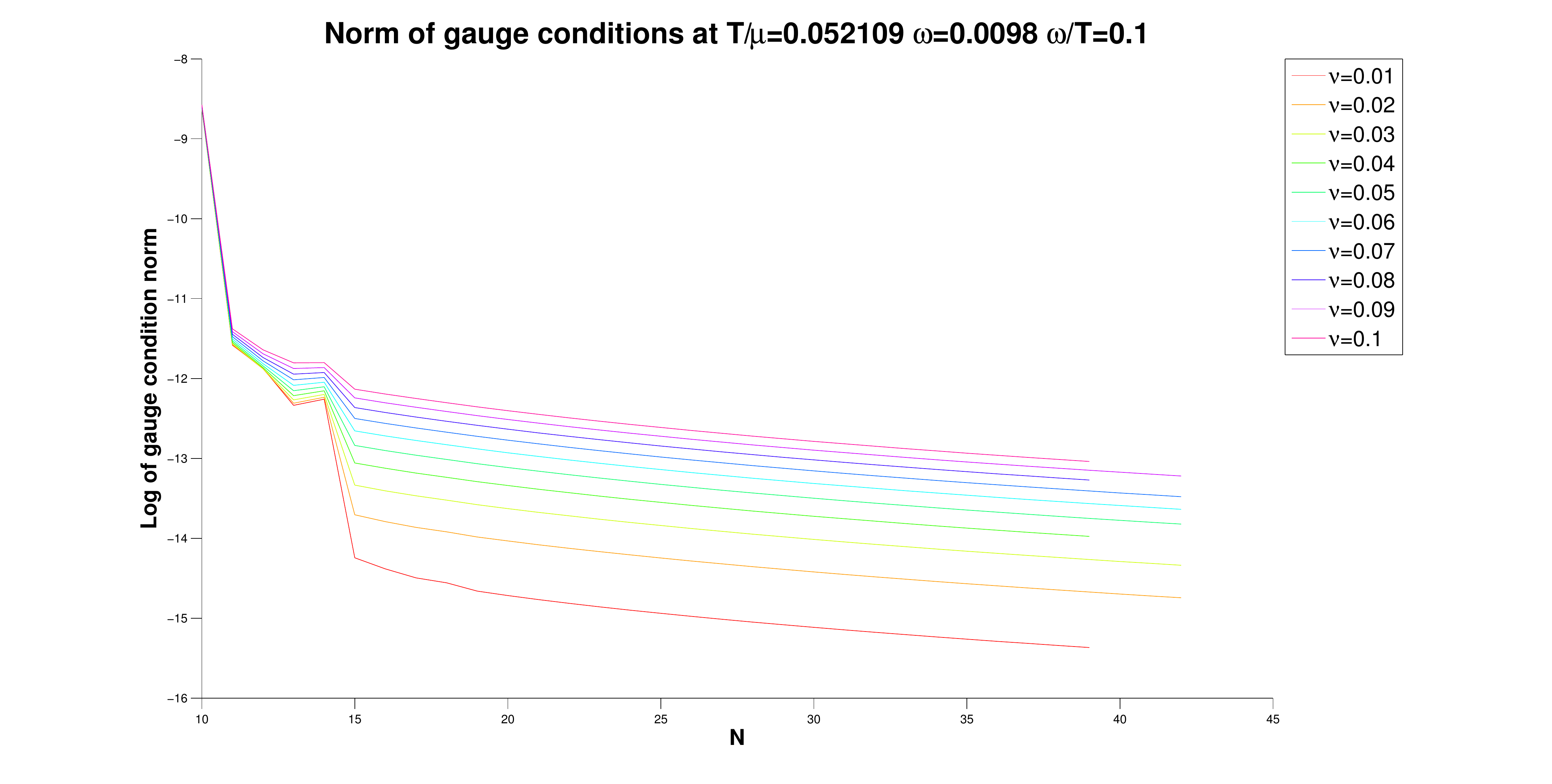}}
\subfigure{\includegraphics[width=0.6\textwidth]{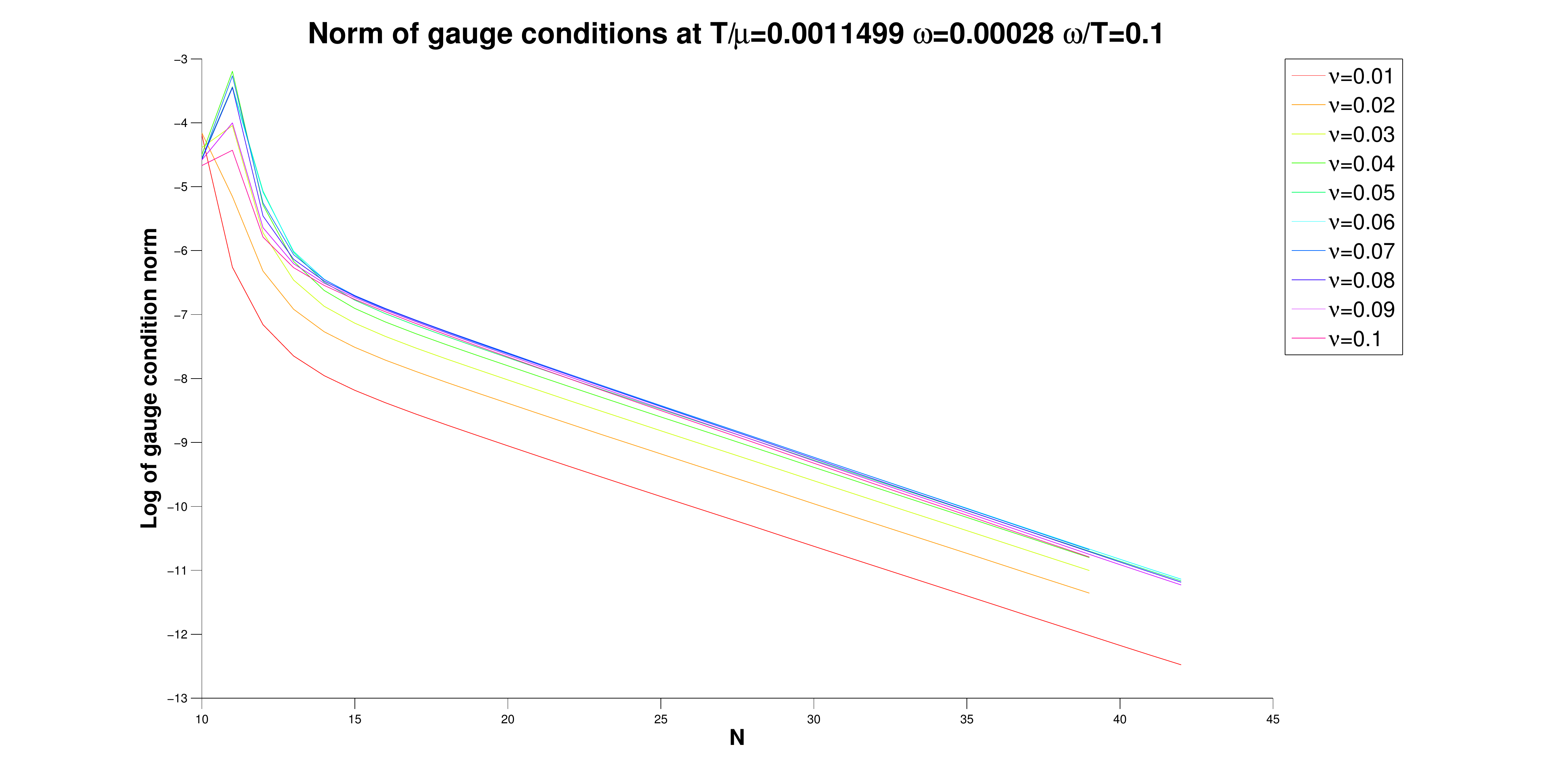}}
}
    \caption[]{Semi-logarithmic plots of the norm of the difference in solutions, and the norm of horizon auxiliary conditions and of the gauge conditions versus $N$ at a low and high temperature for $\frac{w}{T}=0.1$. We note that while still convergent, the resolution of the auxiliary conditions must be monitored closely for small values of  $\frac{T}{\mu}$ and larger values of $\upsilon$. }
\label{fig:low_wT_pert_conv_pt1}
\end{figure}

\begin{figure}
\makebox[\linewidth]{%
\subfigure{\includegraphics[width=0.6\textwidth]{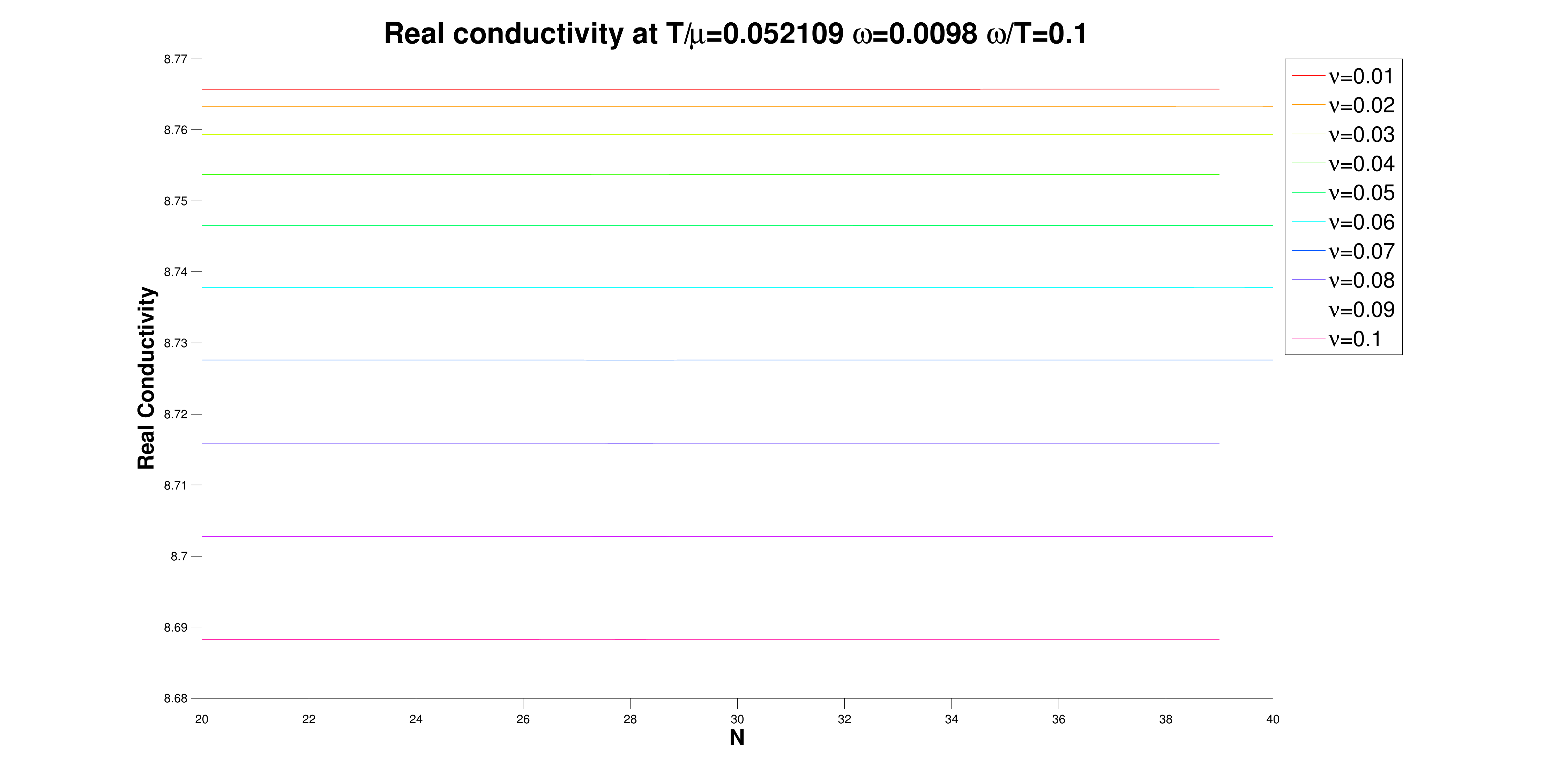}}
\subfigure{\includegraphics[width=0.6\textwidth]{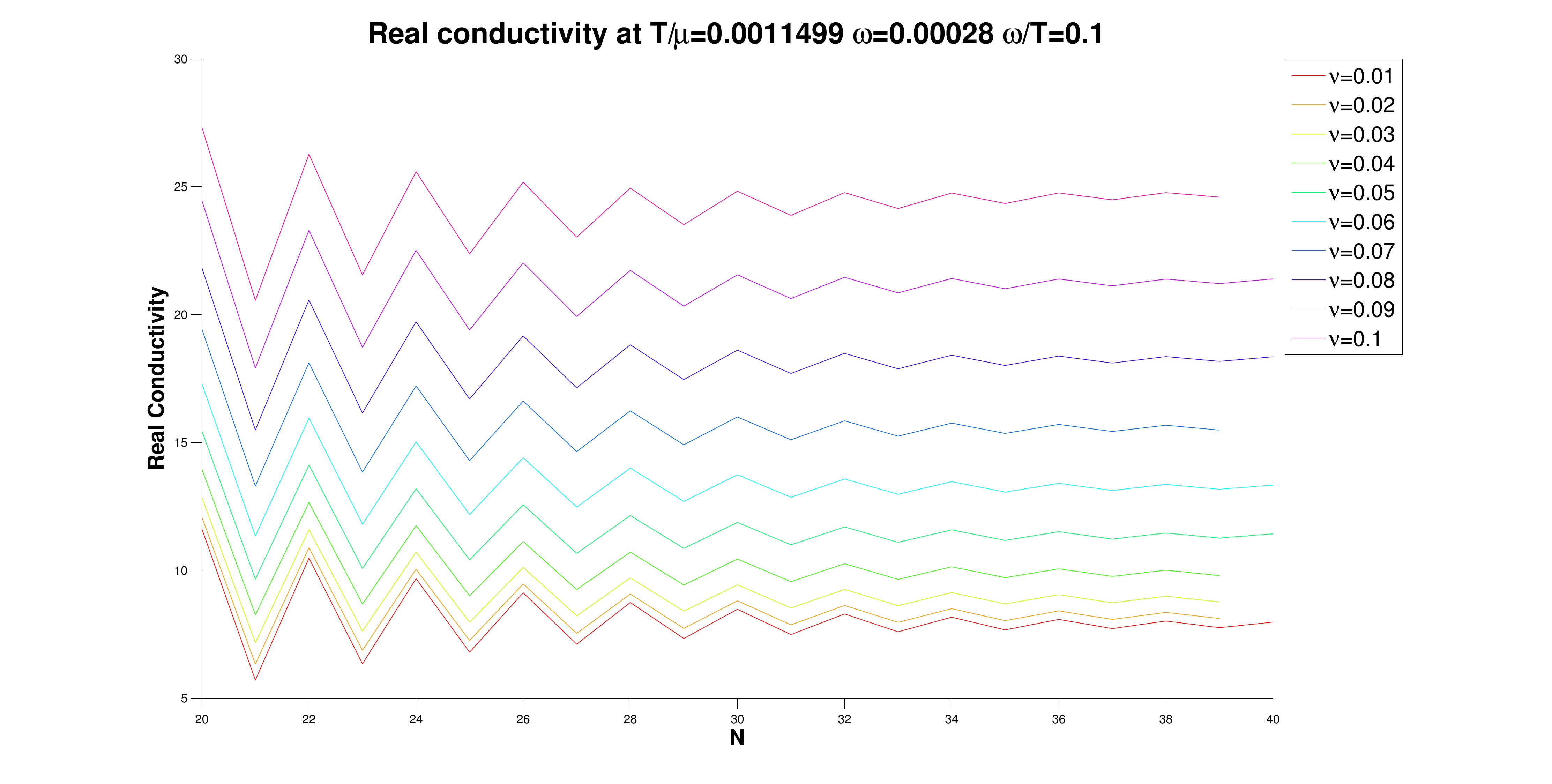}}
}
\makebox[\linewidth]{%
\subfigure{\includegraphics[width=0.6\textwidth]{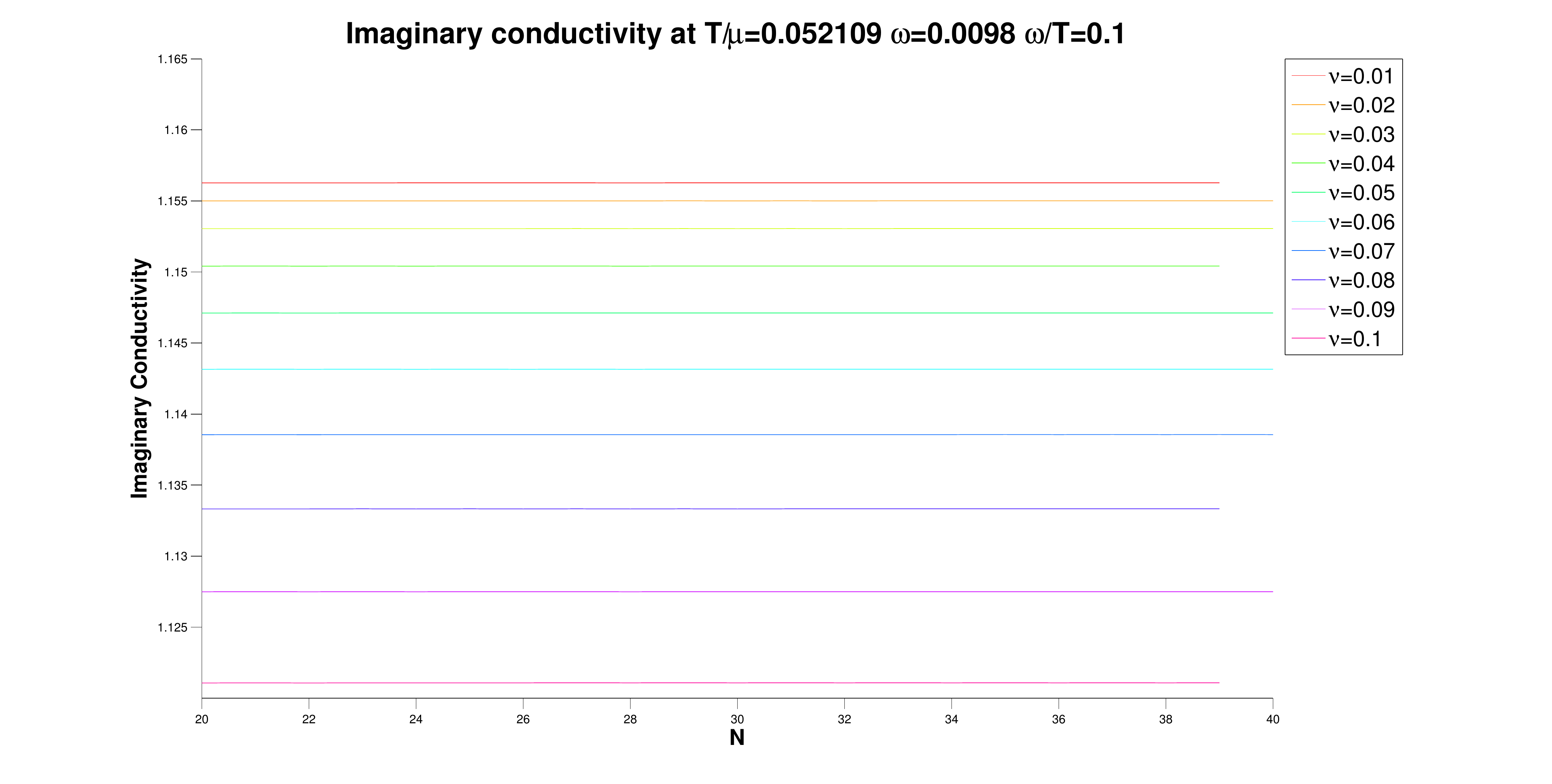}}
\subfigure{\includegraphics[width=0.6\textwidth]{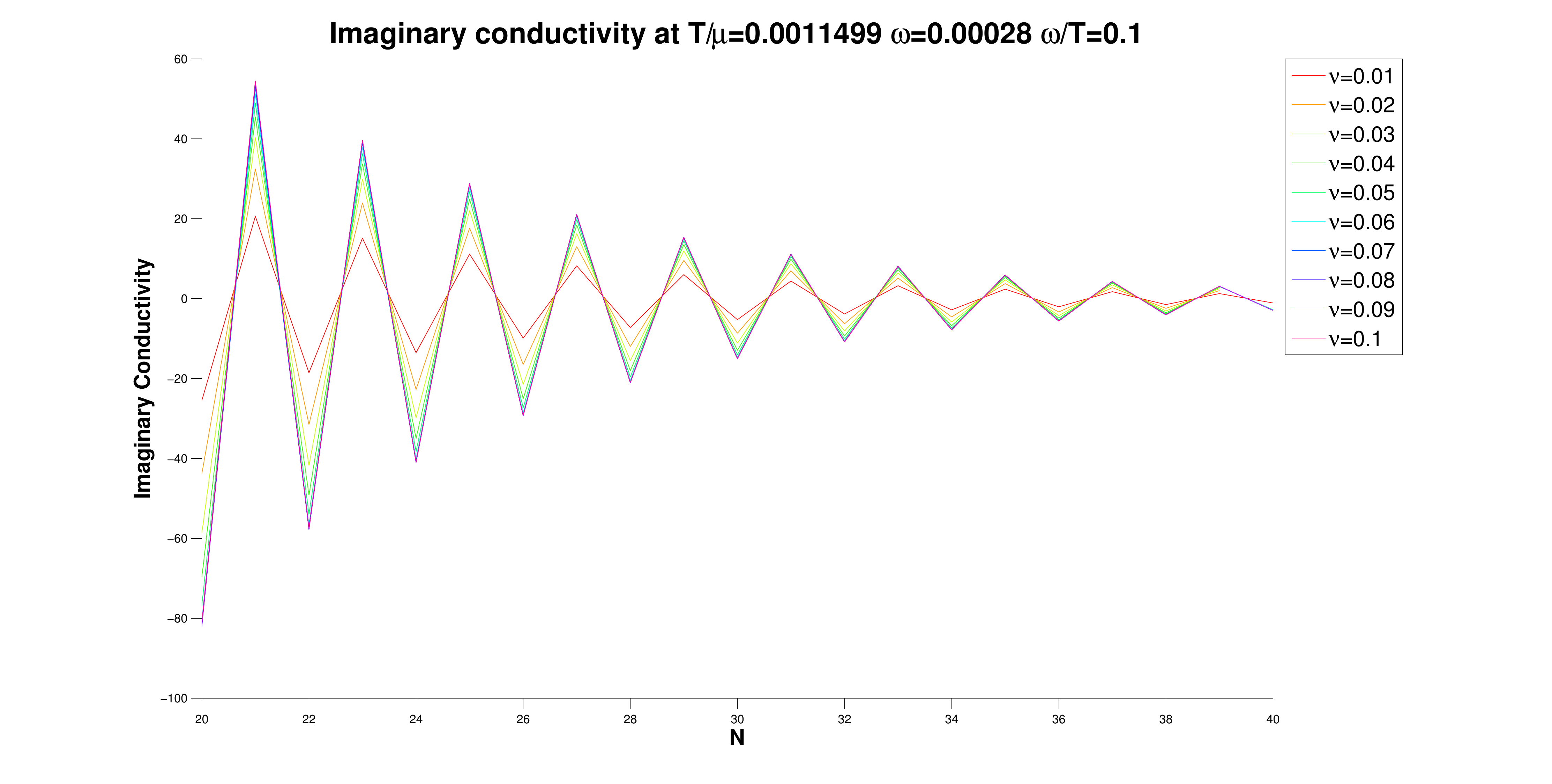}}
}
    \caption[]{Plots of the real and imaginary conductivities versus $N$ for the same choice of parameters as in Fig.~\ref{fig:low_wT_pert_conv_pt1}. We note that, as expected, convergence is slower for larger values of $\upsilon$. This is particularly evident for the imaginary part of the conductivity}
\label{fig:low_wT_pert_conv_pt2}
\end{figure}

 For our purposes there are two ${\mathfrak w}$ regimes where we must examine the convergence of the AC conductivity. These are ${\mathfrak w}\ll1$, which is relevant for comparison to the DC conductivity, and  $5 \sim < \mathfrak w < \sim  10$ which is relevant for examining potential IR scaling regimes. In Fig.~\ref{fig:low_wT_pert_conv_pt1} we display some examples of the convergence of the norm of the difference of the perturbative solutions, and the norm of the gauge and auxiliary conditions as a function of $N$ for ${\mathfrak w}=0.1$. In  Fig.~\ref{fig:low_wT_pert_conv_pt2} we plot the analogous results for the real and imaginary parts of the conductivities themselves.  We note that as $\frac{T}{\mu}$is lowered and $\upsilon$ increased the convergence of the norm of the gauge and auxiliary conditions becomes more strained. We have however checked that for the data displayed in Fig.~\ref{fig:elec_thermoelec_cond} both are on the order of $10^{-5}$ when the DC data points are read off.
\begin{figure}
\makebox[\linewidth]{%
\subfigure{\includegraphics[width=0.6\textwidth]{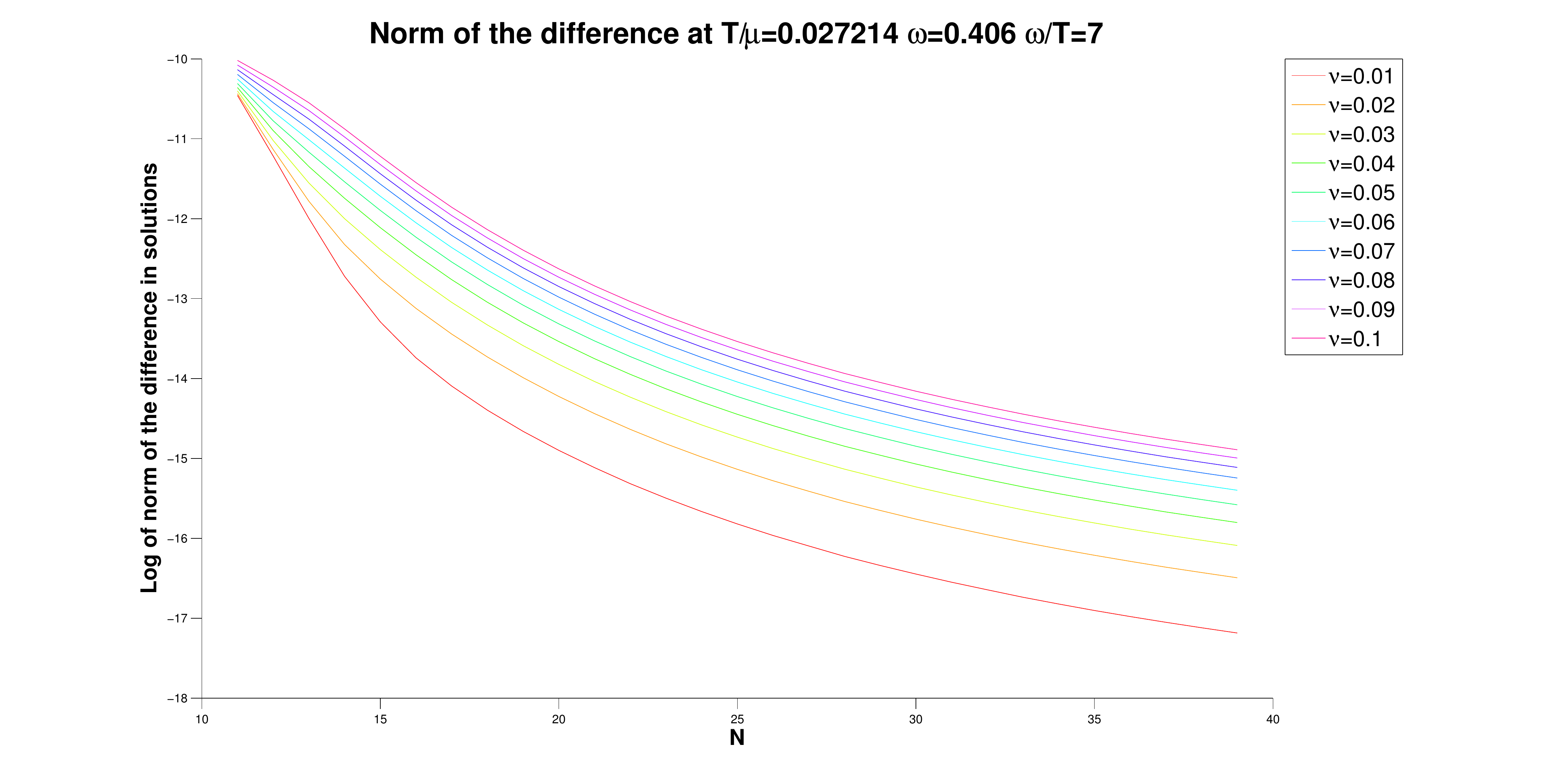}}
\subfigure{\includegraphics[width=0.6\textwidth]{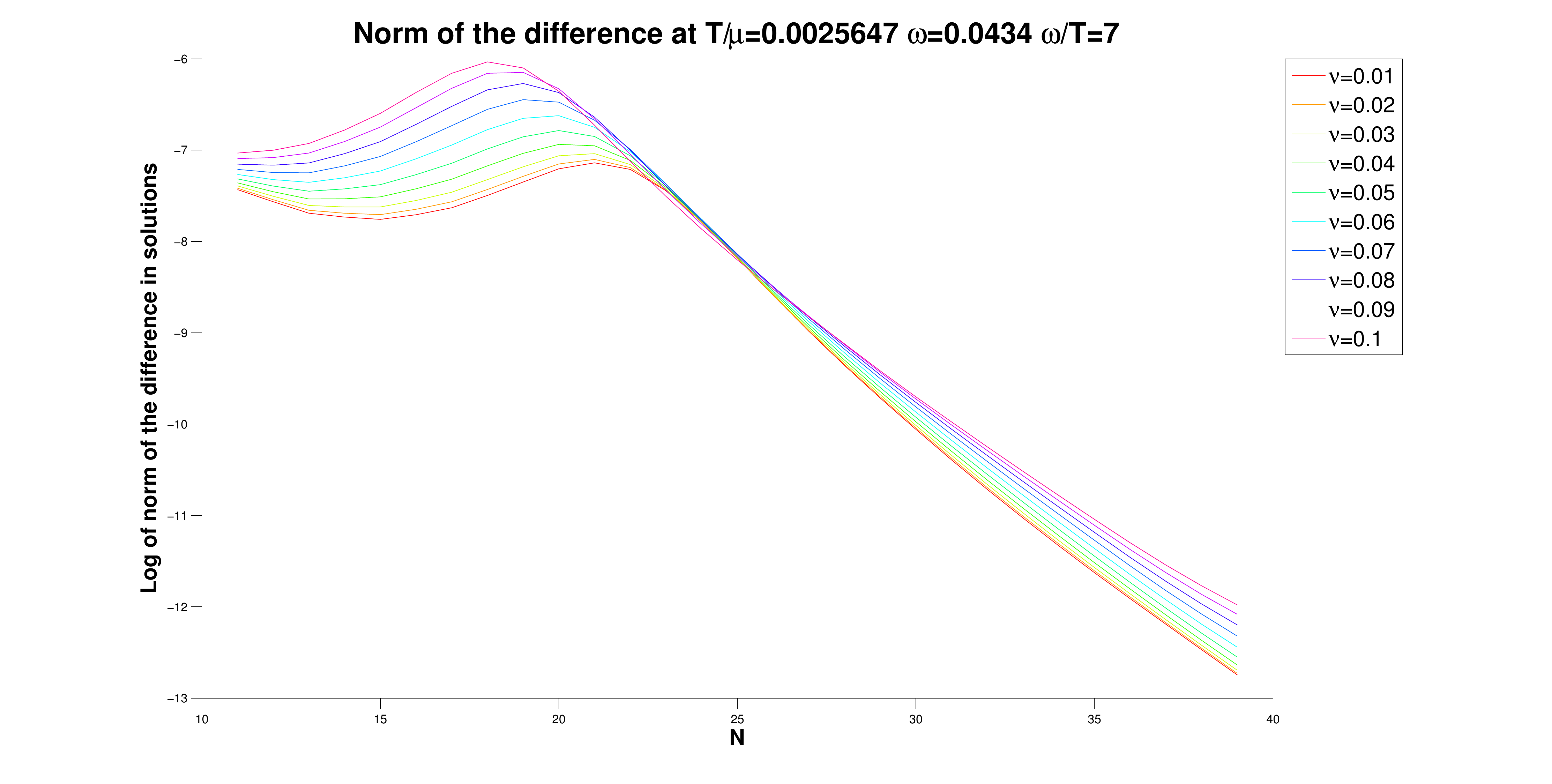}}
}
\makebox[\linewidth]{%
\subfigure{\includegraphics[width=0.6\textwidth]{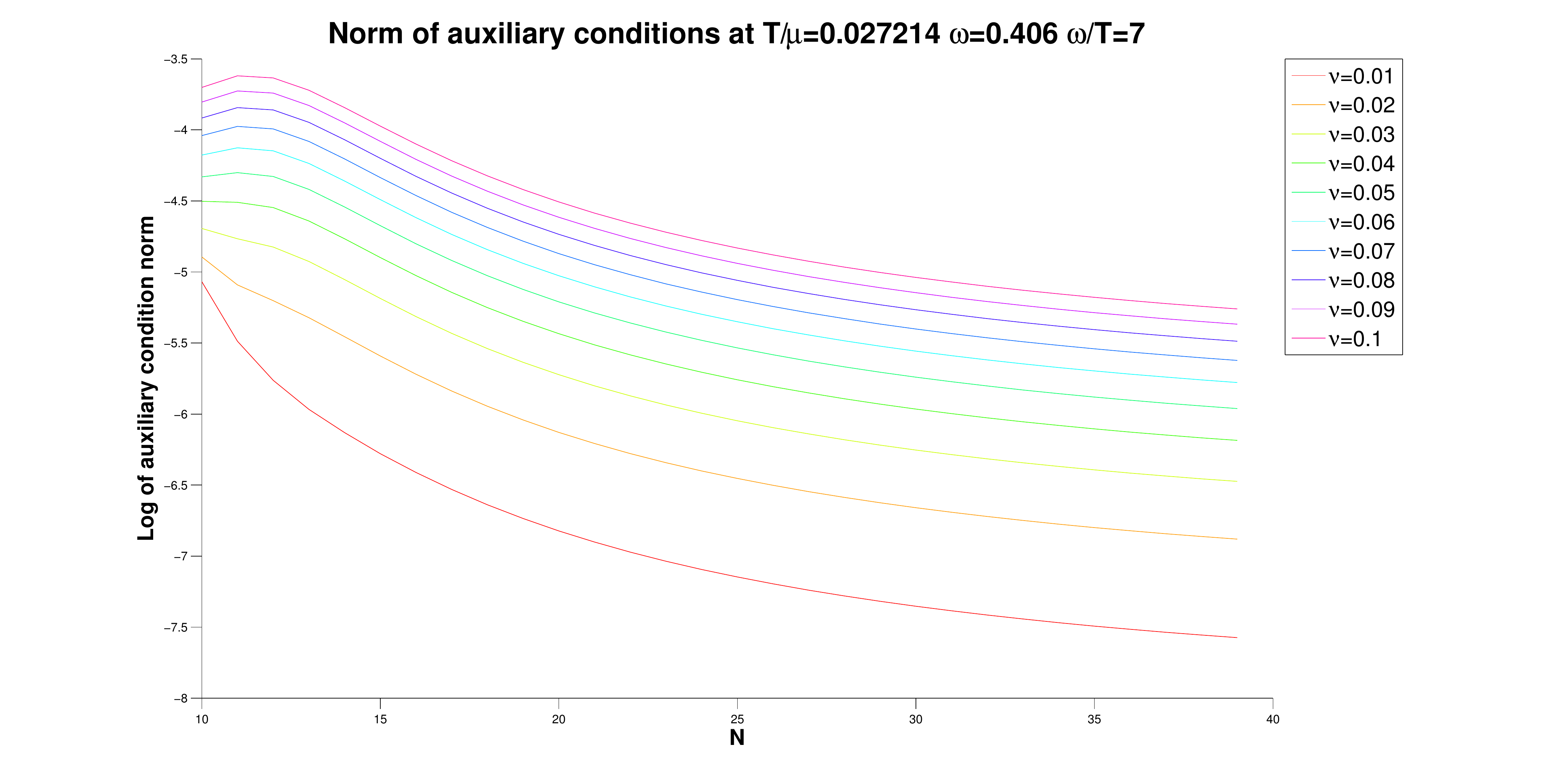}}
\subfigure{\includegraphics[width=0.6\textwidth]{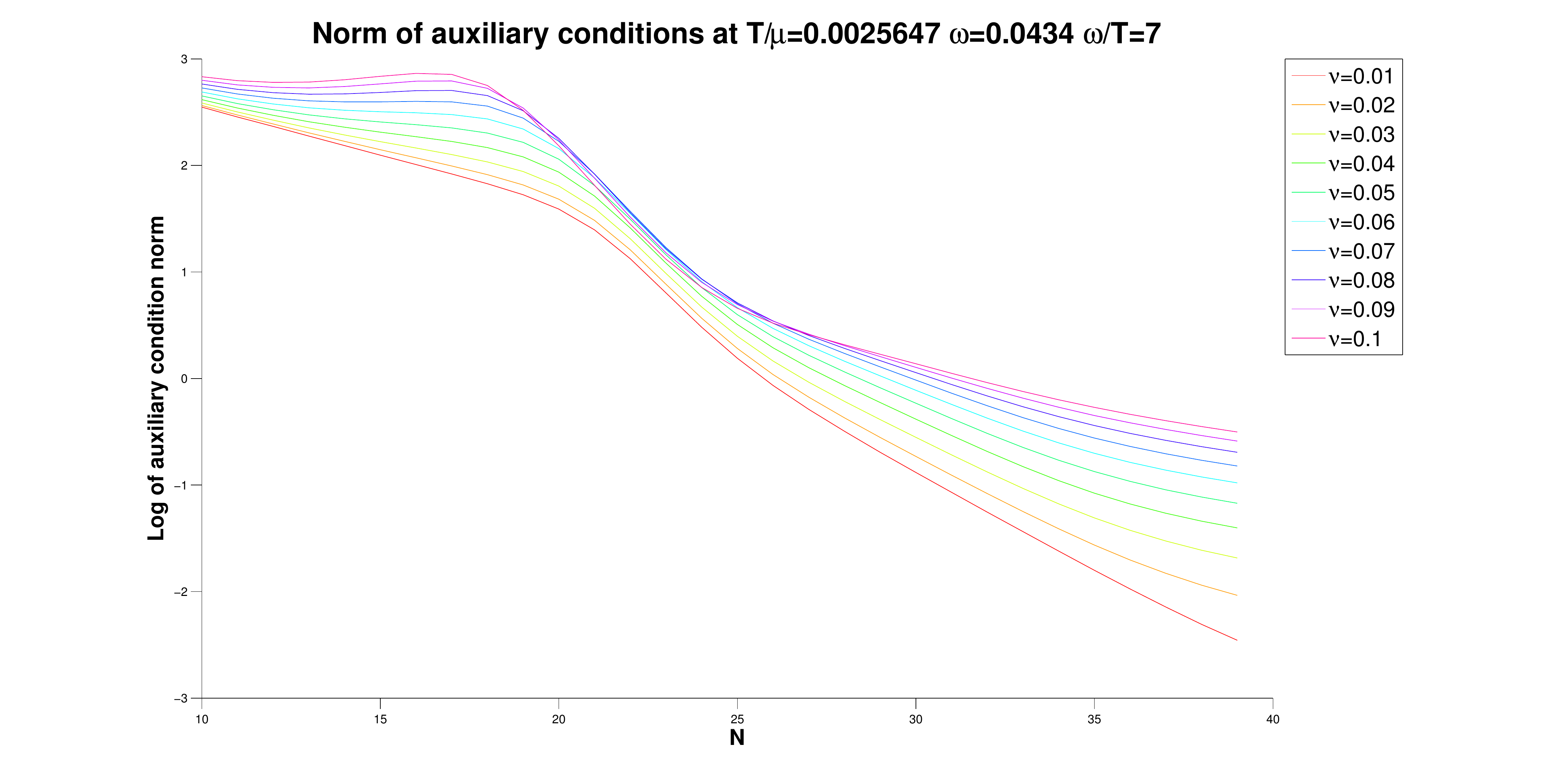}}
}
\makebox[\linewidth]{%
\subfigure{\includegraphics[width=0.6\textwidth]{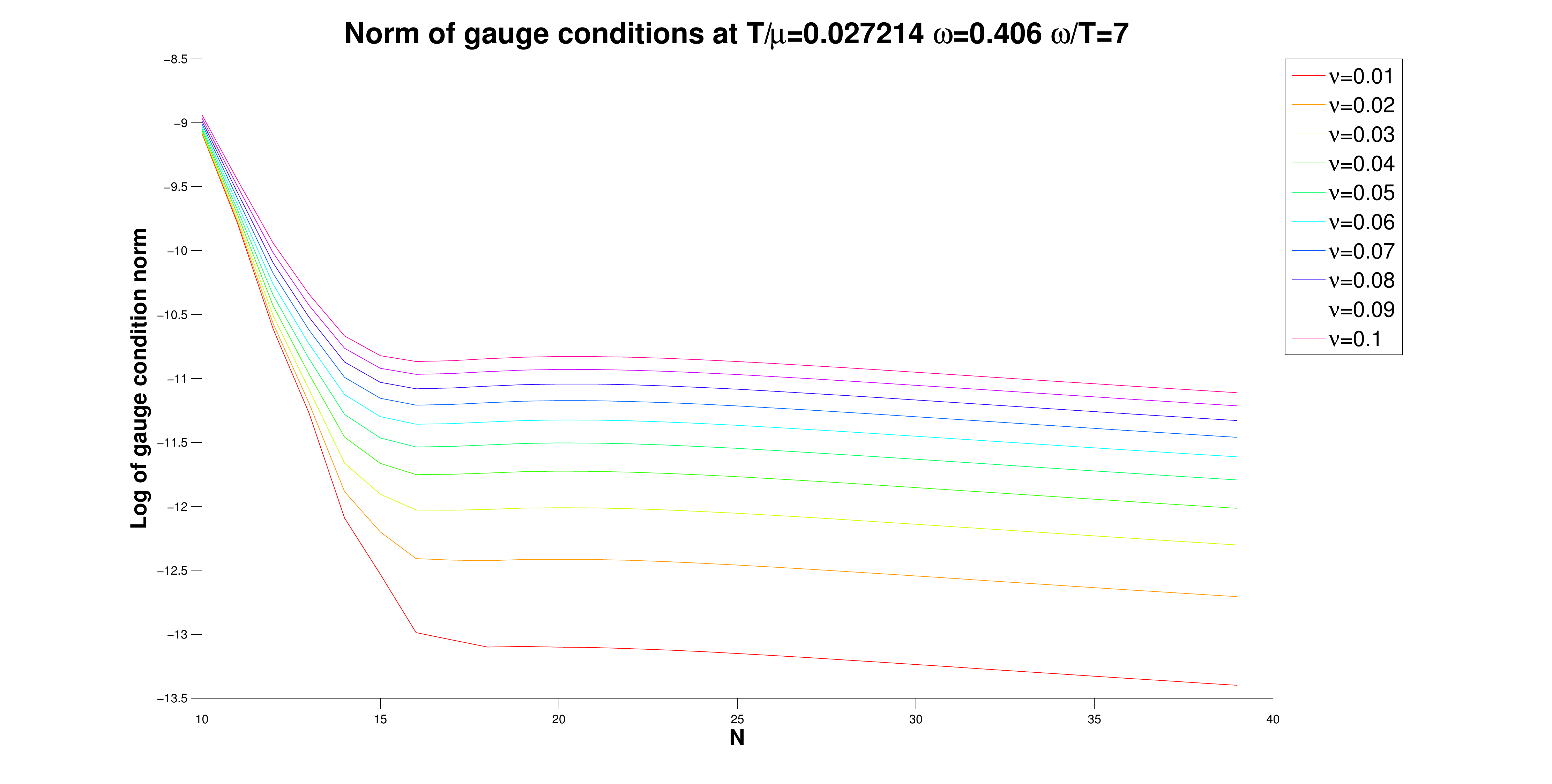}}
\subfigure{\includegraphics[width=0.6\textwidth]{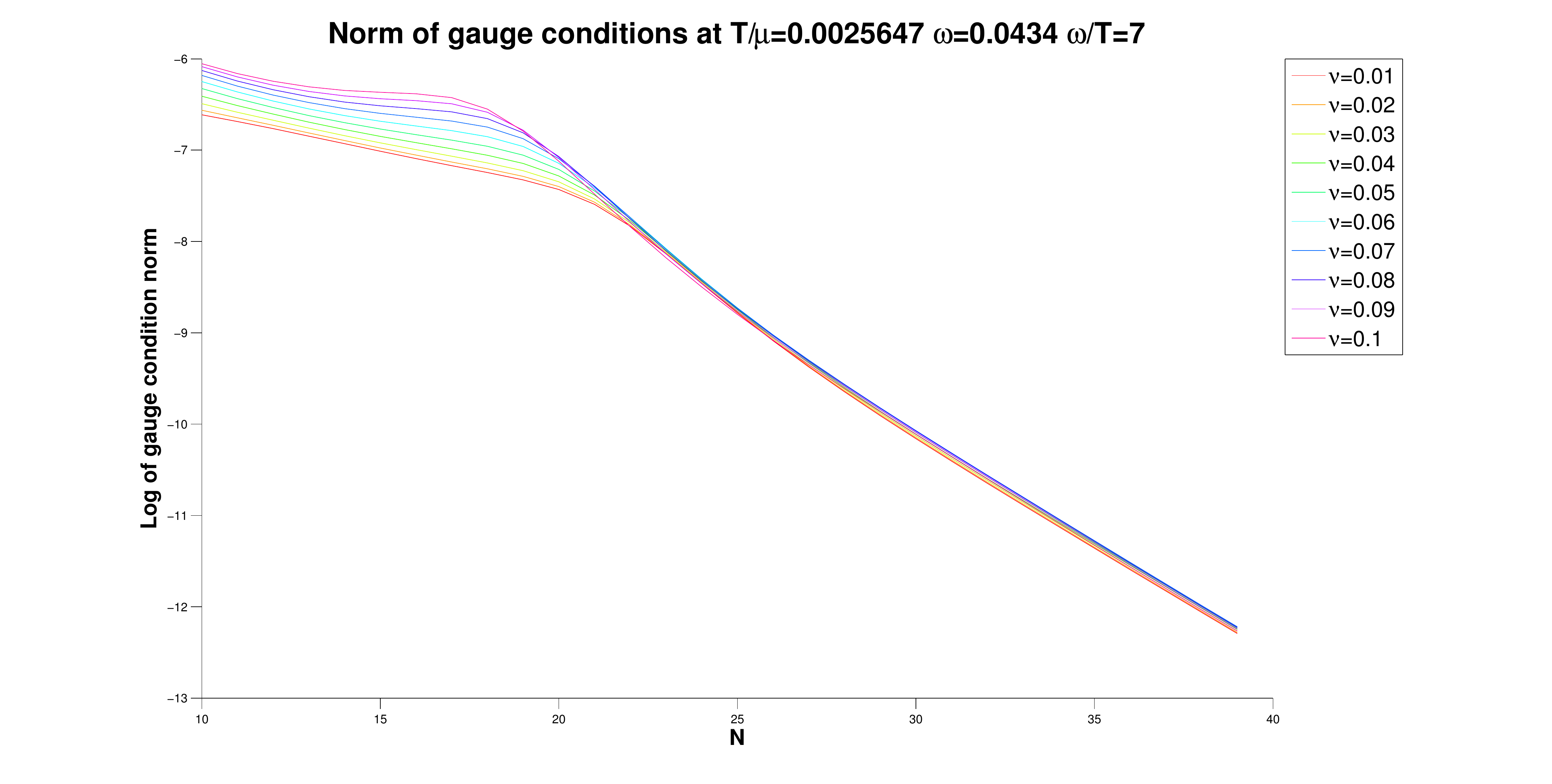}}
}
    \caption[]{Semi-logarithmic plots of the norm of the difference in solutions, and the norm of horizon auxiliary conditions and of the gauge conditions versus $N$ at a low and high temperature for $\frac{w}{T}=7$. We note the unusual dual scaling regimes for the gauge conditions in the higher temperature case. More importantly we note the slow rate of convergence of the auxiliary conditions as the temperature is decreased.}
\label{fig:high_wT_pert_conv_pt1}
\end{figure}  
\begin{figure}
\makebox[\linewidth]{%
\subfigure{\includegraphics[width=0.6\textwidth]{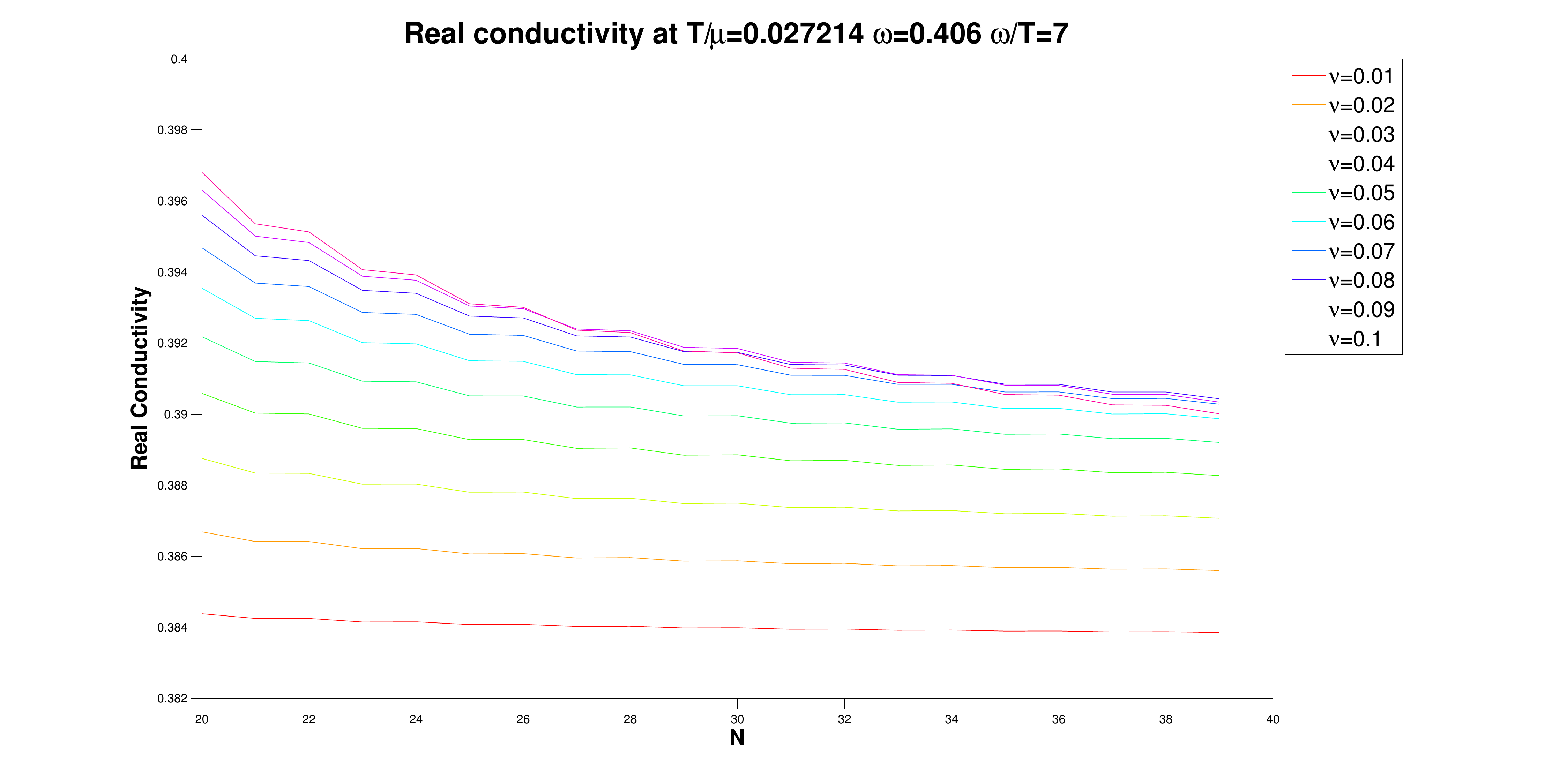}}
\subfigure{\includegraphics[width=0.6\textwidth]{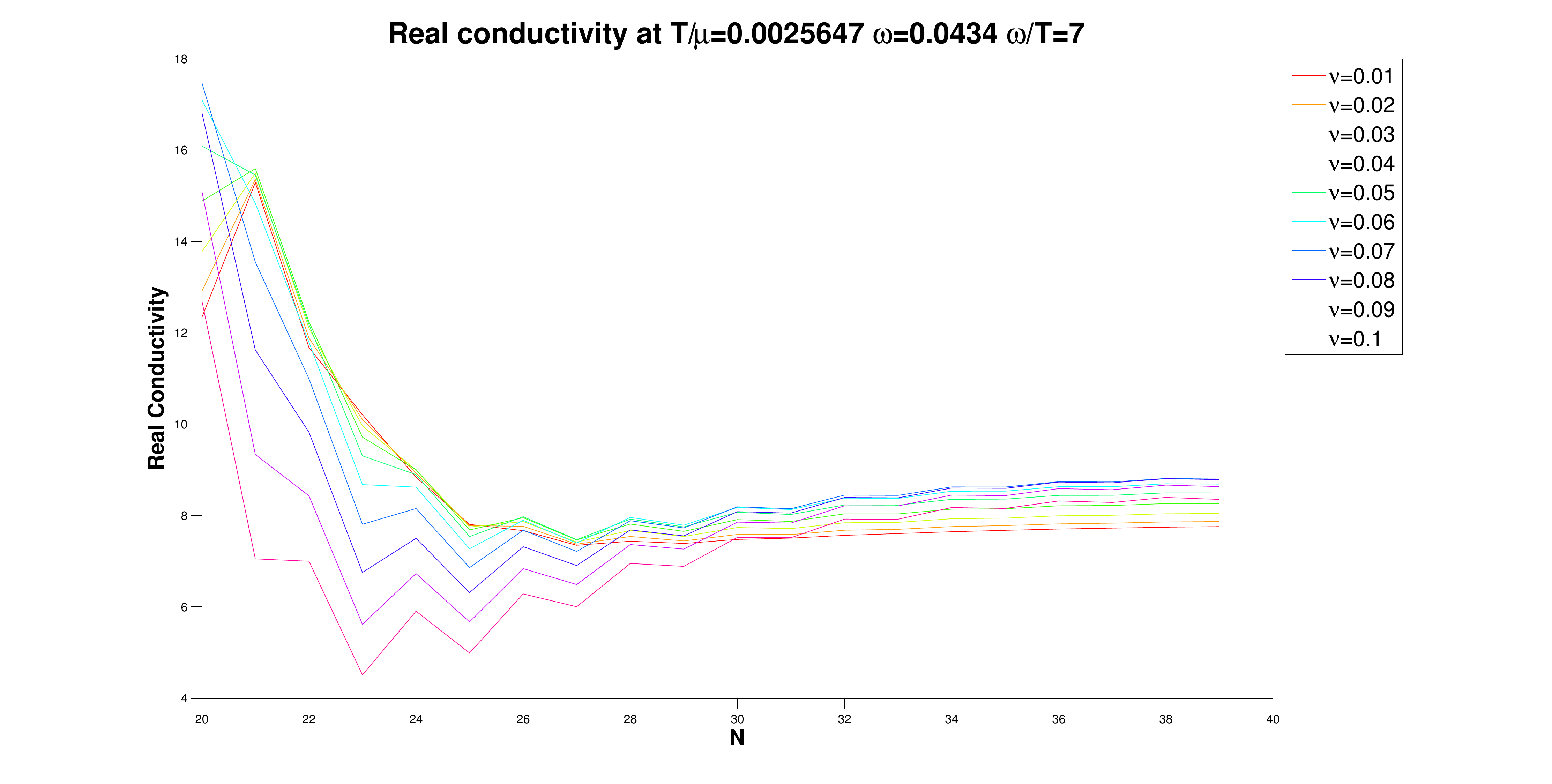}}
}
\makebox[\linewidth]{%
\subfigure{\includegraphics[width=0.6\textwidth]{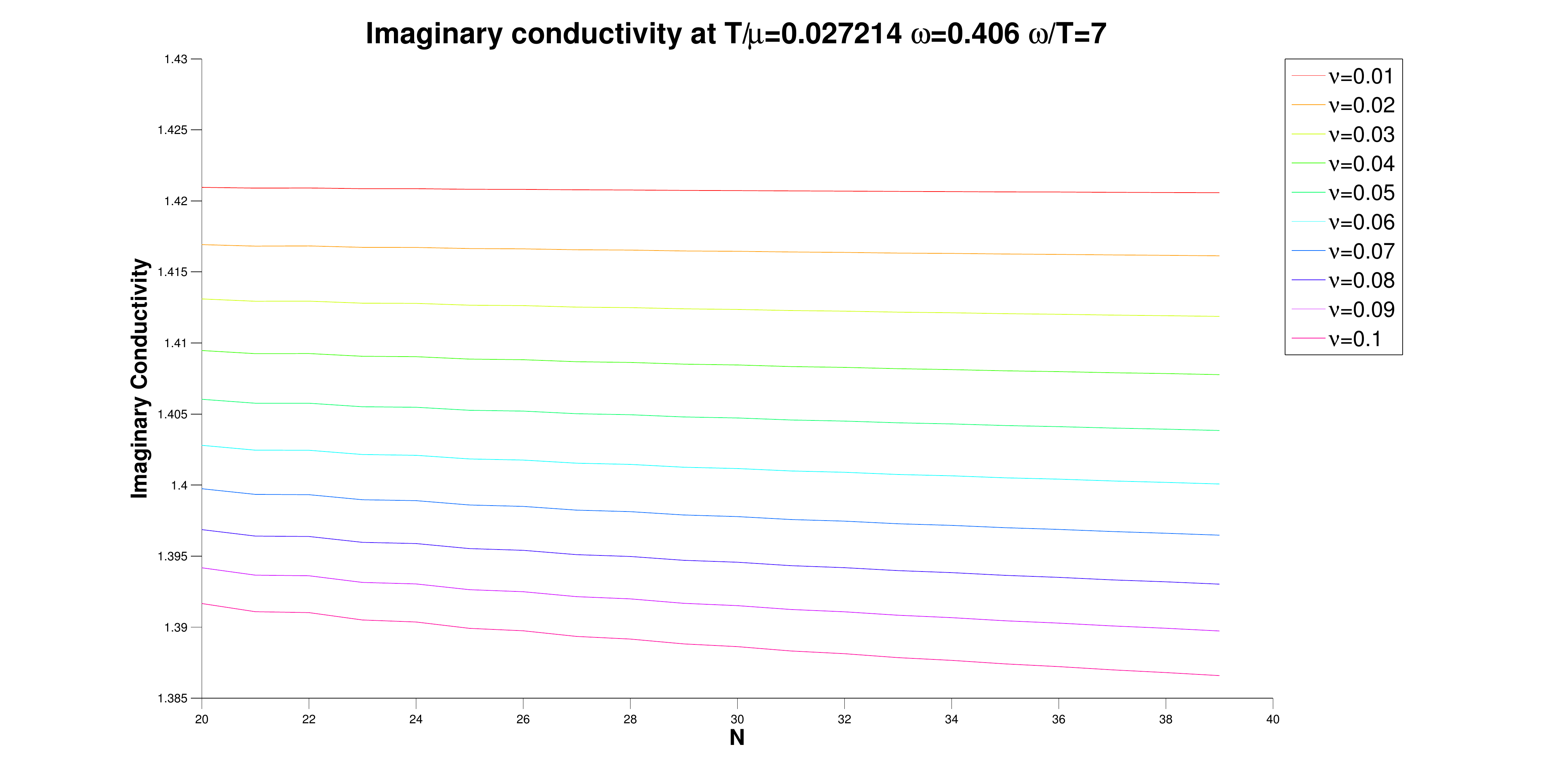}}
\subfigure{\includegraphics[width=0.6\textwidth]{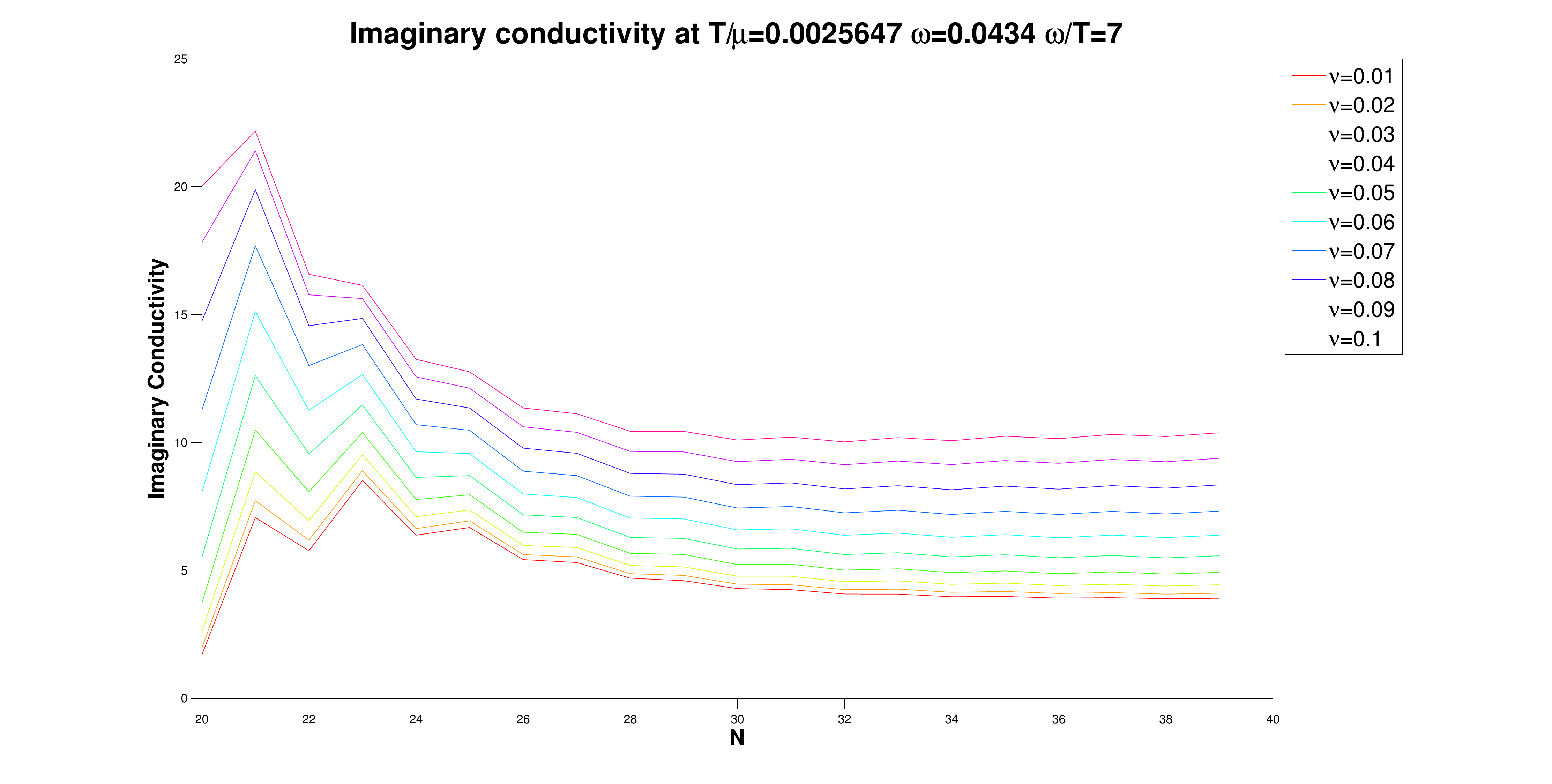}}
}
    \caption[]{Plots of the real and imaginary conductivities versus $N$ for the same choice of parameters as in Fig.~\ref{fig:high_wT_pert_conv_pt1}. Again we note that increasing $\upsilon$ makes the convergence more difficult. In this case the effect is most notable in the real part of the conductivity.}
\label{fig:high_wT_pert_conv_pt2}
\end{figure}  

\begin{figure}
\makebox[\linewidth]{%
\subfigure{\includegraphics[width=0.6\textwidth]{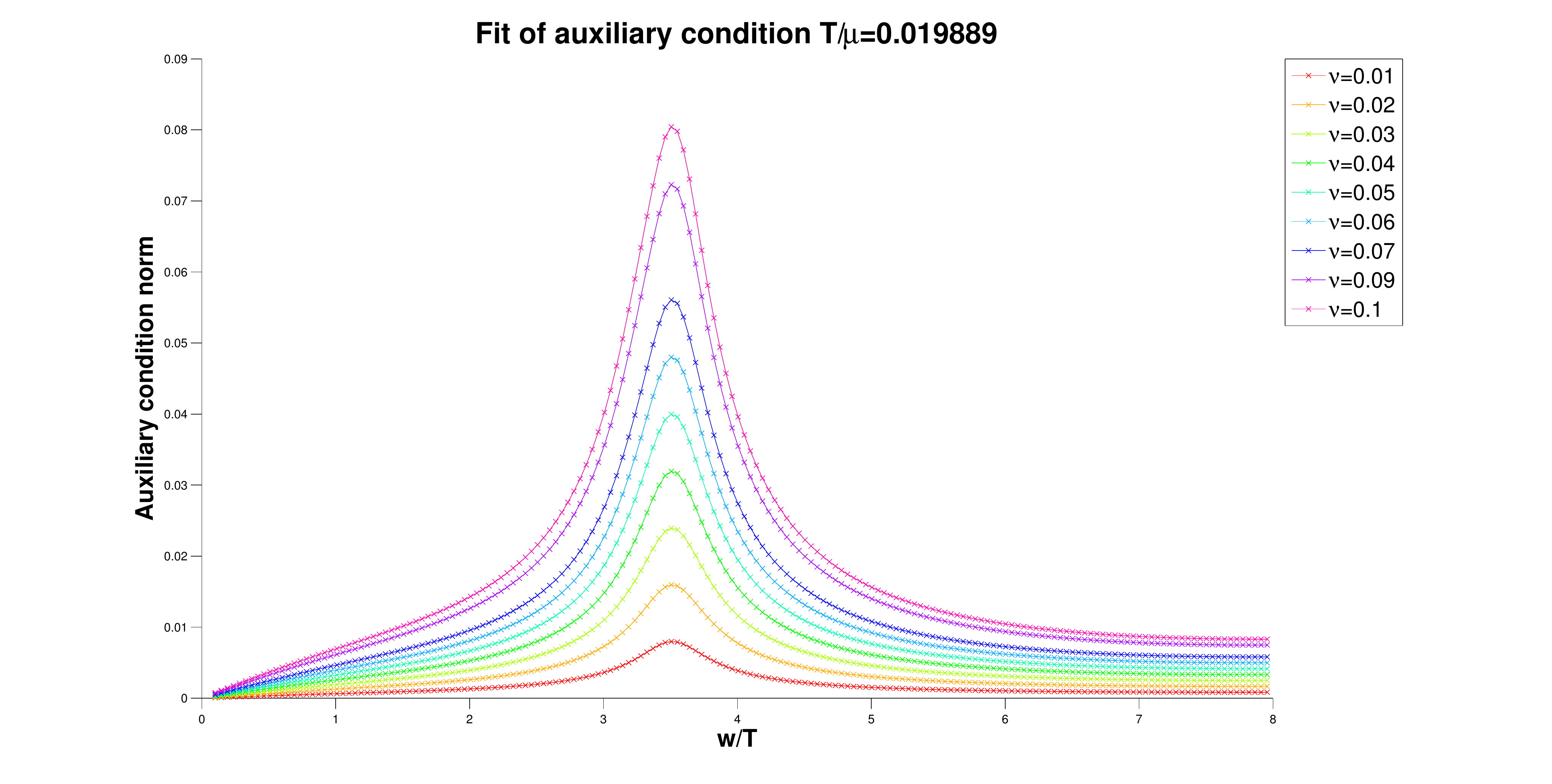}}
\subfigure{\includegraphics[width=0.6\textwidth]{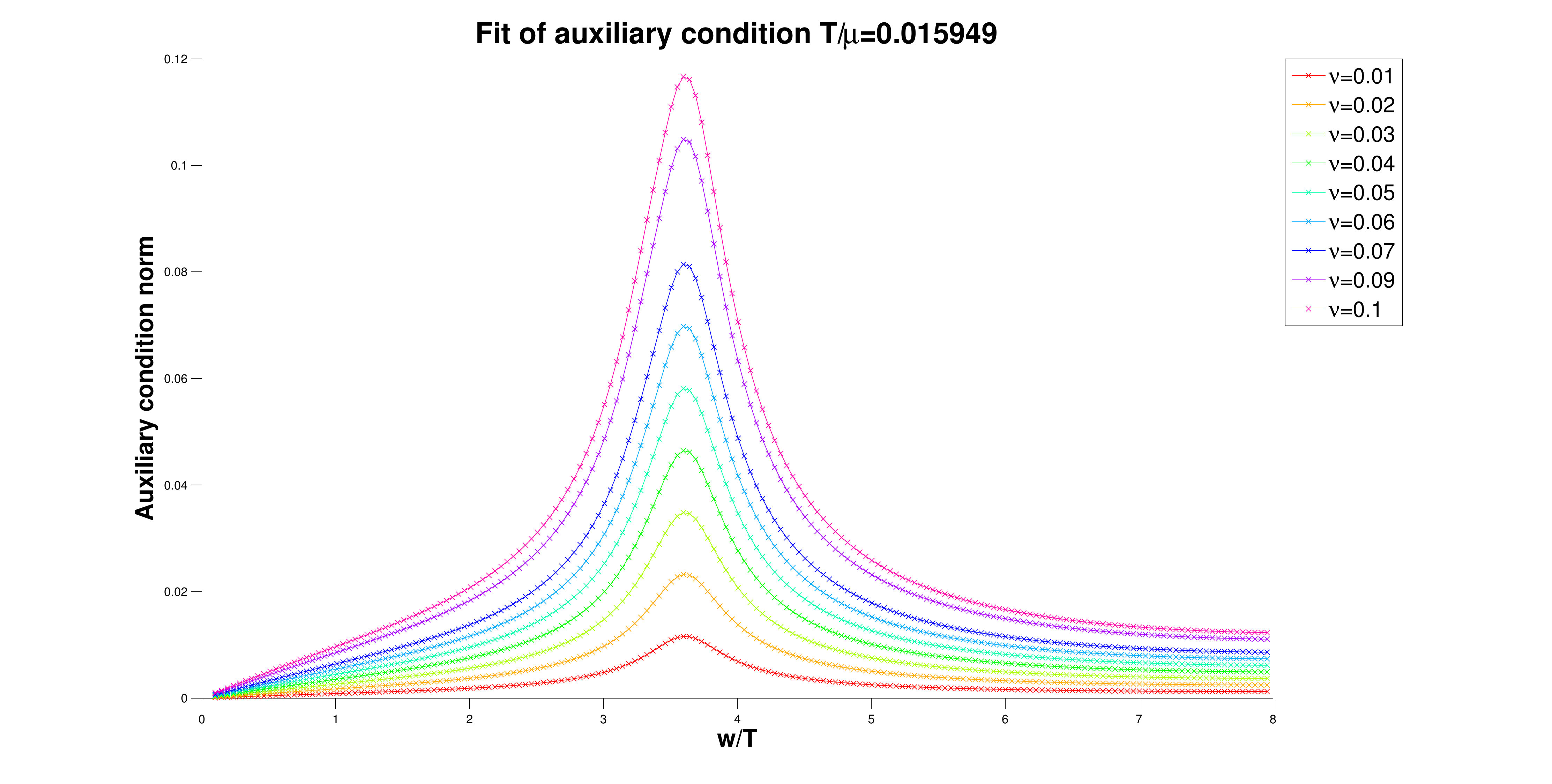}}
}
\makebox[\linewidth]{%
\subfigure{\includegraphics[width=0.6\textwidth]{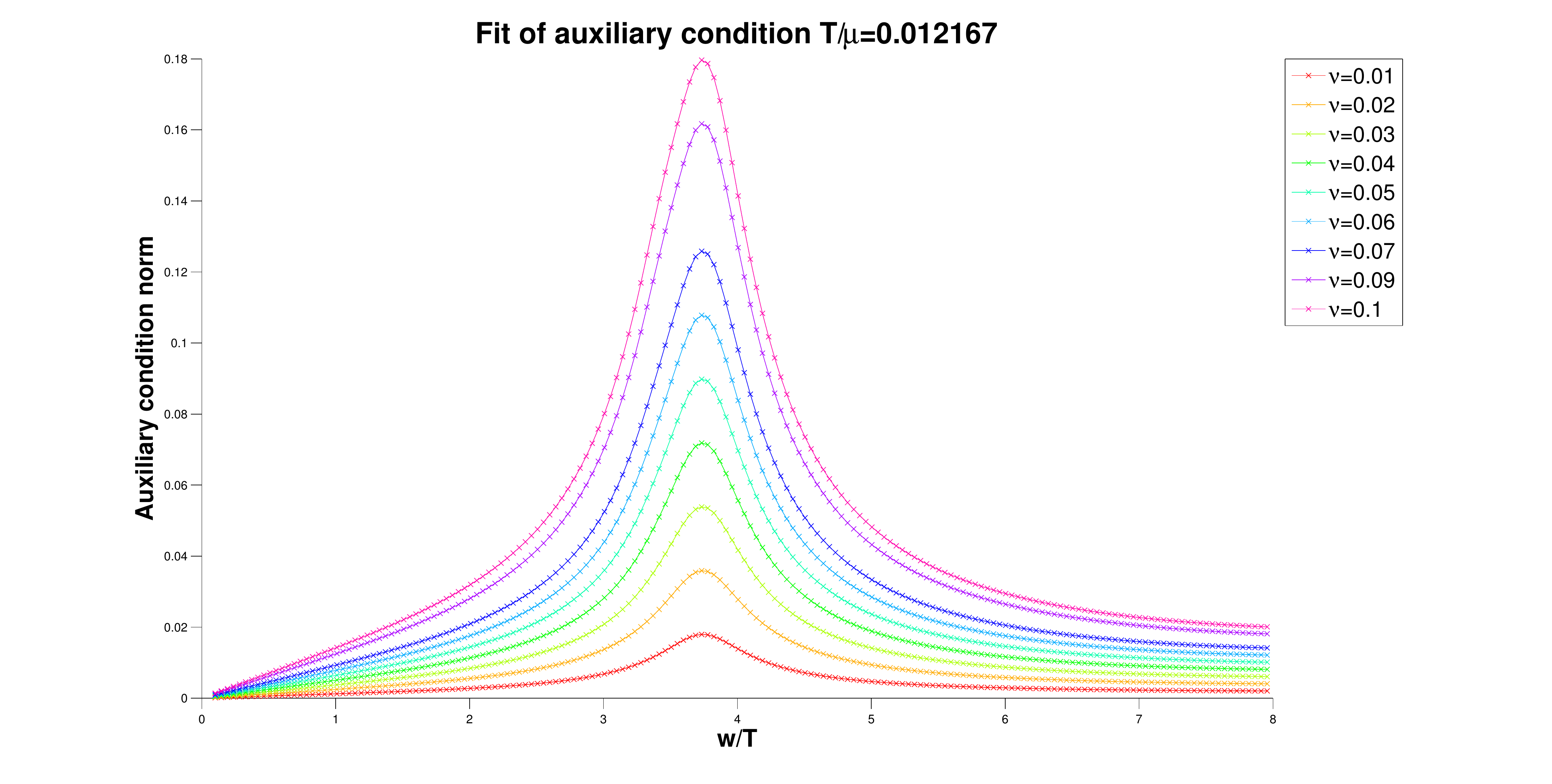}}
\subfigure{\includegraphics[width=0.6\textwidth]{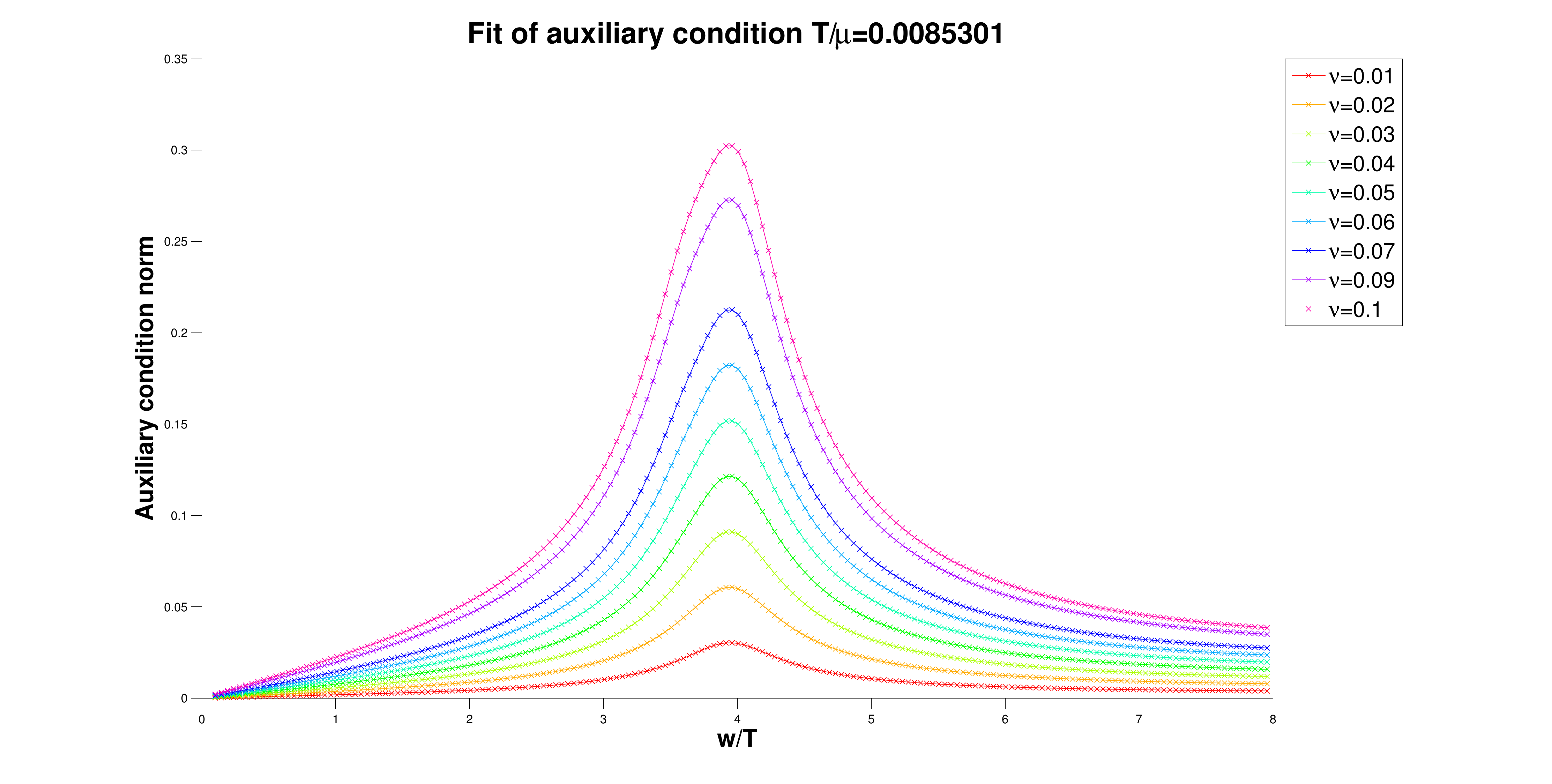}}
}
    \caption[]{The auxiliary horizon constraints corresponding to Fig \ref{fig:IR_scaling_upsilon}. We see that while very well satisfied in the $\mathfrak w \rightarrow 0$ limit there is a regime of significant constraint violation prior to the return of greater accuracy in the mid IR.}
\label{fig:aux_IR_scaling_upsilon}
\end{figure}

\begin{figure}
\makebox[\linewidth]{%
\subfigure{\includegraphics[width=0.6\textwidth]{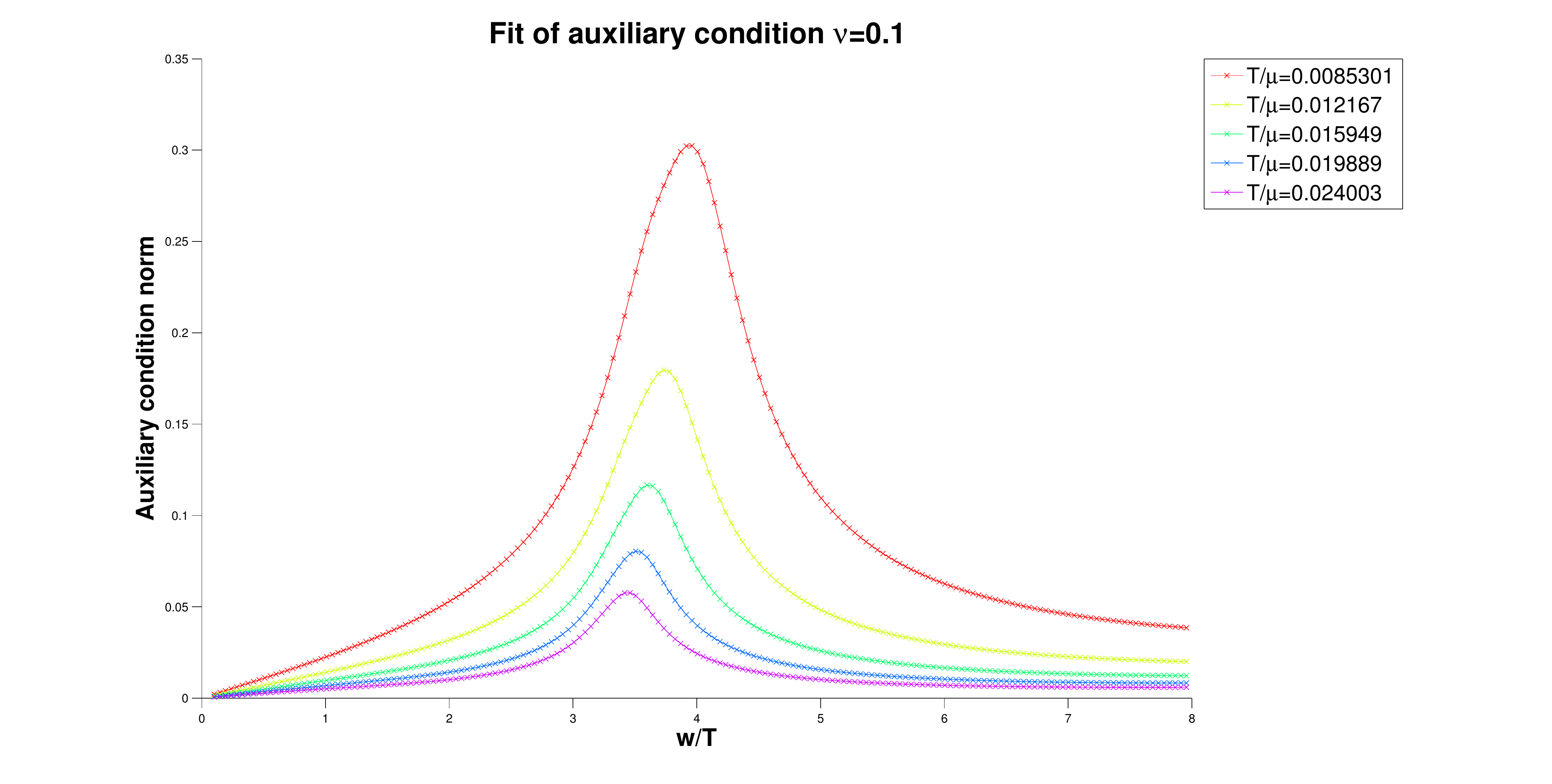}}
\subfigure{\includegraphics[width=0.6\textwidth]{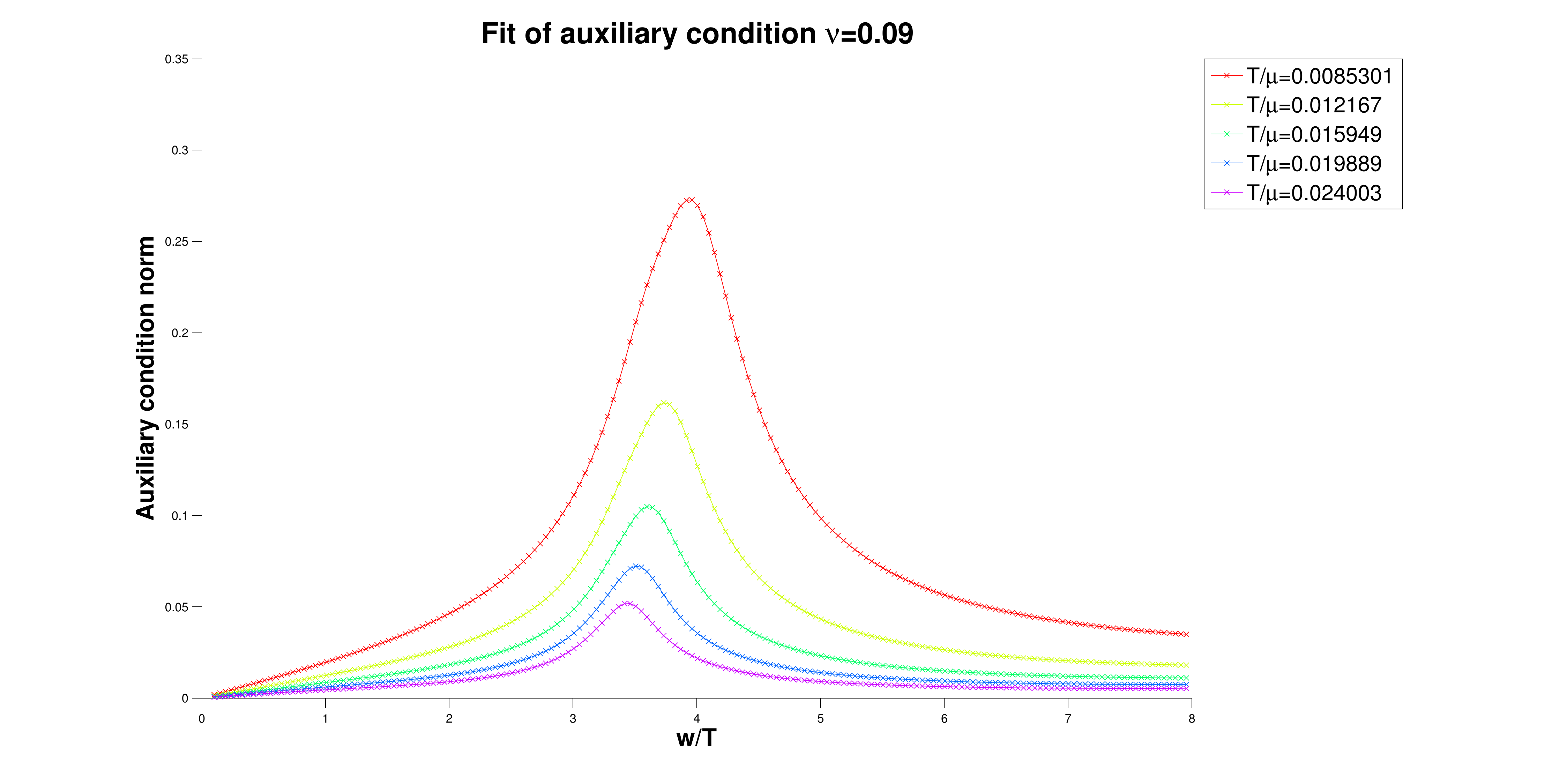}}
}
\makebox[\linewidth]{%
\subfigure{\includegraphics[width=0.6\textwidth]{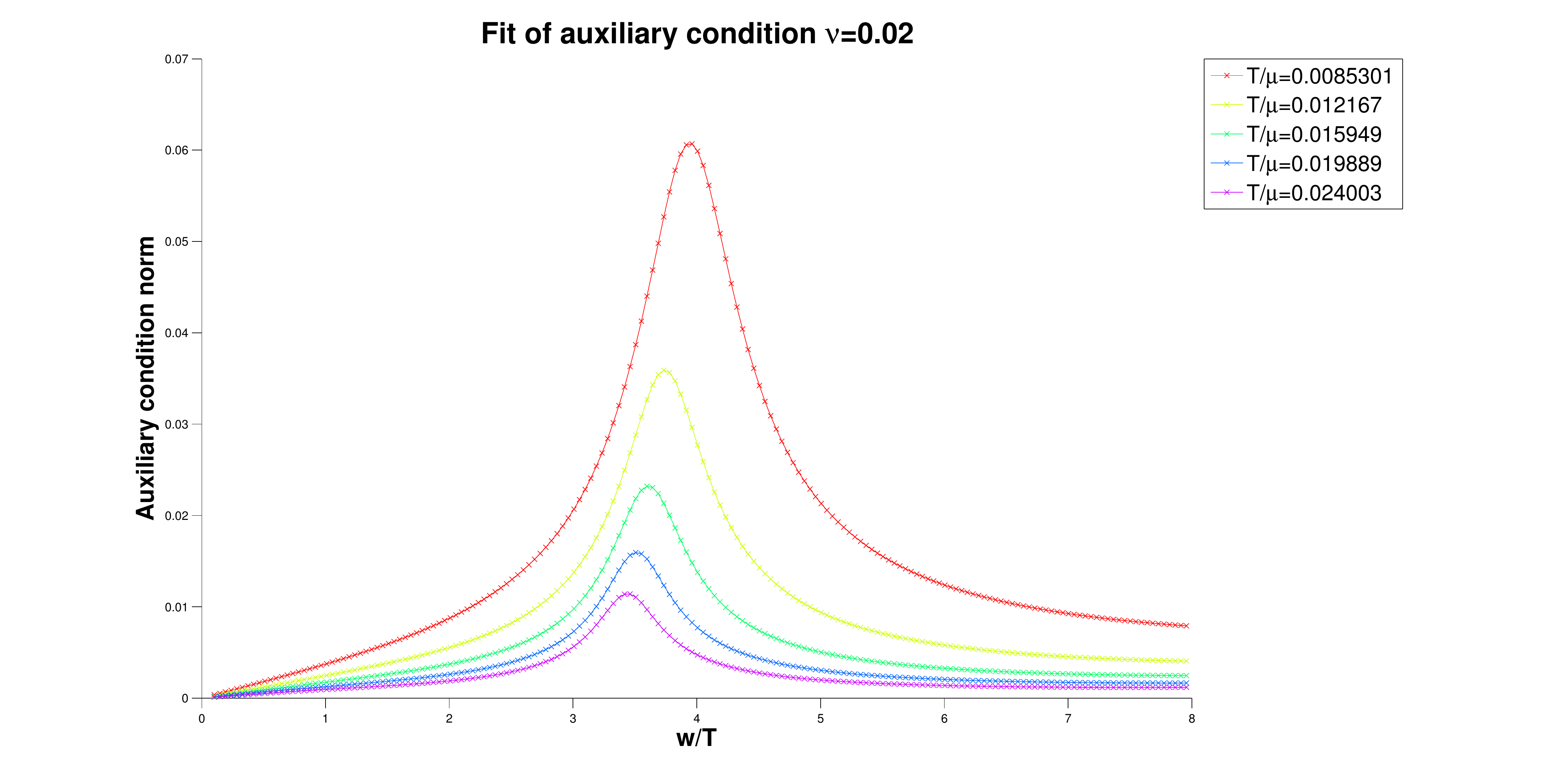}}
\subfigure{\includegraphics[width=0.6\textwidth]{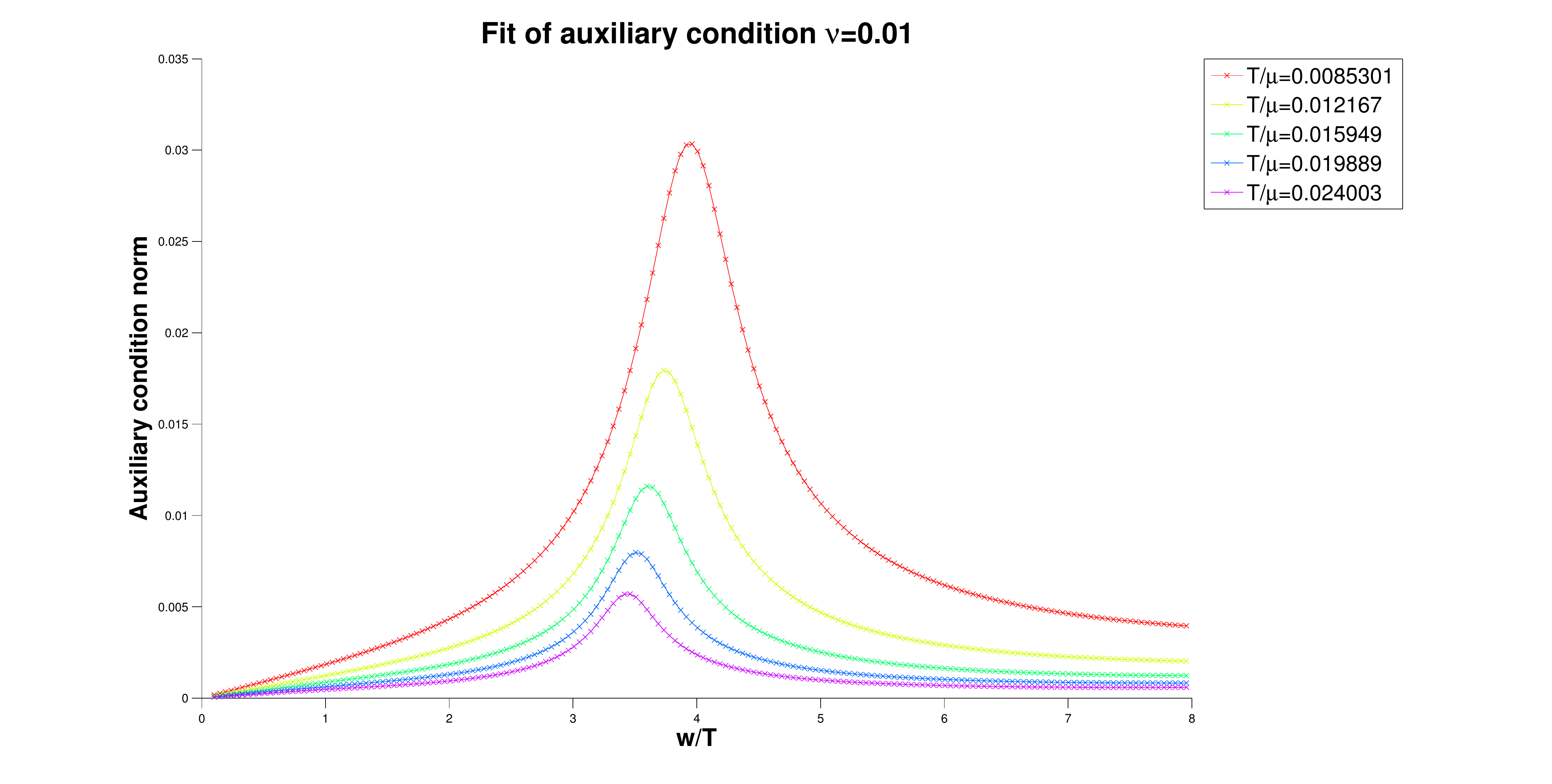}}
}
    \caption[]{The auxiliary horizon constraints corresponding to Fig \ref{fig:IR_scaling_temperature}. Again we observe the non-trivial dependence of the auxiliary constraints on the temperature, $\upsilon$ and $\mathfrak w$ with greater accuracy being obtained at lower and higher values of  $\mathfrak w$.}
\label{fig:aux_IR_scaling_temperature}
\end{figure}

The situation is more complicated in the opposite limit which we examine in Figs.~\ref{fig:high_wT_pert_conv_pt1} and  \ref{fig:high_wT_pert_conv_pt2} where ${\mathfrak w}=7$. We note that while convergence is maintained, the rate of convergence  for the auxiliary horizon constraints becomes problematic as the temperature is lowered. It is instructive to examine the form of these auxiliary constraints corresponding to the results displayed Figs.~\ref{fig:IR_scaling_upsilon} and \ref{fig:IR_scaling_temperature}.This is done in in Figs.~\ref{fig:aux_IR_scaling_upsilon} and \ref{fig:aux_IR_scaling_temperature}. We see that the constraint violation worsens as the temperature is decreased or $\upsilon$ increased, and that the violation is worst in the regime of $3 \sim < \mathfrak w < \sim  5$. It has been observed that as one adjusts parameters further into these regimes numerical artifacts appear in both the imaginary conductivity and the diagnostic function, $F({\mathfrak w})$. However, given that the conditions are under better control for $\mathfrak w >  5$, we believe changed that the results displayed in Figs.~\ref{fig:IR_scaling_upsilon} and \ref{fig:IR_scaling_temperature} are qualitatively correct. This conclusion is supported by the observations that: 
\begin{itemize}
\item As seen in  Figs.~\ref{fig:high_wT_pert_conv_pt1} all quantities,  including the auxiliary horizon constraints, remain in a convergent regime down to temperatures significantly below the lowest used in the construction of Figs.~\ref{fig:IR_scaling_upsilon} and \ref{fig:IR_scaling_temperature}.
\item Even at these lower temperatures the physical variables in which we are interested (the real and imaginary parts of the conductivity) exhibit sensitivity to $N$ only up to a small percentage of their magnitudes. This can be seen in Fig.~\ref{fig:high_wT_pert_conv_pt2} by comparing the degree of fluctuation at larger values of $N$ to the scale of the y-axis.
\end{itemize}

\begin{figure}
\makebox[\linewidth]{%
\subfigure{\includegraphics[width=0.6\textwidth]{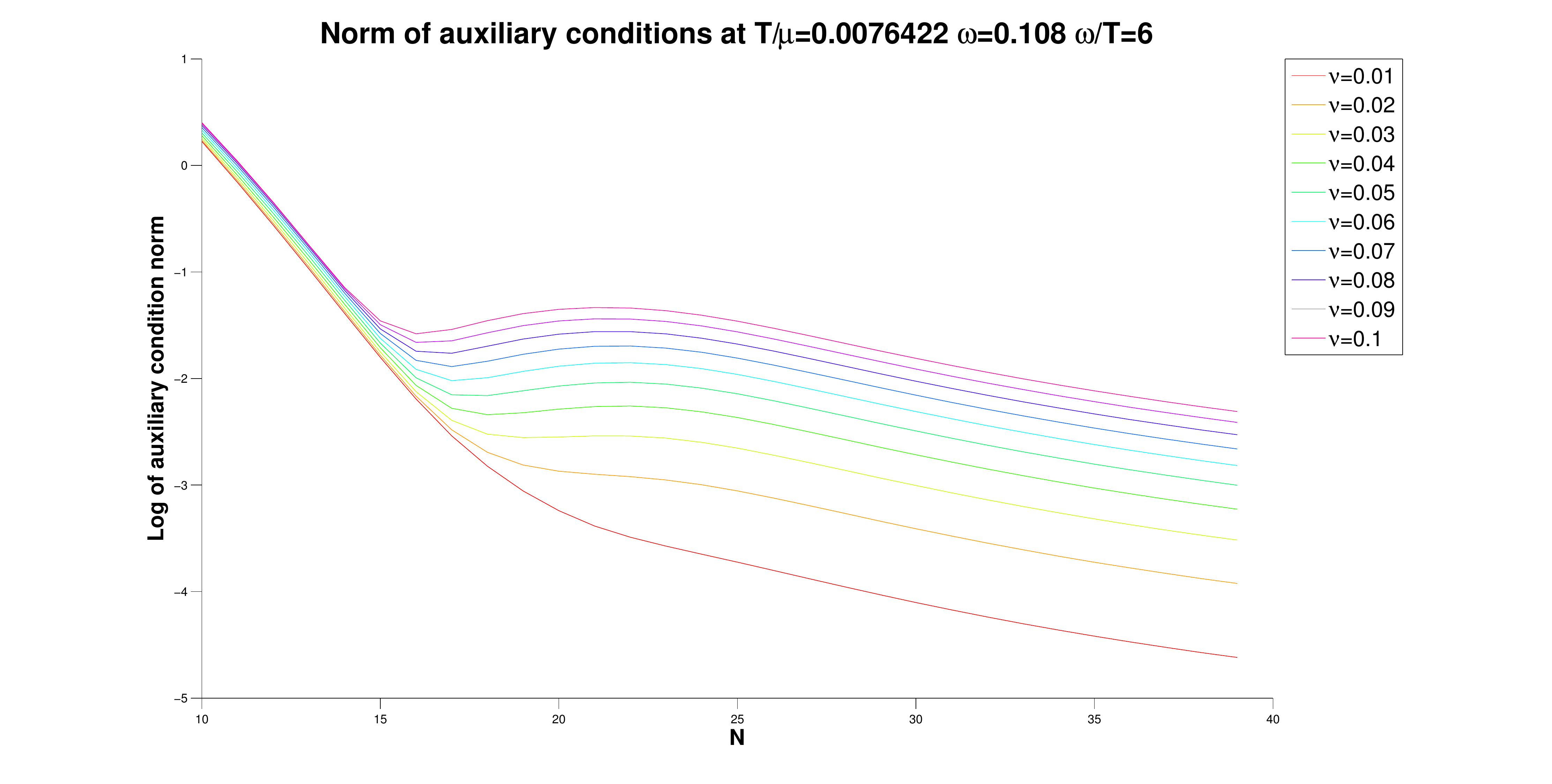}}
\subfigure{\includegraphics[width=0.6\textwidth]{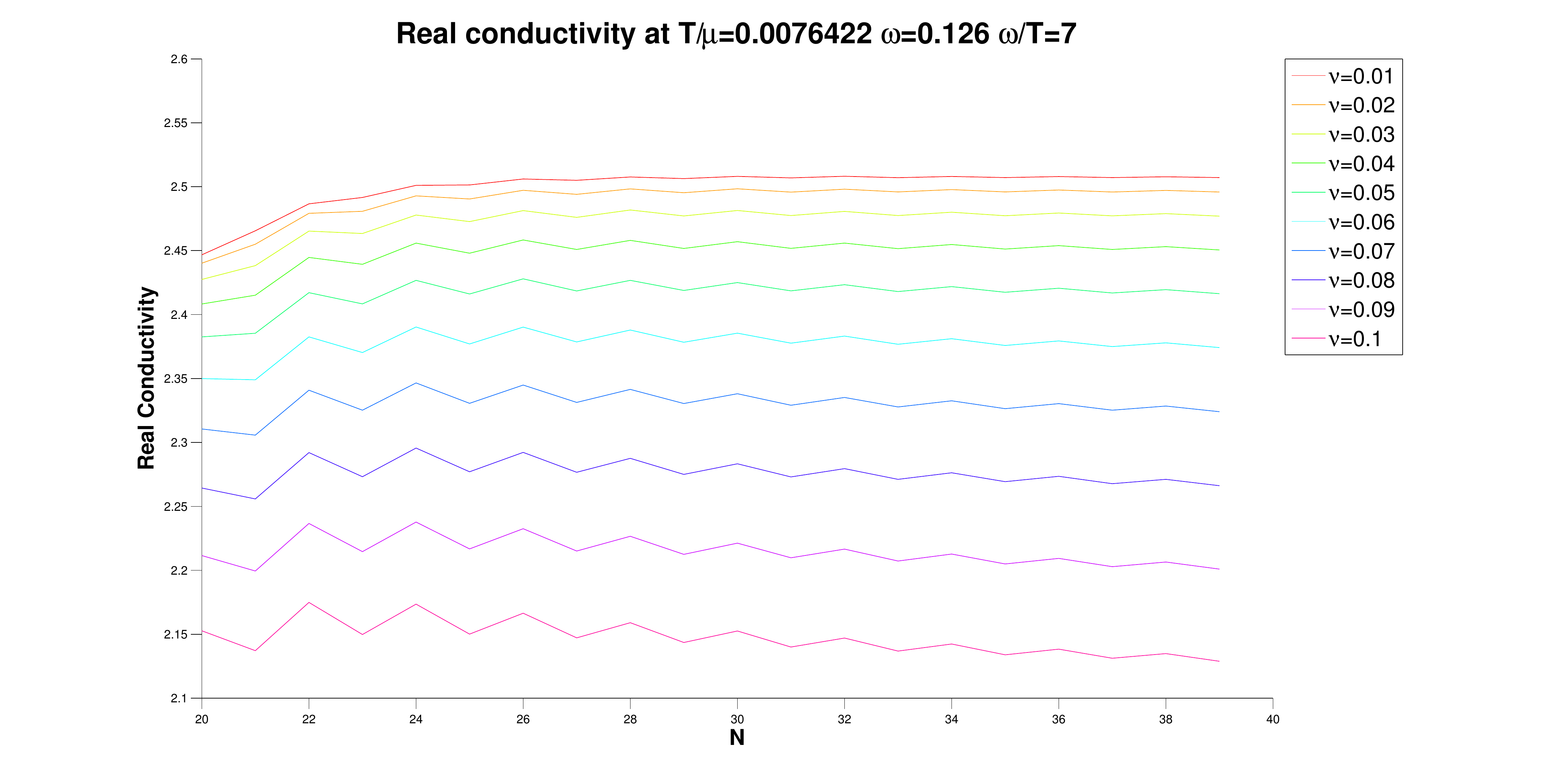}}
}
\makebox[\linewidth]{%
\subfigure{\includegraphics[width=0.6\textwidth]{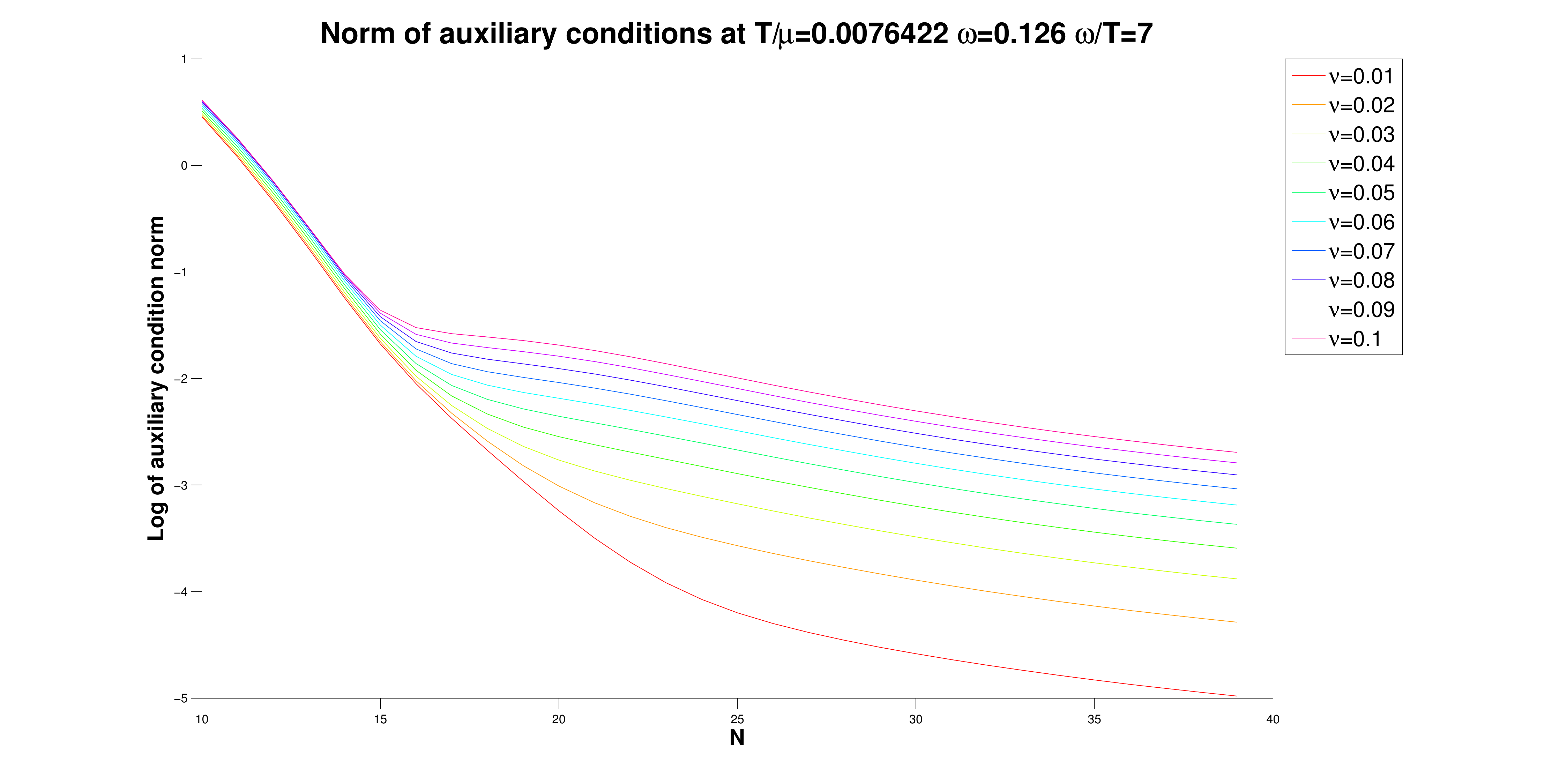}}
\subfigure{\includegraphics[width=0.6\textwidth]{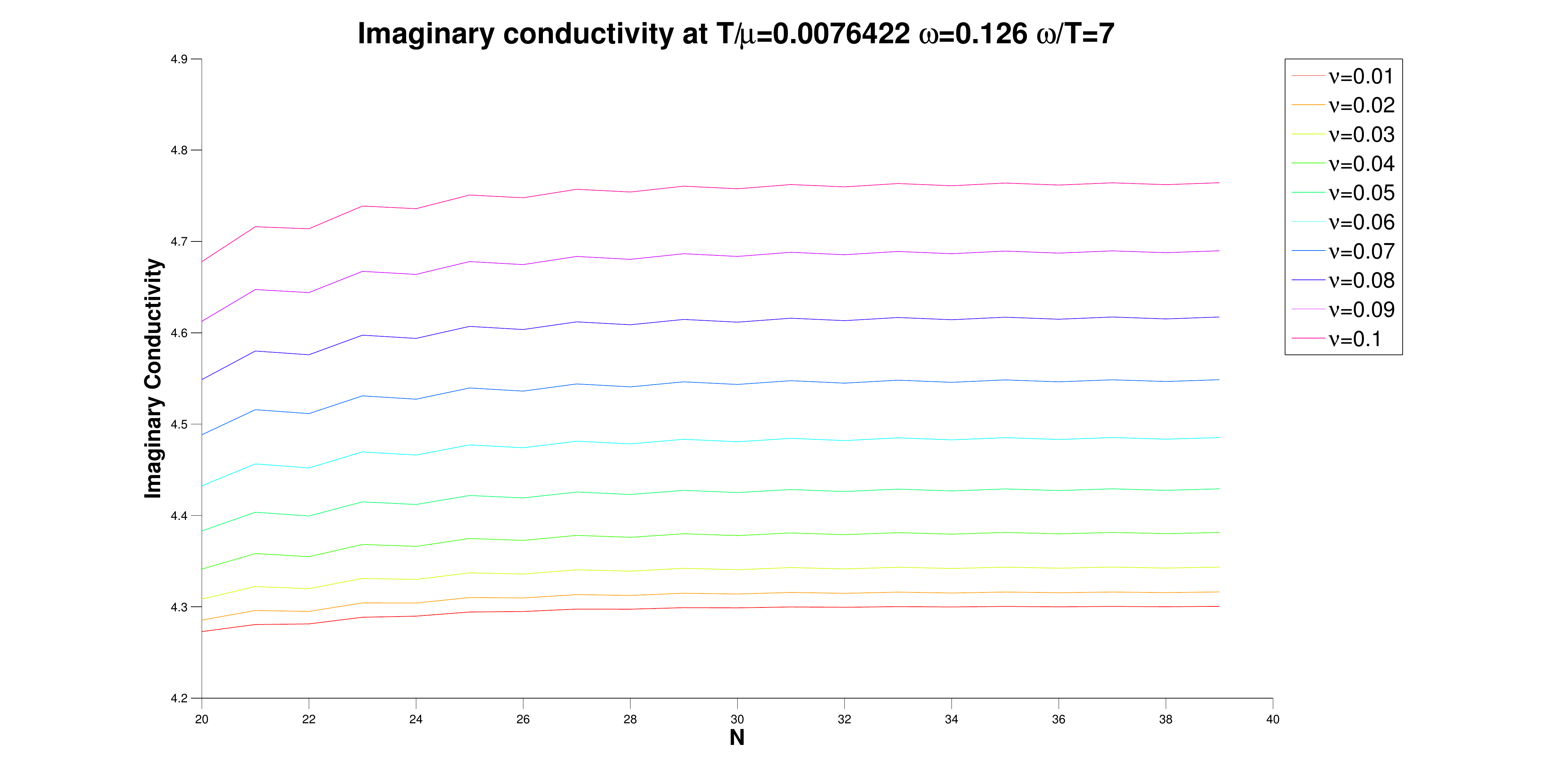}}
}
    \caption[]{An illustration that the auxiliary horizon constraints are well within a convergent regime at temperatures below those used in Figs.~\ref{fig:IR_scaling_upsilon} and \ref{fig:IR_scaling_temperature}. It is also evident that the real and imaginary parts of the conductivity are well converged relative to their magnitudes at these temperatures.}
\label{fig:final_mid_IR_scaling_check}
\end{figure}

As a final illustration of numerical control in the mid-IR regime we plot, in Figs.~\ref{fig:final_mid_IR_scaling_check}, the real and imaginary conductivity, and auxiliary horizon constraints at a temperature slightly below the lowest temperature used in Figs.~\ref{fig:IR_scaling_upsilon} and \ref{fig:IR_scaling_temperature}. We see that while the behaviour is qualitatively similar to the lower temperatures we have just described, the convergence rate of the auxiliary horizon constraints has improved substantially. In addition is is clear that the real and imaginary parts of the conductivity have effectively become independent of $N$ by the time we reach resolutions comparable to those used in our simulations ($\sim N=40 $). Future work in this direction may involve the use of alternative numerical methods (for example finite difference discretization) to tackle these numerically difficult regimes.

  \subsection{Fit to Drude Form}
\label{sec:drude}

\begin{figure}
\makebox[\linewidth]{%
\subfigure{\includegraphics[width=0.6\textwidth]{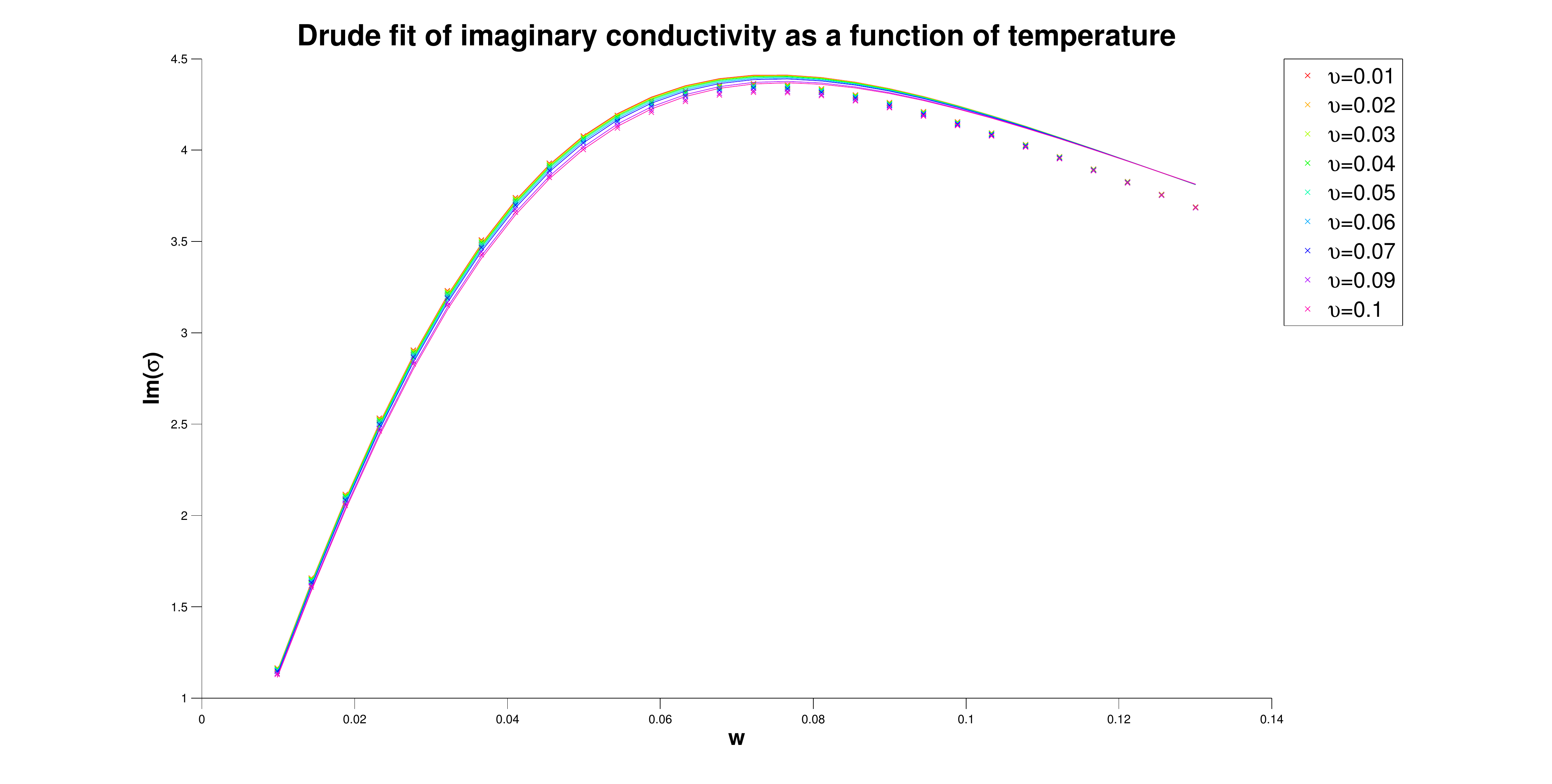}}
\subfigure{\includegraphics[width=0.6\textwidth]{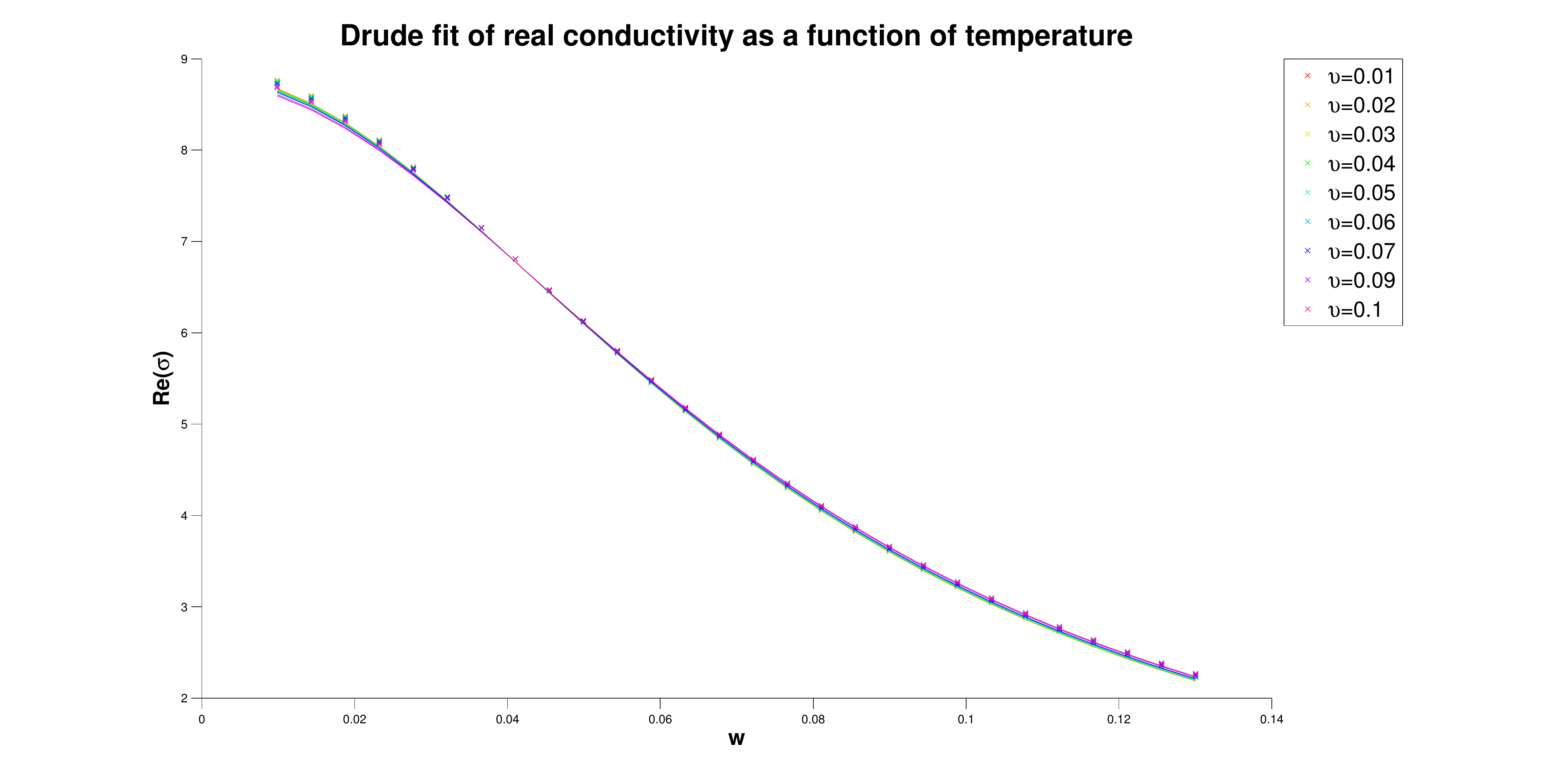}}
}
\makebox[\linewidth]{%
\subfigure{\includegraphics[width=0.6\textwidth]{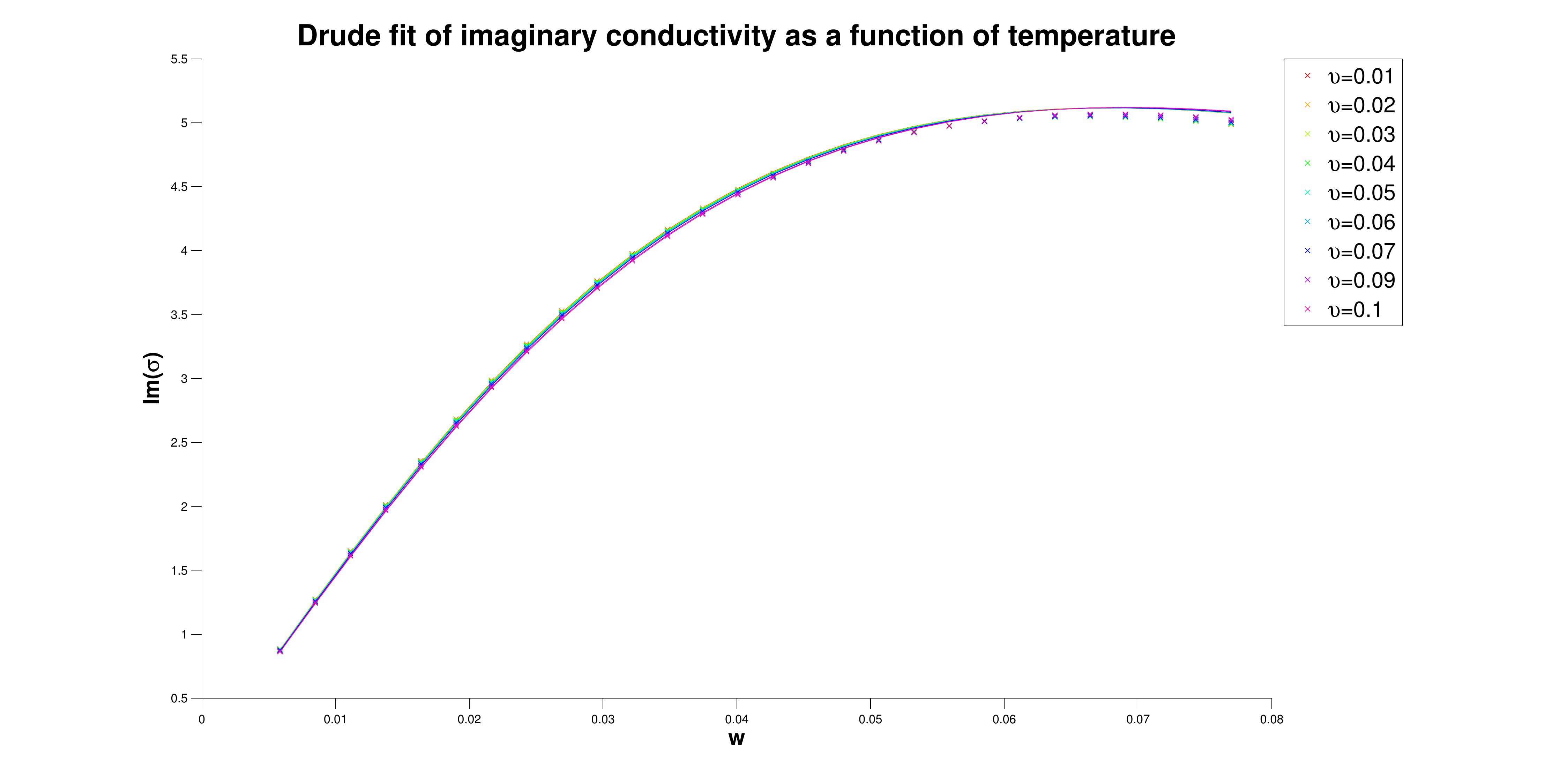}}
\subfigure{\includegraphics[width=0.6\textwidth]{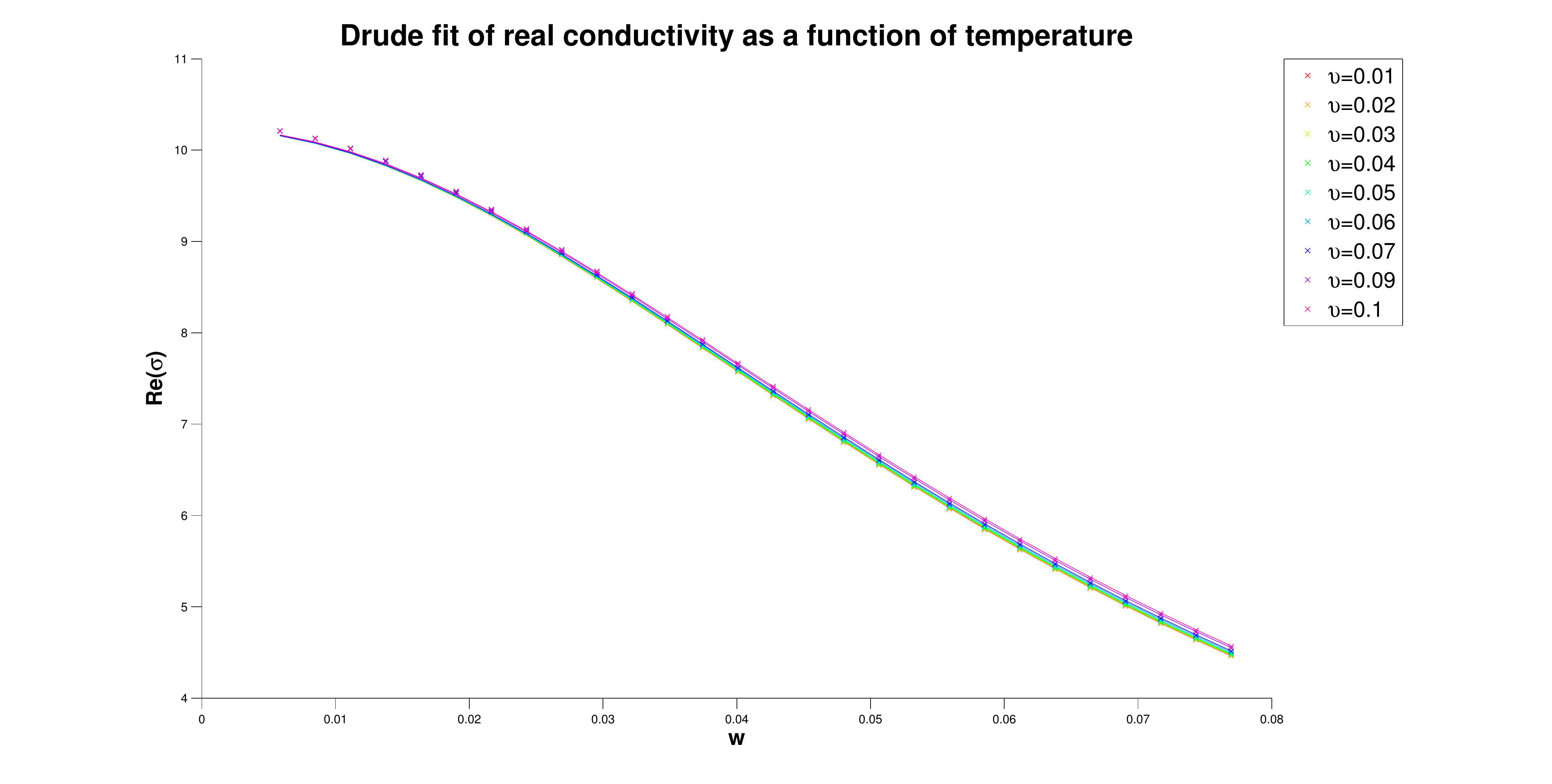}}
}
    \caption[]{The fits of the Drude parameters $\tau, K$ to the AC electric conductivities for two moderate values of $\frac{T}{\mu}$, corresponding to those listed in \autoref{table:goodness_fit}, for $0.01 \geq \upsilon \leq 0.1$ . We see that the fits are good, particularly for lower values of the $\omega$ where the convergence behaviour of the perturbation equations, and associated gauge and auxiliary conditions is best. In both cases the maximum $\frac{\omega}{T}$ values plotted are $\frac{\omega}{T} \sim 1.25$.}
\label{fig:Drude_fit}
\end{figure}

 Another useful test, described in \cite{Horowitz:2012ky}, is comparing  the low frequency behaviour of the AC conductivity to the Drude form of the conductivity, expected on general grounds \cite{Hartnoll:2012rj}:
\begin{align}
\sigma(\omega)=\frac{K\,  \tau}{1-i \,\omega \, \tau}
\end{align} 
where $K$ is a constant. This test is useful only in the metallic phase or above the critical temperature for the onset of the insulating phase. Therefore at moderate values of $\frac{T}{\mu}$  the AC conductivity is well modelled by the Drude behaviour as demonstrated by the examples in Fig.~\ref{fig:Drude_fit}. This may be interpreted as an important additional test of the numerics- the quality of the fit to the Drude form being indicative of good convergence of the perturbation equations over the range of $\omega$ examined.

An additional important test of the Drude behaviour is that the coefficient, $K$, should match with the residue of the zero frequency pole in the imaginary conductivity obtained in the homogeneous case:
\begin{align}
\mathrm{Im} \left(\sigma_\text{hom}(\omega) \right) \rightarrow \frac{K}{\omega}, \quad \text{as} \quad \omega \rightarrow 0
\end{align} 
In order to perform this calculation we solved the ODEs which govern the homogeneous phases of this model at the same point in parameter space as our PDEs \eqref{eq:background_eqns} and extracted the residue of the pole. These ODEs may be derived from the (gauged) PDEs via the following substitution:
\begin{align}
& Q_{xz}(z,x) \rightarrow 0,  \quad Q_{zz}(z,x) \rightarrow p_1(z),  \quad Q_{tt}(z,x) \rightarrow p_2(z),  
\\ \nonumber 
& Q_{yy}(z,x) \rightarrow Q_{xx}(z,x) \rightarrow p_3(z) \quad A_0(z,x) \rightarrow A_0(z), \quad \Phi(z,x) \rightarrow \Phi(z)  
\end{align}   
Once the ODEs have been derived the horizon conditions can again be obtained in the same manner described in  \S\ref{sec:section1} for the PDEs. In this case these boundary conditions relate the value of the fields and their first radial derivative at the horizon. The conformal boundary conditions can also be written as a simple mixture of Dirichlet and Neumann conditions as in the PDE case.
\begin{align}
p_1(0)=1, \quad p_2(0)=1, \quad p_3(0)=1, \quad \Phi'(0)=A, \quad A_0(0)=\mu
\end{align}   

The ODEs were solved via a simple adaption of the Chebychev grid spectral technique described in  
\S\ref{subsec:num_results}. The results also provided an additional sanity check on the solutions of PDE background equations \eqref{eq:background_eqns}. When these equations were solved with the sourced inhomogeneity turned off, $\Phi_{1}(x)=A$, good agreement was obtained with the ODE solutions. 

Considering the linearized perturbation equations around these ODE backgrounds we find that it is sufficient to restrict ourselves to perturbations of the following form:
\begin{align}
h_{xt}= \tilde{h}_{xt}(z)  \, e^{-i \,\omega\, t}, \quad b_{x}=\tilde{b}_{x} \, e^{-i \,\omega\, t}  
\end{align}   
Utilizing the background equations of motion it can be checked that there are two independent equations of motion which can conveniently be decoupled from each other and written as: (i) a first order equation for $\tilde{h}_{xt}$ in terms of the background fields, and (ii) a second order equation for $\tilde{b}_{x}$. We are only concerned with the equation for $\tilde{b}_{x}$. Boundary conditions at the conformal boundary consist of $\tilde{b}_{x}(0)=1$ corresponding to a conveniently normalized external electric field perturbation in the dual QFT. At the horizon regularity in ingoing coordinates necessitates the following leading scaling of the field 
$\tilde{b}_{x}= {\cal P}(z)\, \tilde{b}^{reg}_{x}$. A suitable horizon boundary condition may then be obtained via expanding $\tilde{b}^{reg}_{x}$ to leading order in $(1-z)$. 

Once the solutions of the ODE equations were obtained we extracted the coefficient of the pole and calculated a goodness of fit measure given by $\left(1-\frac{K_\text{hom}}{K_\text{lat}}\right)$. The relevant results are displayed in percentage form in Table~\ref{table:goodness_fit}.  $K_\text{lat}$ is the result obtained for the inhomogeneous sources. 

%
\begin{table}[h]
\centering
\begin{tabular}{ | l | l | l | l | l | l | l | l | l | l |}
 \hline
    $\upsilon$ & $0.01$ & $0.02$ & $0.03$ & $0.04$ & $0.05$ & $0.06$ & $0.07$ & $0.09$ & $0.1$ \\ 
\hline \hline
    $\tfrac{T}{\mu}$=0.052109 & 0.57 & 0.26 & 1.08 & 1.89 & 2.69 & 3.48 & 4.26 & 5.78 & 6.53 \\ \hline
    $\tfrac{T}{\mu}$=0.027214 & 2.67 & 3.53 & 4.36 & 5.16 & 5.93 & 6.67 & 7.38 & 8.7 & 9.32\\ \hline
\end{tabular}
\caption{Testing the fit to the Drude form of the conductivity and checking agreement between the homogeneous and inhomogeneous solutions.}    
    \label{table:goodness_fit}
\end{table}    
 %

 \newpage
 
\bibliographystyle{abbrv}
\bibliography{/Users/darrensmyth/Google_Drive_shared/Google_Drive/Thesis_writing/Thesis_attempts/myrefs_library}

\end{document}